\documentclass[fleqn,usenatbib]{mnras}

\usepackage{newtxtext,newtxmath}
\usepackage[T1]{fontenc}

\DeclareRobustCommand{\VAN}[3]{#2}
\let\VANthebibliography\thebibliography
\def\thebibliography{\DeclareRobustCommand{\VAN}[3]{##3}\VANthebibliography}

\usepackage{graphicx}	% Including figure files
\usepackage{amsmath}	% Advanced maths commands

\newcommand{\SNRfone}{\text{S/N}$_{f_{\rm 1}}$}

\title[Asteroseismology of SPB stars]{Asteroseismology of SPB stars: a comparison of forward asteroseismic modelling results from {\it Kepler} and TESS}

\author[L. J. A. Scott \& D. M. Bowman]{
L. J. A. Scott$^{1}$\thanks{E-mail: laura.scott@newcastle.ac.uk} and
D. M. Bowman$^{1,2}$
\\
$^{1}$School of Mathematics, Statistics and Physics, Newcastle University, Newcastle upon Tyne, NE1 7RU, UK\\
$^{2}$Institute of Astronomy, KU Leuven, Celestijnenlaan 200D, B-3001 Leuven, Belgium
}

\date{Accepted XXX. Received YYY; in original form ZZZ}

\pubyear{\the\year{}}

\begin{document}
\label{firstpage}
\pagerange{\pageref{firstpage}--\pageref{lastpage}}
\maketitle

\begin{abstract}
The slowly pulsating B (SPB) stars are a class of variable star with masses between about 3 and 8~M$_{\odot}$. Their gravity-mode pulsation frequencies are sensitive to the near-core structure, which makes them useful probes of rotation and mixing in the deep stellar interior. Time series photometry, such as from the {\it Kepler} and TESS space telescopes, allows the extraction of their pulsation frequencies and construction of period spacing patterns. Previously, samples of slowly pulsating B stars were observed by the {\it Kepler} mission and underwent forward asteroseismic modelling to retrieve stellar parameters such as mass, age and core mass. However, all of these stars have since been re-observed by the ongoing TESS mission with light curves that are usually shorter and non-continuous, resulting in more difficult frequency extraction and interpretation in terms of constructing period spacing patterns. In this paper we compare the results of forward asteroseismic modelling of a sample of SPB stars using intermittent TESS light curve data to those based on long-duration {\it Kepler} light curves. We show how in some cases that the masses and core masses derived from only a few sectors of TESS data agree well with the 4-yr {\it Kepler} mission results, despite the stars having far fewer significant pulsation frequencies in their TESS light curves. However, some stars yield incompatible results, emphasising the complexities in forward asteroseismic modelling of gravity-mode pulsators with sparsely sampled or short duration TESS light curves.
\end{abstract}


%%%%%%%%%%%%%%%%%%%%%%%%%%%%%%%%%%%%%%%%%%%%%%%%%%

%%%%%%%%%%%%%%%%% BODY OF PAPER %%%%%%%%%%%%%%%%%%

\section{Introduction}
Asteroseismology is the study of stellar oscillations, whose frequencies are used to uncover the structure of the stellar interior \citep{aerts2010, kurtz2022}. In particular, gravity modes (g~modes) are sensitive to the structure near the convective core of intermediate- and high-mass main-sequence stars \citep{aerts2021}. The g~modes in such stars are therefore useful for studying problems such as convective boundary mixing (CBM\footnote{Also sometimes referred to as overshooting in the literature.}; \citealt{anders2023}), rotation and angular momentum transport \citep{aerts2019}.

The slowly pulsating B (SPB) stars are main sequence stars between about 3 and 8~M$_{\odot}$, and pulsate in high-radial order g~modes excited by the heat-engine mechanism \citep{dziembowski1993, miglio2008}. Pulsation frequencies of the same angular degree and azimuthal order, and consecutive radial order form a regular pattern in period following the asymptotic approximation \citep{tassoul1980, miglio2008}. The first gravity-mode period spacing pattern in an SPB star was detected by \citet{degroote2010} using data from the CoRoT mission \citep{auvergne2009}. Later, light curves from the {\it Kepler} mission \citep{borucki2010} yielded a few dozen SPB stars \citep{papics2015, papics2017, szewczuk2021}. Subsequent forward asteroseismic modelling of the period spacing patterns determined their interior rotation rates, masses, ages and mixing properties \citep{moravveji2015, moravveji2016, szewczuk2018, szewczuk2021, szewczuk2022, michielsen2021, michielsen2023, bowman2021, pedersen2021, pedersen2022}, but also constraints on their interior magnetic fields \citep{buysschaert2018,lecoanet2022}.

Between 2009 and 2013, the NASA {\it Kepler} mission \citep{borucki2010, koch2010} assembled high-quality and long-duration light curves for over 200\,000 stars, which were revolutionary for asteroseismology. In its nominal mission, the {\it Kepler} satellite observed a 115~deg$^2$ field of view in the constellations of Cygnus and Lyrae, and assembled almost uninterrupted 4-yr light curves at a cadence of 30~min and a typical photometric precision of order a few ppm. These unprecedentedly high-quality data allowed forward asteroseismic modelling of tens of thousands of pulsating red giants (e.g. \citealt{chaplin2013}), hundreds of intermediate-mass dwarfs, such as $\gamma$~Dor and SPB stars (see \citealt{aerts2021}), but also many different types of pulsators across the Hertzsprung--Russell (HR)~diagram \citep{kurtz2022}. However, the field of view of the nominal {\it Kepler} mission avoided massive stars (i.e. $M \gtrsim 8$~M$_{\odot}$), which means the highest mass stars directly observed by the {\it Kepler} mission were SPB stars (see \citealt{bowman2020} for a review). A handful of massive stars were observed by the previous CoRoT mission \citep{auvergne2009}, and the maximum duration of light curves from different campaigns of the K2 mission \citep{howell2014} were only 80~d, making them generally too short to resolve period spacing patterns of SPB stars (e.g. \citealt{bowman2019}). This means that the paradigm shift of asteroseismology to more massive stars had to await the ongoing TESS mission \citep{burssens2023, bowman2023}.

The NASA Transiting Exoplanet Survey Satellite (TESS) mission is an all-sky survey with a primary goal of detecting Earth-like exoplanets orbiting bright late-type stars \citep{ricker2015}. The field of view of the TESS mission covers a region of $24 \times 96$~deg$^2$ stretching from the ecliptic equator up to an ecliptic pole using four cameras. Each hemisphere is divided into 13 sectors, and each is observed for approximately 28~d. After observations for a sector have been completed, the satellite slews to the next sector's field of view whilst maintaining an overlap near the ecliptic pole, defining a continuous viewing zone (CVZ). Since each of the four cameras has a square field of view, this means there are small regions of overlap between one sector and the next, and on average, a target is viewed for more sectors the larger its ecliptic latitude \citep{ricker2015}. This means that the observing strategy of TESS is entirely complementary to that of the {\it Kepler} mission, with the former scanning the sky and providing relatively short light curves for many pulsating early-type stars, and the latter targeting a relatively small region of the sky for a long time. TESS is currently in its second extended mission, and has provided light curves for millions of stars across the sky, including hundreds of $\beta$~Cep and SPB stars \citep{pedersen2019, burssens2020}, which are yet to be fully exploited.

The goal of this study is to investigate and validate the ability of using short-duration and intermittent light curves from the TESS mission in reproducing {\it Kepler} modelling output parameters for SPB stars using their g-mode period spacing patterns. We use a subsample of SPB stars studied previously by \citet{pedersen2021} as benchmarks to determine what fraction of the period spacing pattern frequencies extracted from {\it Kepler} data can be recovered from the currently available TESS sectors. We then use the g-mode period spacing patterns extracted from TESS data to perform forward asteroseismic modelling and establish if we recover the masses, core masses, and central hydrogen fractions of our sample, and discuss how these compare to the results previously obtained from {\it Kepler} light curves.

\section{Methods and Data}
\label{sec:data}

\subsection{Sample Selection}
\label{sec:sample}

\begin{table*}
    \centering
    \caption{SPB stars studied in this work, with columns of KIC ID, TIC ID, GAIA DR3 ID, Gaia $G$-band magnitude, the amplitude of the dominant pulsation mode as measured in this work (see Appendix~\ref{sec:app_freqs}), the relative percentage error on the derived stellar mass from forward asteroseismic modelling using {\it Kepler} data reported by \citet{pedersen2021}, and a contamination flag defined in this work. The four stars passing all of our selection criteria are listed first, whereas four additional cases with contamination issues are included for comparison purposes.}
    \label{tab:sample}
    
    \begin{tabular}{cccccccc}
    \hline \hline
    & KIC ID & TIC ID & Gaia DR3 ID & G (mag) & $A_{\mathrm{max}}$ (mmag) & $\mathrm{\Delta}M$ (per~cent) & contamination flag \\
    \hline
    good stars   &  4930889     &   138425170   &   2052512437803141760 &   8.8247  &   $6.60\pm0.03$   & 12 & no\\
    &   5941844     &   120319633   &   2104032361519412992 &   9.1546  &   $9.49\pm0.04$  & 3  & no\\
    &   7760680     &   271047903   &   2078179471596942848 &   10.3611 &   $7.00\pm0.01$   & 2  & no\\
    &   8766405     &   272368354   &   2079732295556849536 &   8.7884  &   $3.09\pm0.01$ & 8  & no\\ 
    \hline
    other stars &   3240411     &   137815897   &   2051930078899200768 &   10.2476 &   $0.46\pm0.01$ & 20 & yes\\
    &   3459297     &   139103795   &   2049135601386710144 &   12.4472 &   $1.63\pm0.01$   & 18 & yes\\
    &   6352430     &   121331677   &   2102391512214758144 &   7.8878  &   $5.54\pm0.02$ & 5  & yes\\
    &   11360704    &   27847920    &   2086818132599764224 &   10.6179 &   $1.19\pm0.03$    & 15 & yes\\
    \hline \hline
    \end{tabular}
\end{table*}

To demonstrate the efficacy of TESS for forward asteroseismic modelling of early-type stars, we first require a benchmark sample of SPB stars having undergone forward asteroseismic modelling to compare our TESS analysis with. Therefore, we have selected a subset of the 26 SPB stars studied using {\it Kepler} light curves by \citet{pedersen2021}, which allow us to compare stellar parameter estimations of mass, core mass, and central hydrogen (i.e. as a proxy for age of the main sequence) between the TESS and {\it Kepler} results. These parameters describe the structure of a star but also are related to processes such as CBM (see \citealt{aerts2021}). 

As our overall goal was to determine for which stars we find the `correct' stellar parameters from forward asteroseismic modelling of the TESS light curves, we needed to ensure that only stars with the most reliable results from the \citet{pedersen2021} sample were included in our study. We therefore imposed three criteria to define our sample of SPB stars. First, we excluded stars with \textit{Gaia} $G$-band magnitude fainter than 11~mag to allow for a robust TESS light curve extraction. For example, \citet{Bowman2024} discuss the difficulties in extracting TESS light curves of faint sources (i.e. $V\gtrsim11$~mag). Second, we included only isolated stars, which we define as those with no brighter stars or stars fainter by less than two magnitudes within the apparent point spread function of the target.
This is to ensure that extraction of TESS light curves is not complicated for stars with high crowding and contamination, thus allowing for a fair comparison of {\it Kepler} and TESS. Finally, we used the maximum pulsation amplitude reported by \citet{pedersen2021}, and excluded stars with pulsation amplitudes below $\sim$1~ppt. This was to ensure a sufficiently high signal-to-noise ratio (S/N) of the pulsation frequencies was achieved. This is necessary because the photometric precision of TESS light curves is worse than that of {\it Kepler} data, and coupled with the shorter light curves means low-amplitude pulsation modes may not be detectable with TESS. We also considered imposing a maximum relative error on mass reported by \citet{pedersen2021}, for example 30~per~cent, although all stars passed this criterion. 

The four SPB stars that satisfy all of our criteria span a range of pulsational properties allowing us to test how limited duration TESS light curves manifest in forward asteroseismic modelling under different conditions. We provide information of these four SPB stars in Table~\ref{tab:sample}. For example, KIC\,8766405 is the brightest star of the four with $G = 8.7884$~mag, whereas KIC\,5941844 has the highest pulsation amplitude with $\sim10$~mmag, and KIC\,7760680 has the best relative precision and accuracy on its mass with it being quoted as around 2~per~cent by \citet{pedersen2021} and 3~per~cent by \citet{bowman2021}. Finally, KIC\,4930889 is a detached binary with near-equal primary and secondary B-type components \citep{michielsen2023}. In addition to these four well-studied SPB stars, which we refer to as the `good sample', we added additional targets which only passed two of our three criteria to increase the sample size. These additional SPB stars include KIC 3240411 with its relatively unambiguous period spacing pattern, KIC\,6352430 as the brightest star in the \citet{pedersen2021} sample, and KIC\,11360704 which has the largest number of available TESS sectors. However, all three have nearby contaminating neighbours.  Finally, we included KIC\,3459297, which failed both the contamination and brightness criteria, but also has a relatively unambiguous period spacing pattern (see \citealt{pedersen2021}). Therefore, our final sample of SPB stars includes four good and four other targets, which are provided in Table~\ref{tab:sample}.

\subsection{TESS light curve extraction}
\label{sec:lc} 

Since the launch of the TESS mission, the overall quality and properties of available data products have greatly improved. For example, in the nominal mission (i.e. cycles 1 and 2), full-frame images (FFIs) were provided at a cadence of 30~min, but this was later improved to 10~min and subsequently 200~sec in the first and second extended missions, respectively. Whilst a cadence of 30~min is typically sufficient for studying g~modes in SPB stars, such that the pulsation amplitudes are negligibly affected by the amplitude suppression function caused by smearing of the pulsation signal (see \citealt{bowman2017}), a faster cadence in the available FFI TESS data provides improved precision in the resultant frequency spectrum owing to the larger number of observations within each sector \citep{huber2022}. Random noise caused by jitter is more effectively mitigated in more recent TESS sectors owing to improvements in NASA's data reduction pipelines \citep{jenkins2016}. Moreover, the number of allocated targets for 2-min cadence, and the more recently available 20-sec cadence, for light curve extraction based on NASA's SPOC pipeline \citep{jenkins2016} has also increased since the nominal mission. This means through successful guest investigator (GI) proposals led by members of the community that many more early-type stars have 2-min light curves available (see e.g. \citealt{bowman2022}). Therefore, for all of the above reasons, the quality of TESS data is generally better for more recent sectors.

TESS target pixel files for each star were downloaded from the Mikulsi Archive for Space Telescopes (MAST)\footnote{\url{https://archive.stsci.edu/missions-and-data/tess}} public archive using the \texttt{tesscut} function \citep{brasseur2019, astropy2022} in the \texttt{Python} {\sc lightkurve} software package \citep{lightkurve2018}. Cutouts of size 20 by 20 pixels were used. We extract light curves from TESS FFIs for our SPB targets using simple aperture photometry (SAP). We first assign pixels below the median flux as `background'. The aperture mask of a target in the framework of SAP is typically assigned based on flux-based significance criterion and a watershed-like algorithm applied to a median-stacked image. For studying the relatively large-amplitude pulsations in early-type stars, however, a larger aperture mask compared to that of NASA's SPOC pipeline is usually beneficial for maximising the S/N of stellar variability \citep[see e.g.][]{papics2013,papics2015,bowman2022,burssens2023}. 

\begin{figure}
    \centering
    \includegraphics[width=0.75\linewidth]{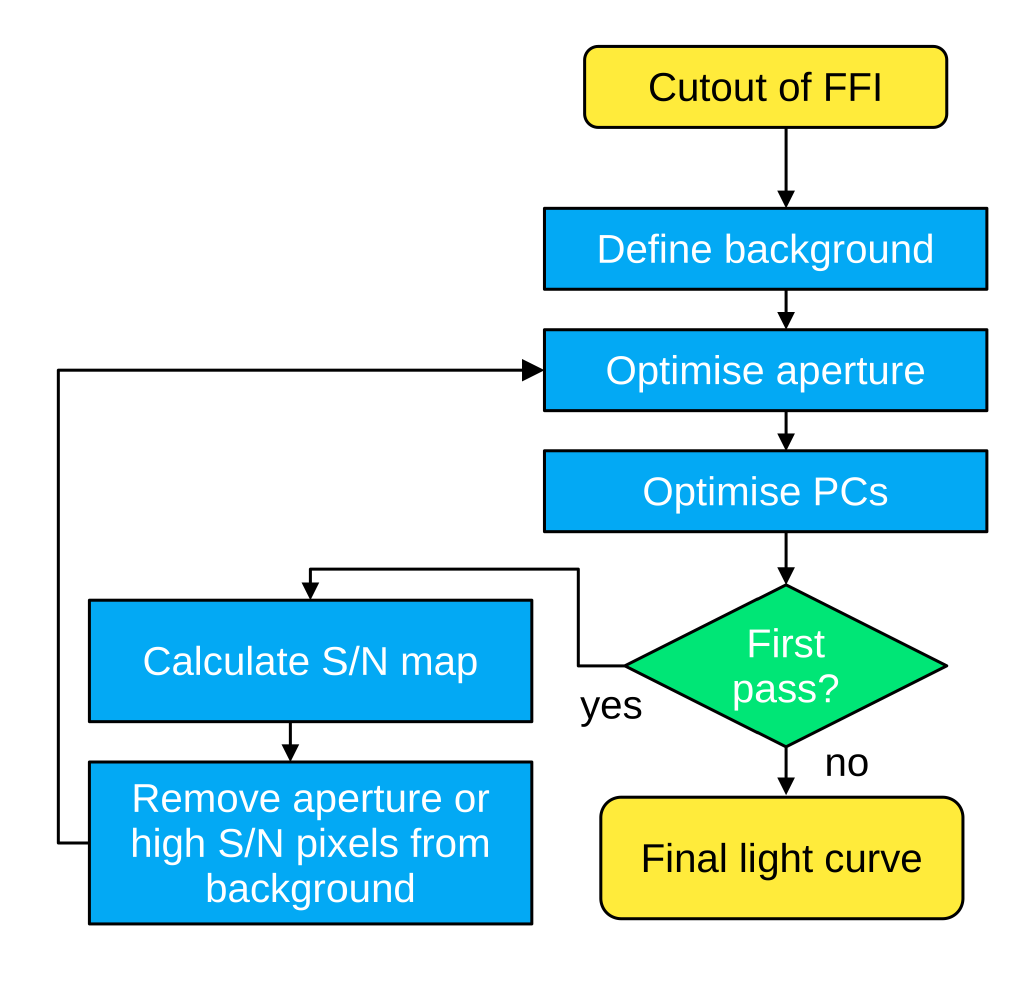}
    \caption{Flowchart describing the light curve extraction process.}
    \label{fig:flowchart}
\end{figure}

In this work we have developed a new method for optimising the target aperture pixel mask of a pulsating early-type star, which uses the S/N of the dominant pulsation mode in the frequency spectrum as the metric for deciding which pixels to include or not. Our methodology is summarised in a schematic flowchart in Fig.~\ref{fig:flowchart}, and individual steps are described in the following subsections. 

\subsubsection{Optimal aperture selection based on pulsation amplitude signal-to-noise ratios}
\label{sec:lc_aperture}

The standard approach of extracting light curves is based on a flux threshold for pixels to be included in a target star's aperture mask (or not). But this does not take into account that pulsations cause flux variability, hence a flux-based mask it is not tailored for asteroseismology of pulsations, which generally operates in the frequency domain rather than the time domain. In this work, similar to \citet{bowman&holdsworth2019}, we use a curve-of-growth method for optimising the size of a target's pixel mask using the S/N of the dominant pulsation mode in a frequency spectrum of the extracted light curve as a metric, \SNRfone. This relies on calculating periodograms for each pixel, in a similar manner to the public {\sc tess\_localize} software package \citep{higgins2023}. In essence, larger aperture masks maximise the extracted flux of a target, hence generally also maximise the signal in the resultant S/N of pulsation mode amplitudes. Yet, over-large target aperture masks increase the noise contribution, which could decrease the overall S/N of pulsation modes, especially for fainter stars. Moreover, this pulsation-optimised mask selection is important for variable stars, such as SPB stars, because they can be high-amplitude multi-periodic pulsators with long-period beating patterns (see e.g. \citealt{papics2015, papics2017}), and appreciably affect the size of the optimal aperture mask from cadence to cadence.

The optimised target aperture mask was decided by testing the S/N of the dominant pulsation mode in the frequency spectra of light curves extracted using different aperture sizes. The S/N is defined as the height of the highest-amplitude peak divided by the mean amplitude within the frequency spectrum ranging from 0 to 6~d$^{-1}$. This frequency range was chosen since prograde dipole modes of SPB stars are typically observed between about 1 and 3~d$^{-1}$ \citep{szewczuk2021, pedersen2021}. Non-linear combination frequencies and harmonics typically occur at higher frequencies, and appear as groups as discussed by \citet{kurtz2015}. The larger sample of {\it Kepler} SPB stars studied by \citet{pedersen2021} also supports this range for the noise window, since the period spacing patterns extracted for such stars are consistently found to have periods ranging from 0.3 to 3\,d (i.e. $0.33 < \nu < 3.3$~d$^{-1}$).

\begin{figure}
    \centering
    \includegraphics[width=\linewidth]{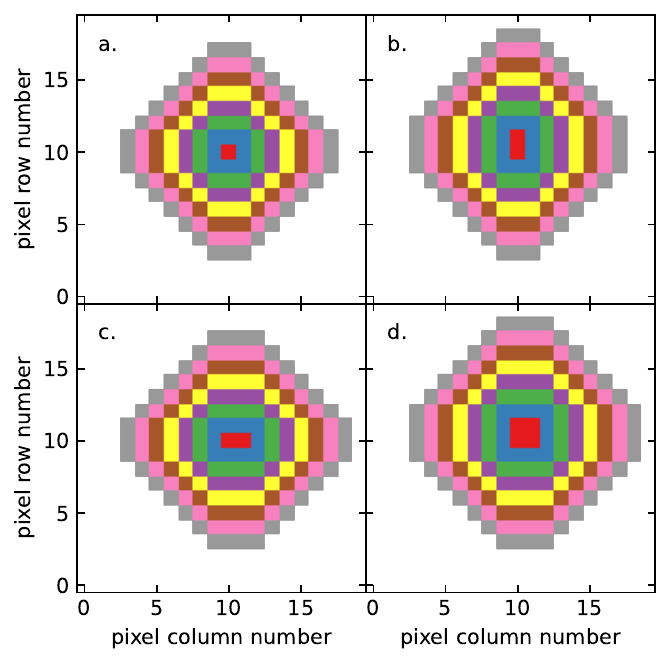}
    \caption{Pixel rings used for iteratively determining each target's optimal aperture size, which are represented in different colours. In this schematic, the target star is located within pixel (10,10). Panel a shows the ring shape for a star located centrally within this pixel. If the star is closer to the pixel edge, the ring shapes are extended vertically, horizontally or both. Panels b, c, and d show ring shapes for stars located closest to the top edge, right edge and top right corner, respectively.}
    \label{fig:rings}
\end{figure}

We begin by first extracting the light curve from only the central pixel containing the target star. The S/N of the dominant pulsation frequency in the resultant periodogram is thus used as a benchmark value to compare against, such that aperture masks with improved S/N are favoured. Next we successively test the change in the resultant S/N of the dominant pulsation frequency in the periodogram for larger aperture masks by adding rings of pixels, the shapes of which are determined by the position of the star within the central pixel, as shown in Fig.\,\ref{fig:rings}. The aperture with the highest \SNRfone was chosen as the optimal aperture. For each target aperture defined, we define the background to exclude any pixels within the target aperture. Finally, to detrend our light curves of any systematic trends, we test the optimal number of principal components (PCs) to remove, also in an iterative fashion, again maximising \SNRfone based on the optimal target aperture mask -- see Sec.\,\ref{sec:lc_detrend} for details of this process. This process is repeated after calculating \SNRfone{} in the remaining background pixels and removing those with \SNRfone{} larger than 3. As our method optimises the target's aperture mask and approach to detrending based on the S/N of a star's pulsation frequencies, it is better suited to the problem of asteroseismology of early-type stars, as opposed to only using a flux-based threshold to define a target aperture mask that does not account for variability.

\subsubsection{Light curve detrending}
\label{sec:lc_detrend}

\begin{figure*}
    \includegraphics[width=0.75\linewidth]{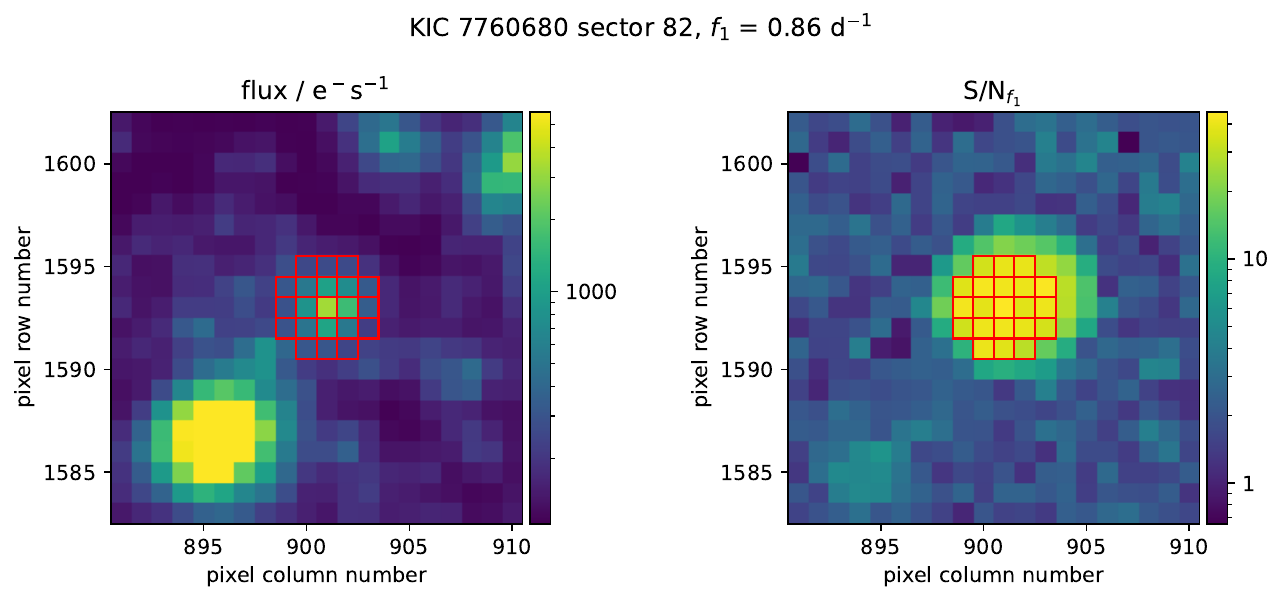}
    \includegraphics[width=0.75\linewidth]{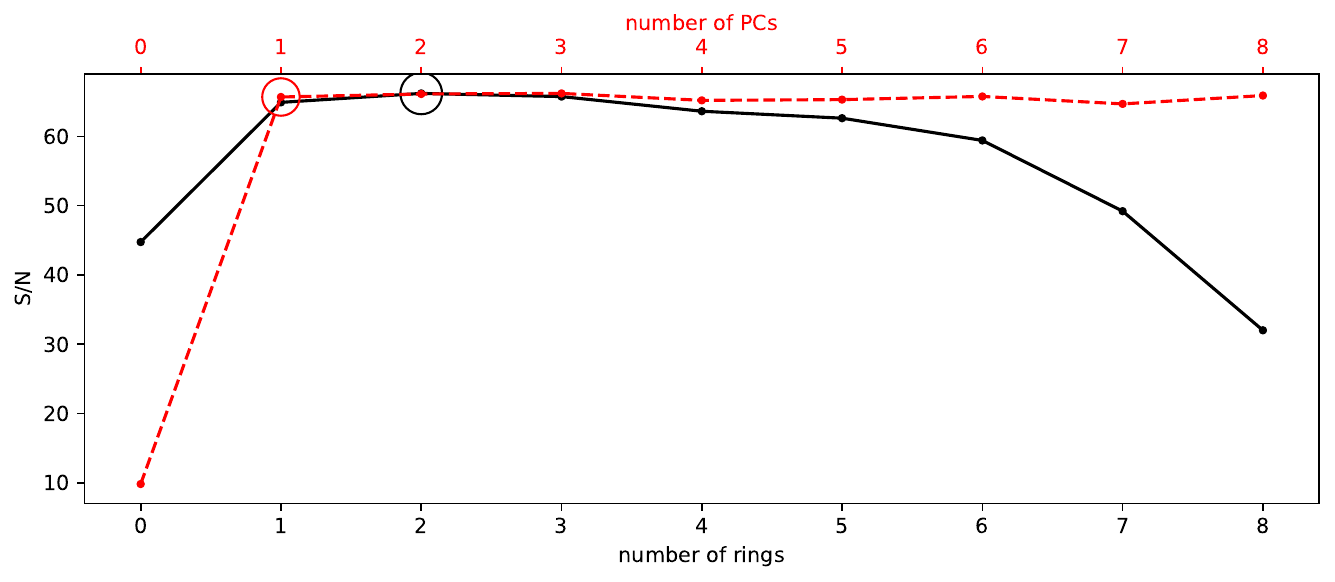}
    \includegraphics[width=0.75\linewidth]{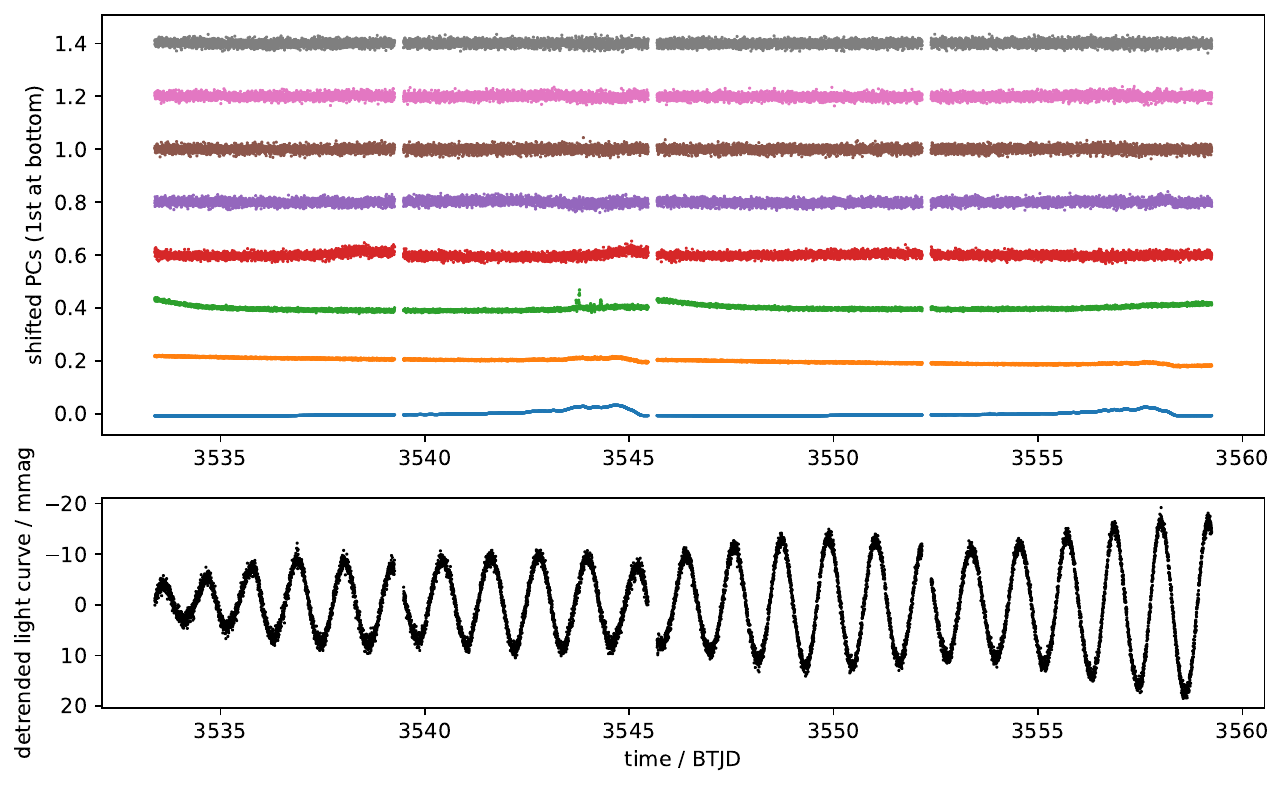}

    \caption{TESS light curve extraction summary of sector 82 for KIC~7760680. The target aperture mask that maximises the S/N of the dominant pulsation mode is highlighted in red in the top two panels, with left- and right-hand panels showing flux and S/N, respectively. The panel below this shows the S/N curves of growth for the number of concentric rings in determining the optimal pixel mask (black) and the number of principal components (PCs, red dashed), with the optimal values circled. The bottom two rows show the PCs, with the first PC at the bottom of the panel, and the final detrended light curve for the sector.}
    \label{fig:summary}
\end{figure*}

To study stellar pulsations in a light curve, they first need to be disentangled from variability caused by instrumental effects. These include the heating or cooling of the detector after and before the telescope reorients itself with respect to the Sun, as well as time-dependent scattered light. These trends can be removed, somewhat subjectively, using methods such as fitting a spline or a low-order polynomial leaving only the astrophysical variability behind \citep[e.g.][]{papics2017,bowman2018}. Alternative methods for detrending light curves include Gaussian process regression (see e.g. \citealt{aigrain2016}) and the increasingly popular approach of principal component analysis (PCA; \citealt{lightkurve2018}).

We have chosen to detrend our light curves using a PCA approach as this has proven powerful for TESS light curves of early-type pulsating stars (see e.g. \citealt{bowman2022, burssens2023}). In our case, the PCs are light curve shapes that when multiplied by a coefficient and summed can reconstruct the background light curve. This model for the background flux can be written as

\begin{equation}
    \label{eq:pca}
    F_{\rm bg}(t) = \sum_n a_nC_n(t),
\end{equation}

\noindent where $F_{\rm bg}(t)$ is the approximated background light curve flux, $C_n(t)$ are the PCs, $a_n$ are their coefficients, and $n$ is the number of PCs used (see \citealt{lightkurve2018}). Each successive PC contains a smaller fraction of the variance in a light curve, and we find that $n\le8$ is consistently sufficient for SPB stars observed in TESS mission data. We calculated the PCs of background pixels using the {\sc lightkurve} software tool \citep{lightkurve2018}, then removed them from the target star's light curve. We emphasize that this detrending was performed before any analysis of a periodogram ––– i.e. before calculating the S/N of pulsation frequencies in the aperture optimisation step (c.f. Section~\ref{sec:lc_aperture}) or for before making S/N maps (e.g. Fig.~\ref{fig:summary}).

For the initial aperture optimisation step, the first three PCs of the background were used as a starting point when detrending. We decided that three PCs was sufficient based on inspection of pixel light curves and the typical structure of the PCs. In this first pass, the background pixels were defined as those with below-median flux (c.f. Section~\ref{sec:lc_aperture}). Then, pixels with \SNRfone{}$>3$ were excluded from the background and the light curve of the target star was extracted. We then repeated the aperture and PC optimisation steps with the new background. Our final choice for the optimal number of PCs was smallest number which produced 99~per~cent of the maximum of \SNRfone{}. 

As a demonstration, we show a summary figure of the application of this multi-step curve-of-growth process to extracting light curves applied to KIC~7760680 in Fig.~\ref{fig:summary}. Here, the top panels show the optimal extraction mask on pixel images of both flux and \SNRfone{}. Below this are the curves of growth showing how S/N changes with the number of pixel rings and PCs. The bottom two panels of Fig.~\ref{fig:summary} show the PCs and the light curve detrended with the optimal number of PCs. Using our extraction method, the S/N of the dominant mode is on average 4~per~cent higher than using NASA's SPOC-pipeline light curve and 18~per~cent higher than the QLP light curve, which are also available on the MAST archive, across our sample of SPB stars.

\subsubsection{Joining TESS sectors}
\label{sec:chunks}

\begin{table}
    \centering
    \caption{The TESS sectors used in defining chunks of light curves}.
    \begin{tabular}{cccc}
    \hline
    KIC ID  & \multicolumn{3}{c}{TESS sectors used} \\
            & chunk 1   & chunk 2   & chunk 3 \\
    \hline
    3240411 & 81, 82& 74    & 54, 55 \\
    3459297 & 81, 82    & 74, 75    & 54, 55 \\
    4930889 & 81, 82    & 74, 75    & 54, 55 \\
    5941844 & 80, 81    & 53, 54    & 40, 41 \\    
    6352430 & 80, 81, 82& 74, 75    & 53, 54, 55 \\
    7760680 & 81, 82    & 74, 75    & 54, 55 \\
    8766405 & 81, 82    & 74, 75    & 54, 55 \\
    11360704& 81, 82, 83& 74, 75, 76& 54, 55, 56 \\
    \hline
    \end{tabular}
    \label{tab:sectors}
\end{table}

Each of the eight SPB stars in our sample was observed by TESS from one to three consecutive sectors, which we refer to as chunks, with then a large gap of 2~yr before TESS re-observes the same star once again. We have concatenated the detrended light curves from individual sectors into longer light curves spanning one, two and three chunks of data. The sectors included in each chunk sometimes differ across our sample because of where a star falls precisely in TESS's field of view, and are listed in Table~\ref{tab:sectors}. We note that some stars are observed in some sectors that were discarded because the target was too close to the edge of the detector, thus yielding unusable data. In the case of KIC\,5941844, this meant that there was a particularly long gap between chunk 1 and chunk 2.

Due to the improvement in the TESS FFI cadence after the end of cycle 2 (i.e. after sector 55) from 10~min to 200~sec, we binned all data to have a 10-min cadence for consistency purposes. This was done before concatenation of different chunks by taking the mean flux in each 10-min bin. This is important because a different (average) cadence defines a different Nyquist frequency and amplitude suppression function \citep{bowman2017}, as well as to avoid having to assign different weights to data points in subsequent Fourier analysis because of the different integration times.

\subsection{Frequency extraction from optimal light curves}
\label{sec:freq}
To investigate how the length of a TESS light curve impacts frequency analysis, we made three light curves for each star: the first using only chunk one, the next using chunks one and two, and the last using all three chunks, as detailed in Table~\ref{tab:sectors}. Each of these light curves was transformed into frequency space by means of a discrete Fourier transform, using the {\sc Period04} software package \citep{lenz2005}. We extracted significant pulsation frequencies following the standard approach of iterative pre-whitening in order of decreasing amplitude until an amplitude S/N threshold was reached, and then we performed a non-linear multi-sinusoid least-squares optimisation (see e.g. \citealt{kurtz2015, bowman2021}). 
We used eq.\,3 of \citet{baran2021} to estimate a S/N threshold of 5.3 for frequency extraction, which corresponds to a false alarm probability of 0.1~per~cent based on simulations of synthetic TESS data. Since our light curves have a cadence of 600\,s and \citet{baran2021} did not simulate this, we used a scaled value between their 120\,s and 1800\,s equations. It should also be noted that eq.\,3 of \citet{baran2021} does not account for gaps in the data. However, in the cases where we have complex spectral windows caused by gaps, we are usually in a regime where our frequency extraction is limited by resolving power rather than S/N.

We defined S/N as the amplitude height of a peak in the frequency spectrum divided by the mean amplitude after pre-whitening the selected peak within a $2\,\mathrm{d}^{-1}$ frequency window. The noise window is therefore smaller compared to the calculation of \SNRfone{}, which encompassed the entire frequency range up to 6\,d$^{-1}$ in order to test the effectiveness of detrending. The frequency extraction process was performed for each of the one-, two- and three-chunk light curves for each star. We provide a list of extracted significant frequencies and mark those that are consistent across the TESS and {\it Kepler} light curve analyses in Tables~\ref{tab:app_3240411}--\ref{tab:app_11360704} in Appendix~\ref{sec:app_freqs}. For uncertainties on our frequencies, amplitudes and phases, we used the least-squares error estimation increased to account for correlation in the residual light curve \citep{degroote2009}. The correction factor for correlation was generally about 3 for our TESS sectors.

\begin{table*}
    \centering
    \caption{Number of pulsation frequencies extracted for each star and the number of frequencies which are part of the period spacing pattern reported by \citet{pedersen2021}. Quantities are shown for each number of chunks in the light curve.}
    \begin{tabular}{ccccccc}
        \hline
        KIC ID & \multicolumn{3}{c}{Number of extracted frequencies} & \multicolumn{3}{c}{Number of extracted frequencies in pattern} \\
                 & 1 chunk & 2 chunks & 3 chunks & 1 chunk & 2 chunks & 3 chunks \\
        \hline
        3240411  & 1  & 3  & 9 & 1 & 2 & 3 \\
        3459297  & 8  & 9  & 11 & 5 & 7 & 9 \\
        4930889  & 12  & 14 & 17 & 7 & 10 & 12 \\
        5941844  & 8  & 9 & 11 & 5 & 5 & 5 \\
        6352430  & 13 & 17 & 23 & 4 & 6 & 9 \\
        7760680  & 11 & 20 & 18 & 6 & 12 & 12 \\
        8766405  & 11 & 17 & 15 & 2 & 2 & 2 \\
        11360704 & 16 & 24 & 18 & 2 & 2 & 2 \\
        \hline
    \end{tabular}
    \label{tab:recovered}
\end{table*}

\begin{figure}
    \centering
    \includegraphics[width=\linewidth]{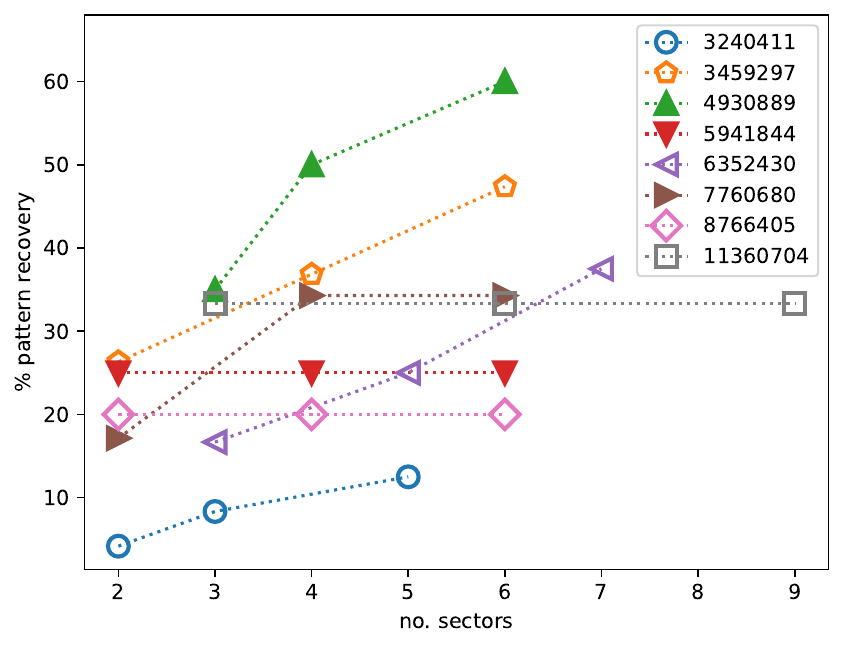}
    \caption{Percentage of frequencies in the {\it Kepler} period spacing patterns reported by \citet{pedersen2021} that were extracted in our analysis of TESS light curves. We were able to find period spacing patterns for the stars with filled symbols. The length and quality of the data was insufficient to find a convincing pattern in the TESS light curves for stars with open symbols.}
    \label{fig:recovery}
\end{figure}

As a first result on the efficacy of using TESS for g-mode asteroseismology, we tested if our extracted frequencies belong to the period spacing pattern identified by \citet{pedersen2021} using {\it Kepler} light curves. In Table~\ref{tab:recovered} we detail the number of frequencies that we extracted for each star, along with the number of those that were also part of the period spacing patterns of \citet{pedersen2021}. In this way, we quantify the agreement between detectable significant pulsation frequencies for short TESS light curves compared to long {\it Kepler} light curves. Our results of how the number of recovered pulsation frequencies in a star's period spacing pattern depends on the amount of TESS data is shown graphically in Fig.~\ref{fig:recovery}. Overall, we find that each SPB star requires a different amount of TESS data to satisfactorily reproduce the period spacing pattern extracted using {\it Kepler} data, if at all. For example, two sectors (i.e. only one chunk) of TESS data for KIC~4930889 yields a 35~per~cent agreement, which rises to  60~per~cent agreement for six sectors (i.e. three chunks) of TESS data. On the other hand, whilst additional chunks increase the number of extracted frequencies for KIC~5941844, KIC~8766405 and KIC~11360704, no additional frequencies from the period spacing patterns of \citet{pedersen2021} are recovered with more sectors. Similarly, there is no increase in the recovery fraction for KIC~7760680 when increasing from two to three chunks. Whilst it is generally the case that more TESS sectors are better for frequency analysis of early-type stars, our results demonstrate that it strongly depends on the duty cycle of the combined light curve and the pulsation frequencies of a star on whether additional sectors allow one to converge (or diverge) from the true period spacing pattern. We discuss all stars individually in Section~\ref{sec:results}.

Up until now, we have been guided in our frequency recovery analysis by heavily relying on the reported period spacing patterns from \citet{pedersen2021} based on 4~yr of {\it Kepler} data. To evaluate potential biases, we next performed an independent analysis to construct period spacing patterns based solely on the TESS light curves, avoiding any knowledge or biases of what the period spacing pattern `should' look like. We do this because the choice of which frequencies to include in the pattern is subjective, and the inclusion of ambiguous frequencies may propagate errors into the final modelling results, as discussed by \citet{michielsen2021} and \citet{bowman2021}.

\subsubsection{Period spacing pattern fitting}
\label{sec:period_spacing_patterns}

It is well established that faster rotation in the near-core region where g~modes are most sensitive, $f_{\rm rot}$, produces a stronger negative gradient in a period spacing pattern of prograde dipole (i.e. $\{\ell,m \} = \{1,1 \}$) g~modes in the observer's inertial reference frame \citep{aerts2021}. This means that the period spacing patterns are generally more difficult to extract for faster rotating stars (see \citealt{papics2017, szewczuk2021}) and benefit from longer light curves. 

\begin{figure}
    \centering
    \includegraphics[width=\linewidth]{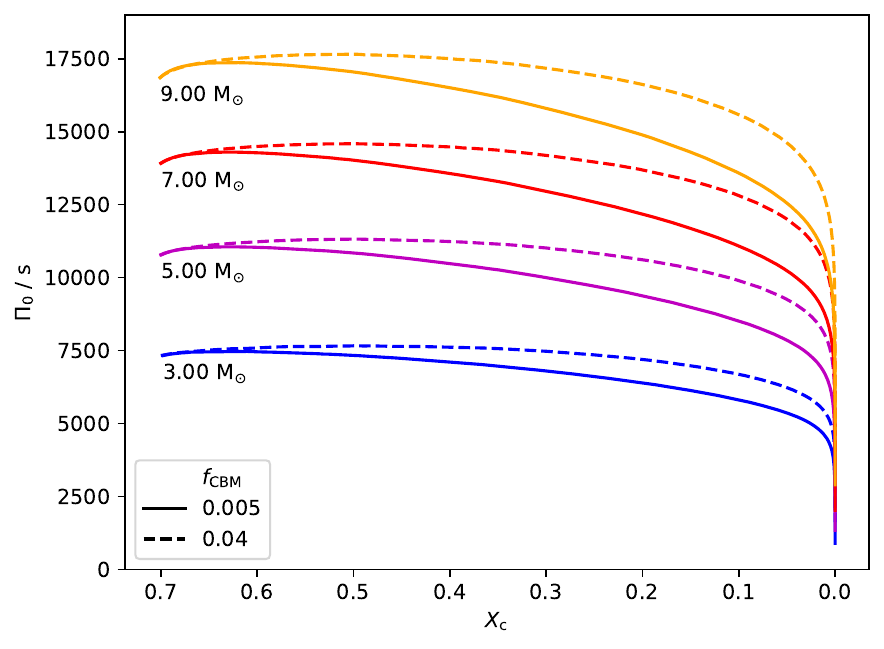}
    \caption{The evolution of the asymptotic period spacing, $\Pi_0$, with central hydrogen mass fraction, $X_{\rm c}$, for selected masses within the grid of structure models used in this work. The value of CBM, $f_{\rm CBM}$, is indicated by either solid or dashed lines. These models have $\log{D_{\rm ext}}=0$; higher $\log{D_{\rm ext}}$ models have similar $\Pi_0$ values.}
    \label{fig:Pi0}
\end{figure}

The asymptotic period spacing value, $\Pi_0$, is related to a star's mass and age, and is defined as
\begin{equation}
    \label{eq:pi0}
    \Pi_{\rm 0}=2\pi^2 \left( \int^{r_2}_{r_1} \frac{N(r)}{r}\rm{dr} \right)^{-1},
\end{equation}

\noindent where $r$ is the radius of a star, $r_1$ and $r_2$ are the radial positions of the g-mode cavity, and $N(r)$ is the Brunt-V\"ais\"al\"a frequency \citep{aerts2021}. The parameter $\Pi_0$ has also been called the buoyancy travel time, and it directly probes the thermal and chemical composition structure in the near-core region through its dependence on $N(r)$. Through a comparison to grids of theoretical models, and the dependence of $\Pi_0$ on $N(r)$, the mass, core mass, and age of a star can be determined. We show values from representative models in our grid of structure models in Fig.~\ref{fig:Pi0}. We used the open-source {\sc amigo}\footnote{\url{https://github.com/TVanReeth/amigo}} software tool \citep{vanreeth2016, vanreeth2018, vanreeth2022} to fit the extracted period spacing patterns of the SPB stars in our sample, and determine the near-core rotation frequency, $f_{\rm{rot}}$, and the asymptotic period spacing, $\Pi_0$. We report the values of $\Pi_0$ and $f_{\rm{rot}}$ in Table~\ref{tab:obs_constraints}.

\subsection{Forward asteroseismic modelling}
\label{sec:model_fitting}

To perform forward asteroseismic modelling, we make use of a subset of the grid of structure models calculated by \citet{johnston2019}, who used $\Pi_0$ as a seismic diagnostic in their study of three eclipsing binaries containing g-mode pulsators. The grid contains 770\,000 structure models calculated using the open-source {\sc mesa}\footnote{\url{https://docs.mesastar.org/en/latest/}} stellar structure and evolution software package \citep{paxton2011,paxton2013,paxton2015,paxton2018}. The grid spans a mass range of 1.2 to 9\,$\rm M_\odot$, with an interval of 0.1\,$\rm M_\odot$ below a mass of 5\,$\rm M_\odot$ and 0.25\,$\rm M_\odot$ above. The models were calculated with a metallicity of $Z=0.014$ and a mixing length theory (MLT) parameter of $\alpha_{\rm MLT}=1.8$, which is expressed in terms of pressure scale heights. The varied parameters were mass, CBM parameter $f_{\rm CBM}$, and envelope mixing parameter $D_{\rm ext}$ \citep{johnston2019}. The CBM parameter takes the form of an exponentially decaying diffusion coefficient \citep{freytag1996}, with $f_{\rm CBM}$ determining the gradient of the efficiency in the near-core mixing region. The envelope mixing is based on a diffusion coefficient that emulates gravity wave mixing with $D(r) \propto \rho^{-1}$ \citep[see][]{rogers2017}, such that the values of $D_{\rm ext}$ is the diffusion coefficient at the interface of the CBM region and the envelope. $f_{\rm CBM}$ ranges between 0.005 to 0.040 in intervals of 0.005 and $\log{(D_{\rm ext}/\rm{cm^2\,s^{-1}})}$ ranges between 0.0 and 2.0 in intervals of 0.5. The models each had 600 time steps, beginning in the pre-main sequence and finishing at the end of the main sequence. The pre-main sequence phase lasted for approximately 250 time steps, but we exclude this in our forward asteroseismic modelling given that \citet{pedersen2021} concluded that our SPB stars are core-hydrogen burning.

For those stars for which we identified a period spacing pattern (listed in Table~\ref{tab:obs_constraints}), we fit our $\Pi_0$ values to the model grid calculated by \citet{johnston2019}. The fit was performed in combination with literature values of effective temperature, $T_{\rm eff}$, and surface gravity, $\log{g}$, as additional observables \citep{papics2013,papics2015,hanes2019,gebruers2021}. 
%We used two methods: either fitting $\Pi_0$ alone but constraining the grid within three $\sigma$ of the observed $T_{\rm eff}$ and $\log{g}$ (henceforth called the $T_{\rm eff}$--$\log{g}$ box), or by fitting all three observables simultaneously. We added the spectroscopic constraints as constraining the solution using only $\Pi_0$ as the sole diagnostic can be highly degenerate (see Fig.~\ref{fig:Pi0}), and a star's location in the HR~diagram is an effective method to delimit the possible parameter space (see \citealt{moravveji2016, buysschaert2018}). To fit the $\Pi_0$ values from our period spacing patterns in the $T_{\rm eff}$--$\log{g}$ box, we calculated $\chi^2$ values,
%\begin{equation}
%    \label{eq:chi_squared}
%    \chi^2 = \left(\frac{\Pi_{\rm 0,obs}-\Pi_{{\rm 0},i}}{\sigma}\right)^2,
%\end{equation}
%\noindent where $\Pi_{\rm 0,obs}$ is the $\Pi_0$ measured from the period spacing pattern (c.f. Section~\ref{sec:sample}), $\Pi_{{\rm 0},i}$ is the value of $\Pi_0$ calculated from each model $i$ using Eq.~(\ref{eq:pi0}), and $\sigma$ is the error on $\Pi_{\rm 0,obs}$ provided by the \texttt{AMiGO} software. We performed $\chi^2$ fitting for $\Pi_0$ both over the whole grid and constrained to within $3\sigma$ of the spectroscopic measurements of $T_{\rm eff}$ and $\log\,g$. 

We used the Mahalanobis distance (${\rm MD}$) as a merit function,
\begin{equation}
    \label{eq:mahalanobis}
    {\rm MD} = (Y_{\rm mod}-Y_{\rm obs})^\top(V+\Lambda)^{-1}(Y_{\rm mod}-Y_{\rm obs}),
\end{equation}
where $Y_{\rm mod}$ and $Y_{\rm obs}$ are vectors containing the model and observed values of $\Pi_0$, $T_{\rm eff}$ and $\log{g}$, respectively, $V$ is the variance-covariance matrix for the fitted quantities calculated from the model grid, and $\Lambda$ is the square matrix containing the squared errors on the fitted quantities in its diagonals \citep{aerts2018}. We use ${\rm MD}$ as a merit function instead of $\chi^2$ in this case since it has the significant advantage of incorporating correlations and degeneracies amongst parameters \citep{aerts2018}. The MD hence disfavours models with atypical combinations of fitted values, unlike $\chi^2$ which treats all parameters as uncorrelated and independent. \citet{michielsen2021} demonstrate the limitations of using $\chi^2$ and benefits of using MD as merit function in forward asteroseismic modelling of g~modes in SPB stars.

We use Bayes' theorem to calculate the probability distribution of models given our observed values, for which we use the likelihood
%\begin{equation}
%    \label{eq:likelihood_chi2}
%    \mathcal{L}(Y_{\rm obs}|\theta) \propto \exp{\left(-0.5\chi^2\right)}
%\end{equation}
%when fitting within the $T_{\rm eff}$--$\log{g}$ box, or when fitting all observables over the whole grid,
\begin{equation}
    \label{eq:likelihood_md}
    \mathcal{L}(Y_{\rm obs}|\theta) \propto \exp{\left(-0.5(\det{(V+\Lambda)}+2k\pi+{\rm MD}\right))} ~ ,
\end{equation}
\noindent where $k=3$ is the number of fitted parameters, and $\theta$ is the set of grid parameters (i.e. mass, $f_{\rm CBM}$, $D_{\rm ext}$ and central hydrogen fraction, $X_{\rm c}$). We assume the unnormalised prior probability, $P(\theta)$, is flat in all dimensions. Using Bayes' theorem, the probability of a set of grid parameters $\theta$ given the observations $Y_{\rm obs}$ is then

\begin{equation}
    \label{eq:probability}
    P(\theta|Y_{\rm obs}) = \frac{P(\theta)\mathcal{L}(Y_{\rm obs}|\theta)}{\Sigma_i{P(\theta)}\mathcal{L}(Y_{\rm obs}|\theta)} ~ .
\end{equation}

\noindent The 1-$\sigma$ confidence intervals for each parameter of the best model can be found using the minimum and maximum values of $\theta$ for those models with  $P(\theta|Y_{\rm obs})$ above the $68^{\rm th}$ percentile. Similarly to \citet{johnston2019}, confidence intervals for the best-fitting mass, age, and core mass of SPB stars based on MD and using $\Pi_0$ in this manner are large, and incorporate a sizeable fraction of the {\sc mesa} grid. This is because marginalising onto an individual parameter typically assigns the entire budget of the best-fitting model to that single parameter, thus resulting in large single-parameter uncertainties despite having a well constrained overall best model. However, since the MD is a maximum likelihood estimation method, the best-fitting values are useful and informative to compare between TESS and {\it Kepler} modelling results.

\section{Results}
\label{sec:results}

Our analysis of the TESS light curves of the eight SPB stars yielded between 1 and 24 extracted frequencies depending on the star, as shown in Table\,\ref{tab:recovered}. This demonstrates that success in the application of iterative pre-whitening to extract coherent pulsation mode frequencies in intermittent TESS light curves is very star dependent. We provide details on the specifics of the extracted frequencies for each star in Appendix\,\ref{sec:app_freqs}, and the frequencies are plotted on pre-whitened spectra in Appendix\,\ref{sec:app_figures}. Due to the reduction in noise, more frequencies were extracted for light curves with more chunks in most cases. However, the gaps between chunks introduced complexity in the spectral window, making the iterative pre-whitening more difficult. This is why fewer significant and independent frequencies were extracted from the three-chunk data sets for KIC\,7760680, KIC\,8766405 and KIC\,11360704 compared to two chunks.

\begin{table*}
    \centering
    \caption{Value of the rotation frequency from \citet{pedersen2021}, $f_{\rm rot,P}$, and our results for $f_{\rm rot}$ and $\Pi_0$ obtained from fitting the period spacing pattern, as well as the literature spectroscopic constraints that were used in the forward asteroseismic modelling. In general, $\Pi_0$ values are derived from the three-chunk light curves in this work, apart from the row marked with `*' that used the two chunk light curve for comparison purposes. The final column provides the literature reference for $T_{\rm eff}$ and $\log{g}$. The period spacing patterns for the last two stars were informed by those of \citet{pedersen2021}, because identifying a reliable period spacing pattern from the TESS data alone was not feasible.}
    \begin{tabular}{ccccccc}
        \hline \hline
        KIC ID & $f_{\rm rot,P}$\,/\,d$^{-1}$ & $f_{\rm rot}$\,/\,d$^{-1}$ & $\Pi_0$\,/\,s & $T_{\rm eff}$\,/\,K & $\log{(g\,/\,\rm{cm\,s^{-2}})}$ & reference \\
        \hline \hline
        4930889 & $0.745\pm0.005$ & $0.69 \pm 0.07$ & $7110\pm1040$ & $15100\pm150$ & $3.95\pm0.1$ & \citet{gebruers2021}\\
        5941844 & $0.988\pm0.067$ & $0.56\pm0.10$ & $22850\pm2380$ & $14025\pm300$ & $4.24\pm0.08$ & \citet{gebruers2021} \\
        *7760680 & $0.455\pm0.022$ & $0.39\pm 0.13$ & $7180\pm1530$ & $11650\pm210$ & $3.97\pm0.08$ & \citet{papics2015} \\
        7760680 & $0.455\pm0.022$ & $0.42\pm0.06$ & $6700\pm1100$ & $11650\pm210$ & $3.97\pm0.08$ & \citet{papics2015} \\
        \hline
        3459297 & $0.627\pm0.026$ & $0.26\pm0.09$ & $2940\pm690$ & $13100\pm600$ & $3.61\pm0.14$ & \citet{hanes2019} \\
        6352430 & $0.681\pm0.050$ & $0.79 \pm 0.08$ & $9940\pm1350$ & $12810\pm200$ & $4.05\pm0.05$ & \citet{papics2013} \\
        \hline
    \end{tabular}
    \label{tab:obs_constraints}
\end{table*}

We examined our extracted frequencies for matches with those in the period spacing patterns of \citet{pedersen2021}. This is shown in the final three columns of Table~\ref{tab:recovered}.
We judged our frequencies to be a match if they were within 0.01~d$^{-1}$ of the \citet{pedersen2019} values. However, if more than one of our frequencies lay within this range, we designated only the closest frequency to be a match.
The percentage of recovery as a function of the number of chunks for each star is shown in Fig.~\ref{fig:recovery}.

\begin{table*}
    \centering
    \caption{Comparison of modelling results for stars with period spacing patterns found in a blind search of frequencies extracted from TESS data. Results marked with `*' used the $\Pi_0$ value from the two chunk light curve, whereas all others used the three chunk light curve.}
    \begin{tabular}{ccccccccc}
    \hline
         & \multicolumn{3}{c}{model fits -- {\it Kepler} \citep{pedersen2021}} & \multicolumn{3}{c}{model fits -- TESS (this work)} \\
         KIC ID & $M/\rm{M}_{\odot}$ & $M_{\rm c}/{\rm M_\odot}$ & $X_{\rm c}/X_{\rm{ini}}$ & $M/{\rm M_\odot}$ & $M_{\rm c}/{\rm M_\odot}$ & $X_{\rm c}/X_{\rm{ini}}$ \\
         \hline
         
        4930889 & $4.135\pm0.493$ & 0.62 & $0.39\pm0.10$ & $4.7^{+4.3}_{-2.6}$ & $0.35^{+1.25}_{-0.35}$ & $0.01^{+0.97}_{-0.01}$ \\
        \\
        
        \vspace{0.1cm}
        
        5941844 & $3.559\pm0.089$ & 0.85 & $0.89\pm0.06$ & $9.0_{-4.4}$ & $2.31^{+0.45}_{-1.52}$ & $0.56^{+0.41}_{-0.49}$ \\
        \\

        \vspace{0.1cm}
        
        7760680 & $3.466\pm0.081$ & 0.62 & $0.64\pm0.17$ & $^*3.2^{+3.3}_{-1.6}$ & $^*0.44^{+0.94}_{-0.44}$ & $^*0.43^{+0.55}_{-0.43}$ \\
        
         &    &    &    & $3.2^{+2.8}_{-1.6}$ & $0.45^{+0.86}_{-0.45}$ & $0.45^{+0.53}_{-0.45}$ \\
        \hline
    \end{tabular}
    \label{tab:modelling}
\end{table*}

\begin{table*}
    \centering
    \caption{Same as Table~\ref{tab:modelling}, but for stars with period spacing patterns informed by the patterns reported by \citet{pedersen2021}.}
    \begin{tabular}{ccccccc}
    \hline
         & \multicolumn{3}{c}{model fits -- {\it Kepler} \citep{pedersen2021}} & \multicolumn{3}{c}{model fits -- TESS (this work)}  \\
         KIC ID & $M/\rm{M}_{\odot}$ & $M_{\rm c}/{\rm M_\odot}$ & $X_{\rm c}/X_{\rm{ini}}$ & $M/{\rm M_\odot}$ & $M_{\rm c}/{\rm M_\odot}$ & $X_{\rm c}/X_{\rm{ini}}$ \\
         \hline
         3459297 & $3.876\pm0.690$ & 0.25 & $0.24\pm0.13$ & $4.0^{+5.0}_{-1.8}$ & $0.01^{+1.07}_{-0.01}$ & $0.04^{+0.11}$ \\
        \\
        6352430 & $3.333\pm0.171$ & 0.40 & $0.36\pm0.21$ & $3.60^{+4.15}_{-1.90}$ & $0.66^{+1.36}_{-0.48}$ & $0.57^{+0.40}_{-0.56}$ \\

         \hline
         
    \end{tabular}
    \label{tab:modelling_may}
\end{table*}

\begin{table*}
    \centering
    \caption{Comparison of modelling results fitting $\Pi_0$ from {\it Kepler} data to the \citet{pedersen2021} results.}
    \begin{tabular}{ccccccccc}
    \hline
         & \multicolumn{3}{c}{model fits -- {\it Kepler} \citep{pedersen2021}} & \multicolumn{3}{c}{model fits -- {\it Kepler} (this work)} \\
         KIC ID & $M/\rm{M}_{\odot}$ & $M_{\rm c}/{\rm M_\odot}$ & $X_{\rm c}/X_{\rm{ini}}$ & $M/{\rm M_\odot}$ & $M_{\rm c}/{\rm M_\odot}$ & $X_{\rm c}/X_{\rm{ini}}$\\
         \hline
         
        4930889 & $4.135\pm0.493$ & 0.62 & $0.39\pm0.10$ & $4.5^{+4.5}_{-2.5}$ & $0.22^{+1.04}_{-0.22}$ & $0.00^{+0.97}$ \\
        \\
        
        \vspace{0.1cm}

        5941844 & $3.559\pm0.089$ & 0.85 & $0.89\pm0.06$ & $3.5^{+2.5}_{-1.5}$ & $0.76^{+0.83}_{-0.48}$ & $0.94^{+0.04}_{-0.63}$ \\
        \\

        \vspace{0.1cm}
        
        7760680 & $3.466\pm0.081$ & 0.62 & $0.64\pm0.17$ & $3.10^{+4.15}_{-1.60}$ & $0.52^{+1.10}_{-0.45}$ & $0.53^{+0.44}_{-0.53}$ \\
        \hline
    \end{tabular}
    \label{tab:modelling_kep}
\end{table*}

Our $\Pi_0$ values and core rotation rates from fitting the period spacing pattern using {\sc amigo} are shown in Table\,\ref{tab:obs_constraints}, along with the spectroscopic constraints used in our fitting. For the 3-chunk data of KIC~7760680, our core rotation rate of $0.42\pm0.06$~d$^{-1}$ is remarkably similar to the value of $0.455\pm0.022$~d$^{-1}$ reported by \citet{pedersen2021},  and to a lesser extent the value of $0.4790^{+0.0066}_{-0.0094}$~d$^{-1}$ reported by \citet{moravveji2016}, both using the 4-yr {\it Kepler} light curves. Our rotation rate for KIC~4930889 also agrees with the \citet{pedersen2021} value, as does the value from the 2-chunk light curve of KIC~7760680, although this confidence interval is the largest for all our stars. The rotation rates of the other stars, whilst having relatively high precision constraints, do not agree within $1\sigma$ between our work and the values reported by \citet{pedersen2021}. This demonstrates how it is possible to achieve high precision but low accuracy in measuring near-core rotation rates from period spacing patterns, because of the potential subjectivity and data quality issues when manually building period spacing patterns. The stars with the best agreement with the {\it Kepler} results were those with the longest period spacing patterns.

The results of forward asteroseismic modelling based on the MD as merit function and a comparison of observed and theoretically predicted $\Pi_0$, $T_{\rm eff}$ and $\log\,g$ parameters for all stars are shown in Tables~\ref{tab:modelling} and \ref{tab:modelling_may}. In these tables, we provide the best-fitting mass, convective core mass, and central hydrogen fraction, along with the $68^{\rm th}$ percentile confidence intervals {(i.e. 1$\sigma$)} calculated using Eq.~\ref{eq:probability}. The values of $f_{\rm CBM}$ and $D_{\rm ext}$ were largely unconstrained in all cases, similar to \citet{johnston2019}, so are not included in Table~\ref{tab:modelling}. The results for individual stars are discussed in Section~\ref{sec:results_best}.

For KIC~4930889, KIC~5941844, and KIC~7760680, we were able to find period spacing patterns using the TESS mission light curves. Our modelling results are compatible to those reported by \citet{pedersen2021} based on 4-yr {\it Kepler} light curves. The results of forward asteroseismic modelling are provided in Table~\ref{tab:modelling}, and the stars are discussed individually in Section~\ref{sec:results_best}. However, we did not extract a robust period spacing pattern using a `blind search' approach for the other five SPB stars in our sample, due to a combination of complex spectral windows and higher noise than the {\it Kepler} light curves. For KIC~3459297 and KIC~6352430 we were able to use the frequencies from the \citet{pedersen2021} patterns to inform our extraction of frequencies in the TESS light curves and successfully determine a compatible $\Pi_0$ value. The results of the fits for these `informed' stars are shown in Table~\ref{tab:modelling_may}. Summary figures of the forward asteroseismic modelling for all stars with successful period spacing pattern extraction are provided in Appendix~\ref{sec:app_figures}. We discuss the modelling results for KIC~3459297 and KIC~6352430 in Section~\ref{sec:results_pedersenpattern}. For KIC~3240411, KIC~8766405 and KIC~11360704, we only recovered two to three frequencies matching the \citet{pedersen2021} patterns (see Table~\ref{tab:recovered} and Fig.~\ref{fig:recovery}), so we were unable to continue our analysis using the TESS data, and thus do not report any forward asteroseismic modelling results. 

\subsection{Reanalysis of Kepler data}
\label{sec:kepler}

In their approach to forward asteroseismic modelling, \citet{pedersen2021} used the MD to fit individual observed frequencies to theoretical counterparts calculated by models. Whereas in this work, owing to the frequency resolution issues of TESS mission light curves, we opt to model the stars using the global asteroseismic parameter, $\Pi_0$. Therefore, to compare like to like, we reanalysed the {\it Kepler} light curves using our light curve extraction method, the only difference being the background detrending. Since the {\it Kepler} target pixel file data usually contain far fewer pixels than the 20 by 20 TESS full-frame image cutouts, we often did not have enough background pixels to perform PCA effectively. Thus we instead detrended the background systematics using a smoothed cubic spline fit. We performed frequency extraction of the {\it Kepler} light curves using iterative pre-whitening in the same way as we did for the TESS data, and determined values of $\Pi_0$ from the same period spacing patterns reported by \citet{pedersen2021} of $5420\pm860$, $7800\pm270$ and $8400\pm800$\,s for KIC\,4930889, KIC\,5941844 and KIC\,7760680, respectively. The best-fitting parameters for these stars are listed in Table~\ref{tab:modelling_kep}. Our mass values for KIC~4930889 and KIC~5941844 lie within the \citet{pedersen2021} confidence interval, but we find a lower mass for KIC~7760680. For central hydrogen fraction, our results agree with the \citet{pedersen2021} values for KIC~5941844 and KIC~7760680, but not KIC~4930889. This demonstrates that a different model grid and the use of $\Pi_0$ rather than individual frequencies for model fitting can produce different results, even with the same light curve data.

\subsection{Stars with clear period spacing patterns}
\label{sec:results_best}

Using a completely `blind' approach, we successfully extracted period spacing patterns for KIC\,4930889, KIC\,5941844 and KIC\,7760680 using TESS mission light curves and compared the masses, ages and core masses to those reported by \citet{pedersen2021} based on 4-yr {\it Kepler} light curves. These stars, being in our good sample, are all bright with high amplitude pulsations. For KIC\,7760680, we found patterns in the two and three chunk data, whereas for KIC\,4930889 and KIC\,5941844, we only identified a pattern in the three chunk data.

When performing forward asteroseimic modelling of KIC~4930889, we discovered that the confidence intervals encompassed nearly the whole range for mass and central hydrogen fraction, $X_{\rm c}$, as shown in Table~\ref{tab:modelling}. The core mass was better constrained, with the upper limit of the $1\sigma$ confidence interval at 1.6\,M$_{\odot}$; the highest core mass in our grid is 2.75\,M$_{\odot}$. Despite our larger uncertainties, our best-fitting mass is close to $1\sigma$ of that reported by \citet{pedersen2021}. It is known that generally large confidence intervals result on forward asteroseismic modelling using $\Pi_0$ and the MD as merit function \citep{johnston2019}, so this aspect is not unexpected. This is because $\Pi_0$ is sensitive to many other parameters (e.g. mass, age and mixing), so is thus degenerate within the HR~diagram.

\begin{figure}
\centering
    \includegraphics[width=\linewidth]{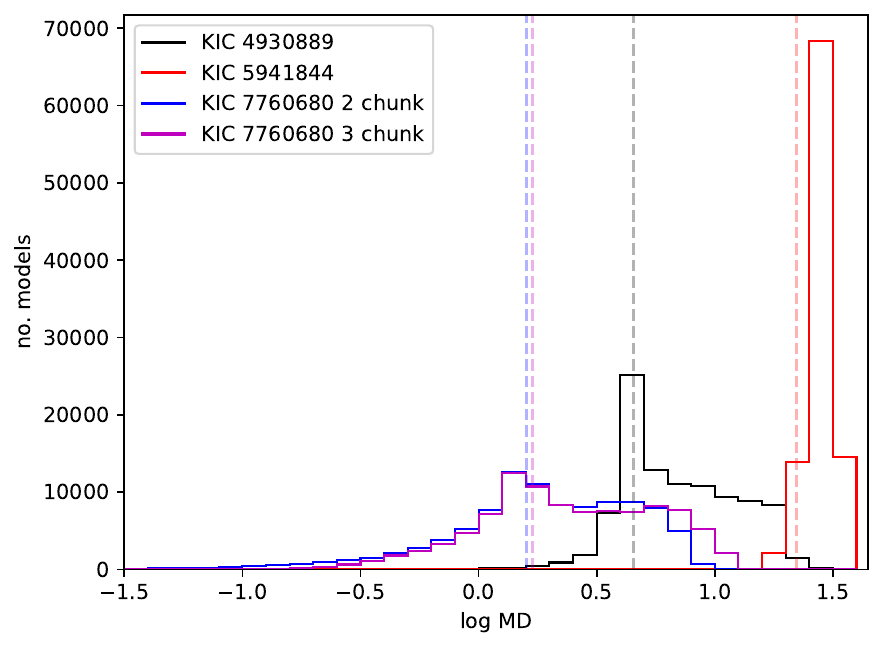}\\
    \caption{Number of grid models at different values of the log Mahalanobis distance MD for the fitting results of KIC~4930889, KIC~5941844 and KIC~7760680. The dashed lines are placed at the MD of the $68^{\rm th}$ percentile of the probability distribution, which we use for the 1-$\sigma$ confidence interval.}
    \label{fig:MD}
\end{figure}

%\subsection{KIC~5941844}
%\label{sec:results_5941844}
For KIC~5941844, we found a very high value of $\Pi_0 = 22850\pm2380$\,s, which is beyond the maximum value in our grid (see Fig.~\ref{fig:Pi0}). Such a high value of $\Pi_0$ is unlikely for a low-mass SPB star (e.g. \citealt{bowman2020}) and would have been given very little weight in the calculation of the MD due to the inclusion of correlations between parameters in the grid. Comparison with \citet{pedersen2021} shows that we were unable to extract lower-amplitude frequencies using the TESS light curves that would have made a pattern with a lower spacing more obvious. In addition, we only identified three period spacings with which to build the pattern, which adds further uncertainty to this value of $\Pi_0$. As a result of this, we derived a much higher mass than \citet{pedersen2021}. Moreover, our best model had a high MD value (see Fig.~\ref{fig:MD}), thus demonstrating that our $\Pi_0$ was not compatible with the spectroscopic observations. Therefore, this star showcases the limitations of performing forward asteroseismic modelling of sparse period spacing patterns given the large disagreement between our results and those of \citet{pedersen2021}.

%\subsection{KIC\,7760680}
%\label{sec:results_7760680}
The final good SPB star in our sample is KIC\,7760680, which is the most well studied SPB star in the {\it Kepler} mission data set. It currently holds the record for the longest period spacing pattern of 36 consecutive radial order g~modes in an SPB star \citep{papics2015,moravveji2016}.
For KIC\,7760680, we successfully measured $\Pi_0$ values of $7180\pm1530$\,s and $6700\pm1100$\,s from the two- and three-chunk data sets, respectively. Our forward asteroseismic modelling yielded a best fit mass that is over $3\sigma$ lower than the \citet{pedersen2021} results, but within $2\sigma$ for $X_{\rm c}$. Using the two-chunk light curve (marked with a `*' in Table~\ref{tab:modelling}) we find a very similar best model as the modelling based on the three chunk light curve, however our confidence intervals are slightly more precise for mass and core mass using the three chunk light curve. The scatter in modelling results for both this analysis and various other literature can be seen in the bottom panel of Fig.~\ref{fig:app_7760680}.

We find large confidence intervals on our best fit model parameters, similar to \citet{johnston2019} who also fit $\Pi_0$ rather than individual frequencies. This is due to the  marginalisation of the probability distribution over individual parameters, which effectively assigns the entire error budget to that parameter. It is possible that using e.g. a $\chi^2$ fit may produce a result that appears better-constrained, but the MD provides a more realistic estimate of the true uncertainties due to its accounting for the correlations between fitting parameters. More detailed information on the constraining power of our fits can be seen in the distributions of MD values, which are shown for KIC~7760680, KIC~4930889 and KIC~5941844 in Fig.~\ref{fig:MD}. For KIC~7760680, the dashed line marking the 68$^{\rm th}$ percentile of the probability distribution is at the lowest out of the three stars, indicating that this fit is the best-constrained. For KIC~5941844, the distribution of MD values is much higher, indicating that the fit is comparatively poor; this, along with the large difference to the {\it Kepler}-derived mass, further indicates that we were unable to measure the correct $\Pi_0$ value from the TESS light curves for this star.

\subsection{Stars with period spacings informed by those reported by \citet{pedersen2021}}
\label{sec:results_pedersenpattern}

We did not initially identify a period spacing pattern in the extracted frequencies for KIC~3459297 and KIC~6352430. For KIC~3459297, we were able to match nine of the frequencies from the three chunk data to the period spacing pattern reported by \citet{pedersen2021} to determine $\Pi_0$. However, due to gaps in the pattern and very few consecutive radial orders identified, and we were only able to use four period spacings in the fit. As in the case of KIC~5941844, this makes the value of $\Pi_0$ less reliable than for those stars with longer patterns. Only our best-fitting mass for this star from forward asteroseismic modelling is within $1\sigma$ of the \citet{pedersen2021} best model, but the core mass and central hydrogen values differ, as shown in Table~\ref{tab:modelling_may}. 

%\subsection{KIC\,6352430}
%\label{sec:results_6352430}
For KIC\,6352430, we also could not find a reliable period spacing pattern with a `blind' approach, but were able to match some frequencies in the TESS light curve to members of the period spacing pattern reported by \citet{pedersen2021} using 4-yr {\it Kepler} data. The recovery fraction of pulsation frequencies in the pattern was relatively high, at nearly 40~per~cent (see Fig.~\ref{fig:recovery}). However, we did not recover the important lower-amplitude frequencies which are needed to fill gaps in the period spacing pattern. We again fitted three period spacings, mostly in the lower period range, and found $\Pi_0 = 9940 \pm 1350$\,s. Our best-fitting model for this star from forward asteroseismic modelling is within $1\sigma$ of the \citet{pedersen2021} best model for mass and central hydrogen fraction, but the confidence intervals on the parameters are large, as shown in Table~\ref{tab:modelling_may}.

\section{Discussion and Conclusions}
\label{sec:disc}

In this work, we have developed and applied a new method to extract TESS light curves of pulsating early-type stars. Our method used the S/N of the dominant pulsation mode to determine the optimum aperture, and often resulted in a different aperture than would be obtained using the commonly employed flux threshold approach (see top panels of Fig.\,\ref{fig:summary}). By optimising aperture mask based on the S/N of pulsations in the resultant frequency spectrum in this way, we maximise the ability to extract as many frequencies as possible from short TESS light curves. This is important for subsequent frequency analysis, as well as determining accurate values of $\Pi_0$ and performing forward asteroseismic modelling. Indeed, asteroseismic observables such as $\Pi_0$ are most constraining when derived from uninterrupted g-mode period spacing patterns spanning a large number of consecutive radial orders.

We extracted frequencies from TESS light curves with different lengths (i.e. 1, 2 and 3 chunks of consecutive TESS sectors) for a sample of eight SPB stars originally studied using {\it Kepler} mission data, and compared the TESS frequencies to those in the period spacing patterns reported by \citet{pedersen2021}. We found between 12 and 60~per~cent of the {\it Kepler} pulsation frequencies within the period spacing patterns were recovered from our 3-chunk light curves. The brightest stars with the highest-amplitude pulsations performed better in the frequency recovery fraction, although adding more chunks of data did not always improve this. In some cases, adding more sectors did not increase the percentage of completeness of the pattern. This demonstrates that whilst it is commonly quoted as `the more data the better' in asteroseismology, it is highly star-dependent on how many TESS sectors are needed to accurately extract a g-mode period spacing pattern. For example, we recovered period spacing patterns in three stars, despite recovering only 60~per~cent of the pattern for KIC\,4930889 and less than 40~per~cent of the pattern for KIC\,5941844 and KIC\,7760680. We find that the overall success of TESS data in providing modelling results similar to {\it Kepler} is highly dependant on the individual star. For KIC~7760680, we can also see that the 3-chunk light curves does not result in an increase in the recovery percentage of the {\it Kepler} period spacing patterns. In this case, sufficient consecutive frequencies were present in the 2-chunk data to produce similar modelling results. However, this is likely due to the exceptionally long period spacing pattern comprising 36 consecutive radial order prograde dipole modes identified in {\it Kepler} data for KIC~7760680. The majority of other SPB stars that have undergone forward asteroseismic modelling have fewer than 20 gravity modes in their pattern, as is the case for 18 of the 26 SPB stars studied by \citet{pedersen2021}.

We determined $\Pi_0$ values from the period spacing patterns for our TESS data, and combined these with spectroscopic constraints from the literature to perform forward asteroseismic modelling. Our best-fitting model results generally lay within 2-$\sigma$ of the parameters reported by \citet{pedersen2021}, although our confidence intervals were much larger and capture a sizeable fraction of the theoretical grid of models. The marginalisation over the probability distribution onto a single parameter to determine the confidence intervals neglects portions of the uncertainty due to the other parameters, thus resulting in large uncertainties despite the best model being well-constrained. This was also expected (see \citealt{johnston2019}) given the degeneracy of $\Pi_0$ across the parameter space (i.e. mass and age) in the HR~diagram. This underpins the power but potential drawbacks of individual frequency fitting in forward asteroseismic modelling. For example, our use of $\Pi_0$ as an observable has some advantages over individual frequency fitting, such that it probes the bulk density profile over the near-core region, making it is less susceptible to sharp local minima in the posterior parameter space compared to individual frequency fitting. On the other hand, such sharp features provide finer detail on the interior physics of stars. In this sense, fitting with $\Pi_0$ could be described as low-precision high accuracy and potentially vice versa for individual frequency fitting --- see discussion by \citet{bowman2021}.

Our results demonstrate the potential biases in the methods of asteroseismology that intermittent TESS data alone can be prone to. For example, those stars for which we found period spacing patterns in TESS data were all bright with no nearby contaminating neighbours. However, KIC\,8766405 was also in our `good' sample, but we were not able to find a period spacing pattern due to the intermittent TESS data not having sensitivity to the presence of low-amplitude modes in the period spacing pattern reported by \citet{pedersen2021}. Selecting bright, isolated stars is a good approach to increase the chance of successful asteroseismology when using only TESS data, but does not guarantee that the true period spacing pattern can be extracted. For example, the fitted value of $\Pi_0$ for KIC~5941844 -- one of the `good' stars -- was unrealistically high, and therefore yielded a very high mass. In this case the limitations of the TESS data meant that the period spacing pattern was not complete enough to yield correct stellar parameters in forward modelling.

To ensure a like-for-like comparison in data analysis methods, as well as forward asteroseismic modelling methodology, we reanalysed the {\it Kepler} light curves of the best three SPB stars in our sample using our new light-curve extraction method. The differences between these results and the \citet{pedersen2021} analysis also using {\it Kepler} data for the same stars show that much of the scatter between different studies may arise from differences in analysis methodology, such as the model grid and the fitting parameters (e.g. $\Pi_0$ versus individual frequencies). In conclusion, our work demonstrates that individual and careful analysis in every aspect from light curve extraction to building period spacing patterns is critical for reliable asteroseismology of g-mode pulsators, since the potential data and methodological biases are highly star dependent when using only TESS data.

\section*{Acknowledgements}
The authors would like to thank M. G. Pedersen for useful discussions on this project. We gratefully acknowledge UK Research and Innovation (UKRI) in the form of a Frontier Research grant under the UK government's ERC Horizon Europe funding guarantee (SYMPHONY; grant number: EP/Y031059/1), and a Royal Society University Research Fellowship (grant number: URF{\textbackslash}R1{\textbackslash}231631). The {\it Kepler} and TESS data presented in this paper were obtained from the Mikulski Archive for Space Telescopes (MAST; \url{https://archive.stsci.edu/missions-and-data/tess}) at the Space Telescope Science Institute (STScI). The Mikulski Archive for Space Telescopes (MAST) is operated by the Association of Universities for Research in Astronomy, Inc., under NASA contract NAS5-26555. Support to MAST for these data is provided by the NASA Office of Space Science via grant NAG5-7584 and by other grants and contracts. Funding for the Kepler/K2 mission was provided by NASA’s Science Mission Directorate, and funding for the TESS mission is provided by the NASA Explorer Program.

%%%%%%%%%%%%%%%%%%%%%%%%%%%%%%%%%%%%%%%%%%%%%%%%%%
\section*{Data Availability}

This research made use of the following open-access software packages: {\sc lightkurve} (\url{https://lightkurve.github.io/lightkurve/}), a \texttt{Python} package for {\it Kepler} and TESS data analysis \citep{lightkurve2018}, {\sc Period04} (\url{https://www.period04.net}) for frequency analysis \citep{lenz2005}, {\sc amigo} (\url{https://github.com/TVanReeth/amigo}) for period spacing pattern analysis \citep{vanreeth2016, vanreeth2018, vanreeth2022}, as well as {\tt astropy} \citep{astropy2022}, {\tt matplotlib} \citep{matplotlib2007} and {\tt numpy} \citep{numpy2006, numpy2011, numpy2020}.

%For the purpose of open access, the authors have applied a CC BY licence to the author accepted manuscript version: \url{https://arxiv.org/abs/TBD}. Data products that support the results in this paper are publicly available via the Zenodo repository: \url{https://zenodo.org/records/TBD}.

%The inclusion of a Data Availability Statement is a requirement for articles published in MNRAS. Data Availability Statements provide a standardised format for readers to understand the availability of data underlying the research results described in the article. The statement may refer to original data generated in the course of the study or to third-party data analysed in the article. The statement should describe and provide means of access, where possible, by linking to the data or providing the required accession numbers for the relevant databases or DOIs.



\bibliographystyle{mnras}
\bibliography{bibby}

\begin{thebibliography}{}
\makeatletter
\relax
\def\mn@urlcharsother{\let\do\@makeother \do\$\do\&\do\#\do\^\do\_\do\%\do\~}
\def\mn@doi{\begingroup\mn@urlcharsother \@ifnextchar [ {\mn@doi@} {\mn@doi@[]}}
\def\mn@doi@[#1]#2{\def\@tempa{#1}\ifx\@tempa\@empty \href {http://dx.doi.org/#2} {doi:#2}\else \href {http://dx.doi.org/#2} {#1}\fi \endgroup}
\def\mn@eprint#1#2{\mn@eprint@#1:#2::\@nil}
\def\mn@eprint@arXiv#1{\href {http://arxiv.org/abs/#1} {{\tt arXiv:#1}}}
\def\mn@eprint@dblp#1{\href {http://dblp.uni-trier.de/rec/bibtex/#1.xml} {dblp:#1}}
\def\mn@eprint@#1:#2:#3:#4\@nil{\def\@tempa {#1}\def\@tempb {#2}\def\@tempc {#3}\ifx \@tempc \@empty \let \@tempc \@tempb \let \@tempb \@tempa \fi \ifx \@tempb \@empty \def\@tempb {arXiv}\fi \@ifundefined {mn@eprint@\@tempb}{\@tempb:\@tempc}{\expandafter \expandafter \csname mn@eprint@\@tempb\endcsname \expandafter{\@tempc}}}

\bibitem[\protect\citeauthoryear{{Aerts}}{{Aerts}}{2021}]{aerts2021}
{Aerts} C.,  2021, \mn@doi [Reviews of Modern Physics] {10.1103/RevModPhys.93.015001}, \href {https://ui.adsabs.harvard.edu/abs/2021RvMP...93a5001A} {93, 015001}

\bibitem[\protect\citeauthoryear{{Aerts}, {Christensen-Dalsgaard}  \& {Kurtz}}{{Aerts} et~al.}{2010}]{aerts2010}
{Aerts} C.,  {Christensen-Dalsgaard} J.,   {Kurtz} D.~W.,  2010, {Asteroseismology}, \mn@doi{10.1007/978-1-4020-5803-5.
}

\bibitem[\protect\citeauthoryear{{Aerts} et~al.,}{{Aerts} et~al.}{2018}]{aerts2018}
{Aerts} C.,  et~al., 2018, \mn@doi [\apjs] {10.3847/1538-4365/aaccfb}, \href {http://adsabs.harvard.edu/abs/2018ApJS..237...15A} {237, 15}

\bibitem[\protect\citeauthoryear{{Aerts}, {Mathis}  \& {Rogers}}{{Aerts} et~al.}{2019}]{aerts2019}
{Aerts} C.,  {Mathis} S.,   {Rogers} T.~M.,  2019, \mn@doi [\araa] {10.1146/annurev-astro-091918-104359}, \href {https://ui.adsabs.harvard.edu/abs/2019ARA&A..57...35A} {57, 35}

\bibitem[\protect\citeauthoryear{{Aigrain}, {Parviainen}  \& {Pope}}{{Aigrain} et~al.}{2016}]{aigrain2016}
{Aigrain} S.,  {Parviainen} H.,   {Pope} B.~J.~S.,  2016, \mn@doi [\mnras] {10.1093/mnras/stw706}, \href {https://ui.adsabs.harvard.edu/abs/2016MNRAS.459.2408A} {459, 2408}

\bibitem[\protect\citeauthoryear{{Anders} \& {Pedersen}}{{Anders} \& {Pedersen}}{2023}]{anders2023}
{Anders} E.~H.,  {Pedersen} M.~G.,  2023, \mn@doi [Galaxies] {10.3390/galaxies11020056}, \href {https://ui.adsabs.harvard.edu/abs/2023Galax..11...56A} {11, 56}

\bibitem[\protect\citeauthoryear{{Astropy Collaboration} et~al.,}{{Astropy Collaboration} et~al.}{2022}]{astropy2022}
{Astropy Collaboration} et~al., 2022, \mn@doi [\apj] {10.3847/1538-4357/ac7c74}, \href {https://ui.adsabs.harvard.edu/abs/2022ApJ...935..167A} {935, 167}

\bibitem[\protect\citeauthoryear{{Auvergne} et~al.,}{{Auvergne} et~al.}{2009}]{auvergne2009}
{Auvergne} M.,  et~al., 2009, \mn@doi [\aap] {10.1051/0004-6361/200810860}, \href {http://adsabs.harvard.edu/abs/2009A%26A...506..411A} {506, 411}

\bibitem[\protect\citeauthoryear{{Baran} \& {Koen}}{{Baran} \& {Koen}}{2021}]{baran2021}
{Baran} A.~S.,  {Koen} C.,  2021, \mn@doi [\actaa] {10.32023/0001-5237/71.2.3}, \href {https://ui.adsabs.harvard.edu/abs/2021AcA....71..113B} {71, 113}

\bibitem[\protect\citeauthoryear{{Borucki} et~al.,}{{Borucki} et~al.}{2010}]{borucki2010}
{Borucki} W.~J.,  et~al., 2010, \mn@doi [Science] {10.1126/science.1185402}, \href {http://adsabs.harvard.edu/abs/2010Sci...327..977B} {327, 977}

\bibitem[\protect\citeauthoryear{{Bowman}}{{Bowman}}{2017}]{bowman2017}
{Bowman} D.~M.,  2017, {Amplitude Modulation of Pulsation Modes in Delta Scuti Stars}.
Springer International Publishing, \mn@doi{10.1007/978-3-319-66649-5}

\bibitem[\protect\citeauthoryear{{Bowman}}{{Bowman}}{2020}]{bowman2020}
{Bowman} D.~M.,  2020, \mn@doi [Frontiers in Astronomy and Space Sciences] {10.3389/fspas.2020.578584}, \href {https://ui.adsabs.harvard.edu/abs/2020FrASS...7...70B} {7, 70}

\bibitem[\protect\citeauthoryear{{Bowman}}{{Bowman}}{2023}]{bowman2023}
{Bowman} D.~M.,  2023, \mn@doi [\apss] {10.1007/s10509-023-04262-7}, \href {https://ui.adsabs.harvard.edu/abs/2023Ap&SS.368..107B} {368, 107}

\bibitem[\protect\citeauthoryear{{Bowman} \& {Holdsworth}}{{Bowman} \& {Holdsworth}}{2019}]{bowman&holdsworth2019}
{Bowman} D.~M.,  {Holdsworth} D.~L.,  2019, \mn@doi [\aap] {10.1051/0004-6361/201935640}, \href {https://ui.adsabs.harvard.edu/abs/2019A&A...629A..21B} {629, A21}

\bibitem[\protect\citeauthoryear{{Bowman} \& {Michielsen}}{{Bowman} \& {Michielsen}}{2021}]{bowman2021}
{Bowman} D.~M.,  {Michielsen} M.,  2021, \mn@doi [\aap] {10.1051/0004-6361/202141726}, \href {https://ui.adsabs.harvard.edu/abs/2021A&A...656A.158B} {656, A158}

\bibitem[\protect\citeauthoryear{{Bowman}, {Buysschaert}, {Neiner}, {P{\'a}pics}, {Oksala}  \& {Aerts}}{{Bowman} et~al.}{2018}]{bowman2018}
{Bowman} D.~M.,  {Buysschaert} B.,  {Neiner} C.,  {P{\'a}pics} P.~I.,  {Oksala} M.~E.,   {Aerts} C.,  2018, \mn@doi [\aap] {10.1051/0004-6361/201833037}, \href {https://ui.adsabs.harvard.edu/abs/2018A&A...616A..77B} {616, A77}

\bibitem[\protect\citeauthoryear{{Bowman} et~al.,}{{Bowman} et~al.}{2019}]{bowman2019}
{Bowman} D.~M.,  et~al., 2019, \mn@doi [Nature Astronomy] {10.1038/s41550-019-0768-1}, \href {https://ui.adsabs.harvard.edu/abs/2019NatAs...3..760B} {3, 760}

\bibitem[\protect\citeauthoryear{{Bowman} et~al.,}{{Bowman} et~al.}{2022}]{bowman2022}
{Bowman} D.~M.,  et~al., 2022, \mn@doi [\aap] {10.1051/0004-6361/202142375}, \href {https://ui.adsabs.harvard.edu/abs/2022A&A...658A..96B} {658, A96}

\bibitem[\protect\citeauthoryear{{Bowman}, {Van Daele}, {Michielsen}  \& {Van Reeth}}{{Bowman} et~al.}{2024}]{Bowman2024}
{Bowman} D.~M.,  {Van Daele} P.,  {Michielsen} M.,   {Van Reeth} T.,  2024, \mn@doi [\aap] {10.1051/0004-6361/202451419}, \href {https://ui.adsabs.harvard.edu/abs/2024A&A...692A..49B} {692, A49}

\bibitem[\protect\citeauthoryear{{Brasseur}, {Phillip}, {Fleming}, {Mullally}  \& {White}}{{Brasseur} et~al.}{2019}]{brasseur2019}
{Brasseur} C.~E.,  {Phillip} C.,  {Fleming} S.~W.,  {Mullally} S.~E.,   {White} R.~L.,  2019, {Astrocut: Tools for creating cutouts of TESS images} (\mn@eprint {ascl} {1905.007})

\bibitem[\protect\citeauthoryear{{Burssens} et~al.,}{{Burssens} et~al.}{2020}]{burssens2020}
{Burssens} S.,  et~al., 2020, \mn@doi [\aap] {10.1051/0004-6361/202037700}, \href {https://ui.adsabs.harvard.edu/abs/2020A&A...639A..81B} {639, A81}

\bibitem[\protect\citeauthoryear{{Burssens} et~al.,}{{Burssens} et~al.}{2023}]{burssens2023}
{Burssens} S.,  et~al., 2023, \mn@doi [Nature Astronomy] {10.1038/s41550-023-01978-y}, \href {https://ui.adsabs.harvard.edu/abs/2023NatAs...7..913B} {7, 913}

\bibitem[\protect\citeauthoryear{{Buysschaert}, {Aerts}, {Bowman}, {Johnston}, {Van Reeth}, {Pedersen}, {Mathis}  \& {Neiner}}{{Buysschaert} et~al.}{2018}]{buysschaert2018}
{Buysschaert} B.,  {Aerts} C.,  {Bowman} D.~M.,  {Johnston} C.,  {Van Reeth} T.,  {Pedersen} M.~G.,  {Mathis} S.,   {Neiner} C.,  2018, \mn@doi [\aap] {10.1051/0004-6361/201832642}, \href {http://adsabs.harvard.edu/abs/2018A%26A...616A.148B} {616, A148}

\bibitem[\protect\citeauthoryear{{Chaplin} \& {Miglio}}{{Chaplin} \& {Miglio}}{2013}]{chaplin2013}
{Chaplin} W.~J.,  {Miglio} A.,  2013, \mn@doi [\araa] {10.1146/annurev-astro-082812-140938}, \href {http://adsabs.harvard.edu/abs/2013ARA%26A..51..353C} {51, 353}

\bibitem[\protect\citeauthoryear{{Degroote} et~al.,}{{Degroote} et~al.}{2009}]{degroote2009}
{Degroote} P.,  et~al., 2009, \mn@doi [\aap] {10.1051/0004-6361/200911782}, \href {https://ui.adsabs.harvard.edu/abs/2009A&A...506..111D} {506, 111}

\bibitem[\protect\citeauthoryear{{Degroote} et~al.,}{{Degroote} et~al.}{2010}]{degroote2010}
{Degroote} P.,  et~al., 2010, \mn@doi [\nat] {10.1038/nature08864}, \href {http://adsabs.harvard.edu/abs/2010Natur.464..259D} {464, 259}

\bibitem[\protect\citeauthoryear{{Dziembowski}, {Moskalik}  \& {Pamyatnykh}}{{Dziembowski} et~al.}{1993}]{dziembowski1993}
{Dziembowski} W.~A.,  {Moskalik} P.,   {Pamyatnykh} A.~A.,  1993, \mnras, \href {http://adsabs.harvard.edu/abs/1993MNRAS.265..588D} {265, 588}

\bibitem[\protect\citeauthoryear{{Freytag}, {Ludwig}  \& {Steffen}}{{Freytag} et~al.}{1996}]{freytag1996}
{Freytag} B.,  {Ludwig} H.~G.,   {Steffen} M.,  1996, \aap, \href {https://ui.adsabs.harvard.edu/abs/1996A&A...313..497F} {313, 497}

\bibitem[\protect\citeauthoryear{{Gebruers} et~al.,}{{Gebruers} et~al.}{2021}]{gebruers2021}
{Gebruers} S.,  et~al., 2021, \mn@doi [\aap] {10.1051/0004-6361/202140466}, \href {https://ui.adsabs.harvard.edu/abs/2021A&A...650A.151G} {650, A151}

\bibitem[\protect\citeauthoryear{{Hanes}, {Waskie}, {Labadie-Bartz}, {Wall}, {Boyer}  \& {McSwain}}{{Hanes} et~al.}{2019}]{hanes2019}
{Hanes} R.~J.,  {Waskie} S.,  {Labadie-Bartz} J.~M.,  {Wall} A.,  {Boyer} A.,   {McSwain} M.~V.,  2019, \mn@doi [\aj] {10.3847/1538-3881/ab064e}, \href {https://ui.adsabs.harvard.edu/abs/2019AJ....157..129H} {157, 129}

\bibitem[\protect\citeauthoryear{Harris et~al.,}{Harris et~al.}{2020}]{numpy2020}
Harris C.~R.,  et~al., 2020, \mn@doi [Nature] {10.1038/s41586-020-2649-2}, 585, 357

\bibitem[\protect\citeauthoryear{{Higgins} \& {Bell}}{{Higgins} \& {Bell}}{2023}]{higgins2023}
{Higgins} M.~E.,  {Bell} K.~J.,  2023, \mn@doi [\aj] {10.3847/1538-3881/acb20c}, \href {https://ui.adsabs.harvard.edu/abs/2023AJ....165..141H} {165, 141}

\bibitem[\protect\citeauthoryear{{Howell} et~al.,}{{Howell} et~al.}{2014}]{howell2014}
{Howell} S.~B.,  et~al., 2014, \mn@doi [\pasp] {10.1086/676406}, \href {http://adsabs.harvard.edu/abs/2014PASP..126..398H} {126, 398}

\bibitem[\protect\citeauthoryear{{Huber} et~al.,}{{Huber} et~al.}{2022}]{huber2022}
{Huber} D.,  et~al., 2022, \mn@doi [\aj] {10.3847/1538-3881/ac3000}, \href {https://ui.adsabs.harvard.edu/abs/2022AJ....163...79H} {163, 79}

\bibitem[\protect\citeauthoryear{{Hunter}}{{Hunter}}{2007}]{matplotlib2007}
{Hunter} J.~D.,  2007, \mn@doi [Computing in Science and Engineering] {10.1109/MCSE.2007.55}, \href {http://adsabs.harvard.edu/abs/2007CSE.....9...90H} {9, 90}

\bibitem[\protect\citeauthoryear{{Jenkins} et~al.,}{{Jenkins} et~al.}{2016}]{jenkins2016}
{Jenkins} J.~M.,  et~al., 2016, in Software and Cyberinfrastructure for Astronomy IV. p. 99133E, \mn@doi{10.1117/12.2233418}

\bibitem[\protect\citeauthoryear{{Johnston}, {Tkachenko}, {Aerts}, {Molenberghs}, {Bowman}, {Pedersen}, {Buysschaert}  \& {P{\'a}pics}}{{Johnston} et~al.}{2019}]{johnston2019}
{Johnston} C.,  {Tkachenko} A.,  {Aerts} C.,  {Molenberghs} G.,  {Bowman} D.~M.,  {Pedersen} M.~G.,  {Buysschaert} B.,   {P{\'a}pics} P.~I.,  2019, \mn@doi [\mnras] {10.1093/mnras/sty2671}, \href {https://ui.adsabs.harvard.edu/abs/2019MNRAS.482.1231J} {482, 1231}

\bibitem[\protect\citeauthoryear{{Koch} et~al.,}{{Koch} et~al.}{2010}]{koch2010}
{Koch} D.~G.,  et~al., 2010, \mn@doi [\apjl] {10.1088/2041-8205/713/2/L79}, \href {http://adsabs.harvard.edu/abs/2010ApJ...713L..79K} {713, L79}

\bibitem[\protect\citeauthoryear{{Kurtz}}{{Kurtz}}{2022}]{kurtz2022}
{Kurtz} D.~W.,  2022, \mn@doi [\araa] {10.1146/annurev-astro-052920-094232}, \href {https://ui.adsabs.harvard.edu/abs/2022ARA&A..60...31K} {60, 31}

\bibitem[\protect\citeauthoryear{{Kurtz}, {Shibahashi}, {Murphy}, {Bedding}  \& {Bowman}}{{Kurtz} et~al.}{2015}]{kurtz2015}
{Kurtz} D.~W.,  {Shibahashi} H.,  {Murphy} S.~J.,  {Bedding} T.~R.,   {Bowman} D.~M.,  2015, \mn@doi [\mnras] {10.1093/mnras/stv868}, \href {https://ui.adsabs.harvard.edu/abs/2015MNRAS.450.3015K} {450, 3015}

\bibitem[\protect\citeauthoryear{{Lecoanet}, {Bowman}  \& {Van Reeth}}{{Lecoanet} et~al.}{2022}]{lecoanet2022}
{Lecoanet} D.,  {Bowman} D.~M.,   {Van Reeth} T.,  2022, \mn@doi [\mnras] {10.1093/mnrasl/slac013}, \href {https://ui.adsabs.harvard.edu/abs/2022MNRAS.512L..16L} {512, L16}

\bibitem[\protect\citeauthoryear{{Lenz} \& {Breger}}{{Lenz} \& {Breger}}{2005}]{lenz2005}
{Lenz} P.,  {Breger} M.,  2005, \mn@doi [Communications in Asteroseismology] {10.1553/cia146s53}, \href {https://ui.adsabs.harvard.edu/abs/2005CoAst.146...53L} {146, 53}

\bibitem[\protect\citeauthoryear{{Lightkurve Collaboration} et~al.,}{{Lightkurve Collaboration} et~al.}{2018}]{lightkurve2018}
{Lightkurve Collaboration} et~al., 2018, {Lightkurve: Kepler and TESS time series analysis in Python}, Astrophysics Source Code Library (\mn@eprint {ascl} {1812.013})

\bibitem[\protect\citeauthoryear{{Michielsen}, {Aerts}  \& {Bowman}}{{Michielsen} et~al.}{2021}]{michielsen2021}
{Michielsen} M.,  {Aerts} C.,   {Bowman} D.~M.,  2021, \mn@doi [\aap] {10.1051/0004-6361/202039926}, \href {https://ui.adsabs.harvard.edu/abs/2021A&A...650A.175M} {650, A175}

\bibitem[\protect\citeauthoryear{{Michielsen}, {Van Reeth}, {Tkachenko}  \& {Aerts}}{{Michielsen} et~al.}{2023}]{michielsen2023}
{Michielsen} M.,  {Van Reeth} T.,  {Tkachenko} A.,   {Aerts} C.,  2023, \mn@doi [\aap] {10.1051/0004-6361/202244192}, \href {https://ui.adsabs.harvard.edu/abs/2023A&A...679A...6M} {679, A6}

\bibitem[\protect\citeauthoryear{{Miglio}, {Montalb{\'a}n}, {Noels}  \& {Eggenberger}}{{Miglio} et~al.}{2008}]{miglio2008}
{Miglio} A.,  {Montalb{\'a}n} J.,  {Noels} A.,   {Eggenberger} P.,  2008, \mn@doi [\mnras] {10.1111/j.1365-2966.2008.13112.x}, \href {https://ui.adsabs.harvard.edu/abs/2008MNRAS.386.1487M} {386, 1487}

\bibitem[\protect\citeauthoryear{{Moravveji}, {Aerts}, {P{\'a}pics}, {Triana}  \& {Vandoren}}{{Moravveji} et~al.}{2015}]{moravveji2015}
{Moravveji} E.,  {Aerts} C.,  {P{\'a}pics} P.~I.,  {Triana} S.~A.,   {Vandoren} B.,  2015, \mn@doi [\aap] {10.1051/0004-6361/201425290}, \href {https://ui.adsabs.harvard.edu/abs/2015A&A...580A..27M} {580, A27}

\bibitem[\protect\citeauthoryear{{Moravveji}, {Townsend}, {Aerts}  \& {Mathis}}{{Moravveji} et~al.}{2016}]{moravveji2016}
{Moravveji} E.,  {Townsend} R. H.~D.,  {Aerts} C.,   {Mathis} S.,  2016, \mn@doi [\apj] {10.3847/0004-637X/823/2/130}, \href {https://ui.adsabs.harvard.edu/abs/2016ApJ...823..130M} {823, 130}

\bibitem[\protect\citeauthoryear{Oliphant}{Oliphant}{2006}]{numpy2006}
Oliphant T.~E.,  2006, A guide to NumPy.
 Vol. 1, Trelgol Publishing USA

\bibitem[\protect\citeauthoryear{{P{\'a}pics} et~al.,}{{P{\'a}pics} et~al.}{2013}]{papics2013}
{P{\'a}pics} P.~I.,  et~al., 2013, \mn@doi [\aap] {10.1051/0004-6361/201321088}, \href {https://ui.adsabs.harvard.edu/abs/2013A&A...553A.127P} {553, A127}

\bibitem[\protect\citeauthoryear{{P{\'a}pics}, {Tkachenko}, {Aerts}, {Van Reeth}, {De Smedt}, {Hillen}, {{\O}stensen}  \& {Moravveji}}{{P{\'a}pics} et~al.}{2015}]{papics2015}
{P{\'a}pics} P.~I.,  {Tkachenko} A.,  {Aerts} C.,  {Van Reeth} T.,  {De Smedt} K.,  {Hillen} M.,  {{\O}stensen} R.,   {Moravveji} E.,  2015, \mn@doi [\apjl] {10.1088/2041-8205/803/2/L25}, \href {https://ui.adsabs.harvard.edu/abs/2015ApJ...803L..25P} {803, L25}

\bibitem[\protect\citeauthoryear{{P{\'a}pics} et~al.,}{{P{\'a}pics} et~al.}{2017}]{papics2017}
{P{\'a}pics} P.~I.,  et~al., 2017, \mn@doi [\aap] {10.1051/0004-6361/201629814}, \href {https://ui.adsabs.harvard.edu/abs/2017A&A...598A..74P} {598, A74}

\bibitem[\protect\citeauthoryear{{Paxton}, {Bildsten}, {Dotter}, {Herwig}, {Lesaffre}  \& {Timmes}}{{Paxton} et~al.}{2011}]{paxton2011}
{Paxton} B.,  {Bildsten} L.,  {Dotter} A.,  {Herwig} F.,  {Lesaffre} P.,   {Timmes} F.,  2011, \mn@doi [\apjs] {10.1088/0067-0049/192/1/3}, \href {https://ui.adsabs.harvard.edu/abs/2011ApJS..192....3P} {192, 3}

\bibitem[\protect\citeauthoryear{{Paxton} et~al.,}{{Paxton} et~al.}{2013}]{paxton2013}
{Paxton} B.,  et~al., 2013, \mn@doi [\apjs] {10.1088/0067-0049/208/1/4}, \href {https://ui.adsabs.harvard.edu/abs/2013ApJS..208....4P} {208, 4}

\bibitem[\protect\citeauthoryear{{Paxton} et~al.,}{{Paxton} et~al.}{2015}]{paxton2015}
{Paxton} B.,  et~al., 2015, \mn@doi [\apjs] {10.1088/0067-0049/220/1/15}, \href {https://ui.adsabs.harvard.edu/abs/2015ApJS..220...15P} {220, 15}

\bibitem[\protect\citeauthoryear{{Paxton} et~al.,}{{Paxton} et~al.}{2018}]{paxton2018}
{Paxton} B.,  et~al., 2018, \mn@doi [\apjs] {10.3847/1538-4365/aaa5a8}, \href {https://ui.adsabs.harvard.edu/abs/2018ApJS..234...34P} {234, 34}

\bibitem[\protect\citeauthoryear{{Pedersen}}{{Pedersen}}{2022}]{pedersen2022}
{Pedersen} M.~G.,  2022, \mn@doi [\apj] {10.3847/1538-4357/ac5b05}, \href {https://ui.adsabs.harvard.edu/abs/2022ApJ...930...94P} {930, 94}

\bibitem[\protect\citeauthoryear{{Pedersen} et~al.,}{{Pedersen} et~al.}{2019}]{pedersen2019}
{Pedersen} M.~G.,  et~al., 2019, \mn@doi [\apjl] {10.3847/2041-8213/ab01e1}, \href {http://adsabs.harvard.edu/abs/2019ApJ...872L...9P} {872, L9}

\bibitem[\protect\citeauthoryear{{Pedersen} et~al.,}{{Pedersen} et~al.}{2021}]{pedersen2021}
{Pedersen} M.~G.,  et~al., 2021, \mn@doi [Nature Astronomy] {10.1038/s41550-021-01351-x}, \href {https://ui.adsabs.harvard.edu/abs/2021NatAs...5..715P} {5, 715}

\bibitem[\protect\citeauthoryear{{Ricker} et~al.,}{{Ricker} et~al.}{2015}]{ricker2015}
{Ricker} G.~R.,  et~al., 2015, \mn@doi [Journal of Astronomical Telescopes, Instruments, and Systems] {10.1117/1.JATIS.1.1.014003}, \href {http://adsabs.harvard.edu/abs/2015JATIS...1a4003R} {1, 014003}

\bibitem[\protect\citeauthoryear{{Rogers} \& {McElwaine}}{{Rogers} \& {McElwaine}}{2017}]{rogers2017}
{Rogers} T.~M.,  {McElwaine} J.~N.,  2017, \mn@doi [\apjl] {10.3847/2041-8213/aa8d13}, \href {https://ui.adsabs.harvard.edu/abs/2017ApJ...848L...1R} {848, L1}

\bibitem[\protect\citeauthoryear{{Szewczuk} \& {Daszy{\'n}ska-Daszkiewicz}}{{Szewczuk} \& {Daszy{\'n}ska-Daszkiewicz}}{2018}]{szewczuk2018}
{Szewczuk} W.,  {Daszy{\'n}ska-Daszkiewicz} J.,  2018, \mn@doi [\mnras] {10.1093/mnras/sty1126}, \href {http://adsabs.harvard.edu/abs/2018MNRAS.478.2243S} {478, 2243}

\bibitem[\protect\citeauthoryear{{Szewczuk}, {Walczak}  \& {Daszy{\'n}ska-Daszkiewicz}}{{Szewczuk} et~al.}{2021}]{szewczuk2021}
{Szewczuk} W.,  {Walczak} P.,   {Daszy{\'n}ska-Daszkiewicz} J.,  2021, \mn@doi [\mnras] {10.1093/mnras/stab683}, \href {https://ui.adsabs.harvard.edu/abs/2021MNRAS.503.5894S} {503, 5894}

\bibitem[\protect\citeauthoryear{{Szewczuk}, {Walczak}, {Daszy{\'n}ska-Daszkiewicz}  \& {Mo{\'z}dzierski}}{{Szewczuk} et~al.}{2022}]{szewczuk2022}
{Szewczuk} W.,  {Walczak} P.,  {Daszy{\'n}ska-Daszkiewicz} J.,   {Mo{\'z}dzierski} D.,  2022, \mn@doi [\mnras] {10.1093/mnras/stac168}, \href {https://ui.adsabs.harvard.edu/abs/2022MNRAS.511.1529S} {511, 1529}

\bibitem[\protect\citeauthoryear{{Tassoul}}{{Tassoul}}{1980}]{tassoul1980}
{Tassoul} M.,  1980, \mn@doi [\apjs] {10.1086/190678}, \href {http://adsabs.harvard.edu/abs/1980ApJS...43..469T} {43, 469}

\bibitem[\protect\citeauthoryear{{Van Reeth}, {Tkachenko}  \& {Aerts}}{{Van Reeth} et~al.}{2016}]{vanreeth2016}
{Van Reeth} T.,  {Tkachenko} A.,   {Aerts} C.,  2016, \mn@doi [\aap] {10.1051/0004-6361/201628616}, \href {https://ui.adsabs.harvard.edu/abs/2016A&A...593A.120V} {593, A120}

\bibitem[\protect\citeauthoryear{{Van Reeth} et~al.,}{{Van Reeth} et~al.}{2018}]{vanreeth2018}
{Van Reeth} T.,  et~al., 2018, \mn@doi [\aap] {10.1051/0004-6361/201832718}, \href {https://ui.adsabs.harvard.edu/abs/2018A&A...618A..24V} {618, A24}

\bibitem[\protect\citeauthoryear{{Van Reeth} et~al.,}{{Van Reeth} et~al.}{2022}]{vanreeth2022}
{Van Reeth} T.,  et~al., 2022, \mn@doi [\aap] {10.1051/0004-6361/202142921}, \href {https://ui.adsabs.harvard.edu/abs/2022A&A...662A..58V} {662, A58}

\bibitem[\protect\citeauthoryear{{van der Walt}, {Colbert}  \& {Varoquaux}}{{van der Walt} et~al.}{2011}]{numpy2011}
{van der Walt} S.,  {Colbert} S.~C.,   {Varoquaux} G.,  2011, Computing in Science Engineering, 13, 22

\makeatother
\end{thebibliography}

%\begin{thebibliography}{99}
%\end{thebibliography}

%%%%%%%%%%%%%%%%%%%%%%%%%%%%%%%%%%%%%%%%%%%%%%%%%%

%%%%%%%%%%%%%%%%% APPENDICES %%%%%%%%%%%%%%%%%%%%%

%\newpage

\appendix

\section{Frequency extraction summary}
\label{sec:app_freqs}

Tables~\ref{tab:app_3240411} to \ref{tab:app_11360704} list the frequencies, amplitudes, and phases, as well as their respective uncertainties extracted from our TESS light curves using {\sc Period04} \citep{lenz2005}.

\begin{table*}
    \caption{Extracted frequencies for KIC~3240411. The number of light curve chunks is shown in the first column. The next five columns list the frequency, period, amplitude, phase, and signal to noise ratio of each extracted frequency. The last column(s) mark which frequencies we used to build our period spacing patterns (where possible) and those we judged to be a match with \citet{pedersen2021}. The `x--' mark signifies which frequencies are followed by a gap in the pattern.}
    \centering
    \begin{tabular}{ccccccc}
    \hline
    no. chunks & $f$ / d$^{-1}$ & $P$ / d & $A$ / mmag & $\phi$ & S/N & Pedersen pattern \\
    \hline
    1 & $0.9144\pm0.0001$ & $1.0936\pm0.0001$ & $0.46\pm0.02$ & $0.9\pm0.4$ & 7.8 & x \\
    \\
	2 & $0.91428\pm0.00005$ & $1.09376\pm0.00006$ & $0.41\pm0.01$ & $0.5\pm0.2$ & 8.5 & x--\\
	& $1.1872\pm0.0001$ & $0.84231\pm0.00008$ & $0.17\pm0.01$ & $0.4\pm0.4$ & 5.7 & x\\
	& $1.2372\pm0.0001$ & $0.80826\pm0.00007$ & $0.19\pm0.01$ & $0.8\pm0.4$ & 5.8 & \\
	& $1.5111\pm0.0001$ & $0.66177\pm0.00006$ & $0.16\pm0.01$ & $0.2\pm0.5$ & 5.8 & \\
    \\
	3 & $0.11970\pm0.00002$ & $8.354\pm0.002$ & $0.22\pm0.01$ & $0.04\pm0.07$ & 5.5 & \\
	& $0.72156\pm0.00002$ & $1.38588\pm0.00005$ & $0.20\pm0.01$ & $0.51\pm0.08$ & 5.4 & \\
	& $0.91487\pm0.00001$ & $1.09305\pm0.00001$ & $0.46\pm0.01$ & $0.44\pm0.04$ & 11.7 & x-- \\
	& $1.19156\pm0.00003$ & $0.83923\pm0.00002$ & $0.22\pm0.01$ & $0.96\pm0.09$ & 7.9 & x-- \\
	& $1.19917\pm0.00004$ & $0.83391\pm0.00003$ & $0.15\pm0.01$ & $0.6\pm0.1$ & 6.3 & \\
	& $1.23712\pm0.00002$ & $0.80833\pm0.00001$ & $0.23\pm0.01$ & $0.13\pm0.07$ & 7.8 & \\
	& $1.45311\pm0.00003$ & $0.68818\pm0.00001$ & $0.19\pm0.01$ & $0.49\pm0.09$ & 7.4 & x \\
	& $1.51156\pm0.00003$ & $0.66157\pm0.00001$ & $0.15\pm0.01$ & $0.5\pm0.1$ & 6.2 & \\
	& $2.70490\pm0.00006$ & $0.369700\pm0.000009$ & $0.08\pm0.01$ & $0.2\pm0.2$ & 6.9 & \\
    \hline
    \end{tabular}
    \label{tab:app_3240411}
\end{table*}

\begin{table*}
    \caption{Same as Table~\ref{tab:app_3240411} but for KIC~3459297.}
    \centering
    \begin{tabular}{ccccccc}
    \hline
    no. chunks & $f$ / d$^{-1}$ & $P$ / d & $A$ / mmag & $\phi$ & SNR & Pedersen pattern\\
    \hline
	1 & $0.8453\pm0.0004$ & $1.1830\pm0.0005$ & $0.20\pm0.03$ & $0\pm1$ & 5.6 & \\
	& $0.8736\pm0.0003$ & $1.1447\pm0.0004$ & $0.24\pm0.03$ & $0\pm1$ & 5.8 & \\
	& $0.9806\pm0.0001$ & $1.0198\pm0.0001$ & $0.69\pm0.03$ & $0.3\pm0.4$ & 15.0 & x \\
	& $0.99110\pm0.00004$ & $1.00898\pm0.00004$ & $1.83\pm0.03$ & $0.4\pm0.1$ & 20.3 & x--\\
	& $1.0272\pm0.0002$ & $0.9735\pm0.0002$ & $0.40\pm0.02$ & $0.3\pm0.6$ & 10.0 & x--\\
	& $1.0590\pm0.0001$ & $0.9443\pm0.0001$ & $0.55\pm0.03$ & $0.1\pm0.5$ & 21.3 & \\
	& $1.08476\pm0.00005$ & $0.92186\pm0.00004$ & $1.57\pm0.02$ & $0.0\pm0.2$ & 20.8 & x \\
	& $1.1085\pm0.0002$ & $0.9021\pm0.0002$ & $0.39\pm0.02$ & $0.2\pm0.7$ & 10.4 & x \\
    \\
	2 & $0.87863\pm0.00009$ & $1.1381\pm0.0001$ & $0.25\pm0.02$ & $0.6\pm0.3$ & 7.9 & \\
	& $0.8982\pm0.0001$ & $1.1133\pm0.0002$ & $0.16\pm0.02$ & $0.9\pm0.5$ & 5.3 & x--\\
	& $0.97336\pm0.00005$ & $1.02737\pm0.00006$ & $0.47\pm0.02$ & $0.7\pm0.2$ & 11.9 & x\\
	& $0.98967\pm0.00002$ & $1.01044\pm0.00002$ & $1.43\pm0.02$ & $0.37\pm0.08$ & 30.1 & x--\\
	& $0.9971\pm0.0001$ & $1.0029\pm0.0001$ & $0.38\pm0.02$ & $0.4\pm0.3$ & 11.9 & \\
	& $1.02701\pm0.00008$ & $0.97370\pm0.00008$ & $0.28\pm0.02$ & $0.1\pm0.3$ & 9.4 & x--\\
	& $1.06612\pm0.00005$ & $0.93798\pm0.00005$ & $0.43\pm0.02$ & $0.9\pm0.2$ & 11.6 & x\\
	& $1.08541\pm0.00002$ & $0.92131\pm0.00002$ & $1.08\pm0.02$ & $0.75\pm0.08$ & 26.1 & x\\
	& $1.10190\pm0.00007$ & $0.90753\pm0.00005$ & $0.35\pm0.02$ & $0.5\pm0.2$ & 10.8 & x\\
    \\
	3 & $0.83488\pm0.00005$ & $1.19778\pm0.00007$ & $0.15\pm0.01$ & $0.4\pm0.2$ & 5.4 & \\
	& $0.87856\pm0.00003$ & $1.13822\pm0.00003$ & $0.27\pm0.01$ & $0.81\pm0.09$ & 9.3 & \\
	& $0.89831\pm0.00005$ & $1.11320\pm0.00006$ & $0.14\pm0.01$ & $0.5\pm0.2$ & 5.4 & x--\\
	& $0.94196\pm0.00005$ & $1.06161\pm0.00006$ & $0.15\pm0.01$ & $0.6\pm0.2$ & 5.8 & x\\
	& $0.95813\pm0.00004$ & $1.04369\pm0.00004$ & $0.21\pm0.01$ & $0.0\pm0.1$ & 6.9 & x\\
	& $0.97290\pm0.00001$ & $1.02785\pm0.00001$ & $0.61\pm0.01$ & $0.27\pm0.04$ & 16.5 & x--\\
	& $0.989971\pm0.000004$ & $1.010131\pm0.000005$ & $1.63\pm0.01$ & $0.32\pm0.01$ & 35.8 & x--\\
	& $1.03211\pm0.00002$ & $0.96888\pm0.00002$ & $0.31\pm0.01$ & $0.05\pm0.07$ & 10.3 & x--\\
	& $1.06604\pm0.00002$ & $0.93805\pm0.00001$ & $0.43\pm0.01$ & $0.12\pm0.05$ & 13.9 & x\\
	& $1.085563\pm0.000006$ & $0.921181\pm0.000005$ & $1.15\pm0.01$ & $0.22\pm0.02$ & 31.1 & x\\
	& $1.10685\pm0.00002$ & $0.90346\pm0.00002$ & $0.31\pm0.01$ & $0.01\pm0.07$ & 11.4 & x\\
    \hline
    \end{tabular}
    \label{tab:app_3459297}
\end{table*}

\begin{table*}
    \caption{Same as Table~\ref{tab:app_3240411} but for KIC\,4930889.}
    \centering
    \begin{tabular}{cccccccc}
    \hline
    no. chunks & $f$ / d$^{-1}$ & $P$ / d & $A$ / mmag & $\phi$ & SNR & our pattern & Pedersen pattern\\
    \hline
	1 & $1.0781\pm0.0002$ & $0.9275\pm0.0002$ & $0.51\pm0.04$ & $0.8\pm0.8$ & 6.7 && x--\\
	& $1.10417\pm0.00005$ & $0.90566\pm0.00004$ & $2.22\pm0.04$ & $0.6\pm0.2$ & 12.4 && x--\\
	& $1.1291\pm0.0001$ & $0.8857\pm0.0001$ & $0.81\pm0.04$ & $0.5\pm0.5$ & 6.0 && \\
	& $1.16948\pm0.00009$ & $0.85508\pm0.00006$ & $1.33\pm0.04$ & $0.8\pm0.3$ & 14.5 && x--\\
	& $1.20305\pm0.00004$ & $0.83122\pm0.00003$ & $2.87\pm0.04$ & $0.1\pm0.1$ & 10.0 && \\
	& $1.22215\pm0.00002$ & $0.81823\pm0.00002$ & $5.00\pm0.04$ & $0.58\pm0.08$ & 32.7 && x\\
	& $1.24137\pm0.00002$ & $0.80556\pm0.00001$ & $5.83\pm0.04$ & $0.65\pm0.07$ & 21.8 && x--\\
	& $1.2976\pm0.0003$ & $0.7707\pm0.0002$ & $0.42\pm0.04$ & $1.0\pm0.9$ & 6.2 && x--\\
	& $1.3728\pm0.0003$ & $0.7284\pm0.0001$ & $0.40\pm0.04$ & $0\pm1$ & 6.2 && x--\\
	& $1.7180\pm0.0003$ & $0.58206\pm0.00009$ & $0.42\pm0.04$ & $1.0\pm0.9$ & 5.8 && \\
	& $2.3424\pm0.0005$ & $0.42690\pm0.00009$ & $0.22\pm0.04$ & $0\pm1$ & 6.1 && \\
	& $2.4790\pm0.0004$ & $0.40339\pm0.00006$ & $0.31\pm0.04$ & $1\pm1$ & 8.3 && \\
    \\
	2 & $1.0675\pm0.0001$ & $0.9368\pm0.0001$ & $0.48\pm0.04$ & $0.5\pm0.5$ & 5.6 && x--\\
	& $1.0894\pm0.0001$ & $0.9180\pm0.0001$ & $0.46\pm0.04$ & $0.0\pm0.5$ & 5.3 && x\\
	& $1.10680\pm0.00003$ & $0.90350\pm0.00002$ & $2.16\pm0.04$ & $0.3\pm0.1$ & 18.2 && x--\\
	& $1.1311\pm0.0002$ & $0.8841\pm0.0001$ & $0.39\pm0.05$ & $0.4\pm0.6$ & 5.3 && x--\\
	& $1.17236\pm0.00007$ & $0.85298\pm0.00005$ & $0.97\pm0.04$ & $0.7\pm0.2$ & 11.4 && x\\
	& $1.19969\pm0.00004$ & $0.83355\pm0.00003$ & $1.56\pm0.04$ & $0.9\pm0.1$ & 16.6 && x\\
	& $1.21645\pm0.00001$ & $0.822066\pm0.000009$ & $4.95\pm0.04$ & $0.74\pm0.05$ & 40.9 && x\\
	& $1.23984\pm0.00001$ & $0.806553\pm0.000006$ & $6.64\pm0.04$ & $0.04\pm0.03$ & 35.9 && x\\
	& $1.2672\pm0.0002$ & $0.7891\pm0.0001$ & $0.32\pm0.05$ & $0.4\pm0.7$ & 5.8 && x--\\
	& $1.3726\pm0.0002$ & $0.7286\pm0.0001$ & $0.34\pm0.04$ & $0.2\pm0.6$ & 5.7 && x\\
	& $1.7177\pm0.0001$ & $0.58218\pm0.00005$ & $0.45\pm0.04$ & $0.2\pm0.5$ & 6.4 && \\
	& $2.3467\pm0.0003$ & $0.42612\pm0.00006$ & $0.19\pm0.04$ & $0\pm1$ & 7.2 && \\
	& $2.4563\pm0.0004$ & $0.40711\pm0.00006$ & $0.17\pm0.04$ & $0\pm1$ & 7.1 && \\
	& $2.4797\pm0.0002$ & $0.40327\pm0.00004$ & $0.28\pm0.04$ & $0.1\pm0.8$ & 10.7 && \\
    \\
	3 & $0.17913\pm0.00002$ & $5.5826\pm0.0008$ & $0.66\pm0.03$ & $0.56\pm0.08$ & 6.3 &  & \\
	& $1.06808\pm0.00003$ & $0.93626\pm0.00003$ & $0.57\pm0.03$ & $0.46\pm0.09$ & 7.6 &  & x--\\
	& $1.08891\pm0.00004$ & $0.91835\pm0.00003$ & $0.40\pm0.03$ & $0.6\pm0.1$ & 6.7 &  & x\\
	& $1.106954\pm0.000008$ & $0.903380\pm0.000006$ & $2.12\pm0.03$ & $0.81\pm0.03$ & 22.8 &  & x\\
	& $1.12602\pm0.00003$ & $0.88808\pm0.00003$ & $0.51\pm0.03$ & $0.3\pm0.1$ & 6.7 &  & x--\\
	& $1.15907\pm0.00004$ & $0.86276\pm0.00003$ & $0.45\pm0.03$ & $0.5\pm0.1$ & 7.2 & x & x\\
	& $1.17291\pm0.00002$ & $0.85258\pm0.00001$ & $0.94\pm0.03$ & $0.77\pm0.06$ & 13.5 & x & x\\
	& $1.19462\pm0.00001$ & $0.837088\pm0.000008$ & $1.42\pm0.03$ & $0.86\pm0.04$ & 17.8 & x & x\\
	& $1.216334\pm0.000003$ & $0.822142\pm0.000002$ & $5.15\pm0.03$ & $0.13\pm0.01$ & 49.1 & x & x\\
	& $1.239868\pm0.000003$ & $0.806537\pm0.000002$ & $6.60\pm0.03$ & $0.961\pm0.008$ & 45.6 & x & x\\
	& $1.26740\pm0.00005$ & $0.78902\pm0.00003$ & $0.34\pm0.03$ & $0.9\pm0.2$ & 6.3 & x & x\\
	& $1.29426\pm0.00006$ & $0.77264\pm0.00003$ & $0.29\pm0.03$ & $0.6\pm0.2$ & 5.8 & x & x--\\
	& $1.37273\pm0.00005$ & $0.72848\pm0.00003$ & $0.30\pm0.03$ & $0.6\pm0.2$ & 6.0 &  & x\\
	& $1.71772\pm0.00004$ & $0.58217\pm0.00001$ & $0.45\pm0.03$ & $0.1\pm0.1$ & 7.5 &  & \\
	& $2.34685\pm0.00008$ & $0.42610\pm0.00001$ & $0.19\pm0.03$ & $0.9\pm0.3$ & 8.7 &  & \\
	& $2.4562\pm0.0001$ & $0.40713\pm0.00002$ & $0.16\pm0.03$ & $0.4\pm0.3$ & 8.2 &  & \\
	& $2.47976\pm0.00006$ & $0.40327\pm0.00001$ & $0.26\pm0.03$ & $1.0\pm0.2$ & 12.2 &  & \\
    \hline
    \end{tabular}
    \label{tab:app_4930889}
\end{table*}

\begin{table*}
    \caption{Same as Table~\ref{tab:app_3240411} but for KIC~5941844.}
    \centering
    \begin{tabular}{cccccccc}
    \hline
    no. chunks & $f$ / d$^{-1}$ & $P$ / d & $A$ / mmag & $\phi$ & SNR & our pattern & Pedersen pattern\\
    \hline
    1 & $0.7678\pm0.0001$ & $1.3024\pm0.0002$ & $1.79\pm0.06$ & $0.9\pm0.4$ & 10.1 &  & \\
	& $0.9001\pm0.0001$ & $1.1110\pm0.0002$ & $1.51\pm0.06$ & $0.8\pm0.4$ & 10.1 &  & \\
	& $1.17935\pm0.00003$ & $0.84793\pm0.00002$ & $5.47\pm0.06$ & $0.5\pm0.1$ & 17.5 &  & x\\
	& $1.21108\pm0.00008$ & $0.82571\pm0.00006$ & $2.23\pm0.06$ & $0.1\pm0.3$ & 10.3 &  & x--\\
	& $1.2832\pm0.0002$ & $0.7793\pm0.0001$ & $1.02\pm0.06$ & $0.3\pm0.6$ & 8.6 &  & \\
	& $1.30938\pm0.00002$ & $0.76372\pm0.00001$ & $9.94\pm0.06$ & $0.14\pm0.07$ & 26.7 &  & x--\\
	& $1.43636\pm0.00007$ & $0.69621\pm0.00003$ & $2.76\pm0.06$ & $1.0\pm0.2$ & 11.6 &  & x--\\
	& $1.7204\pm0.0001$ & $0.58127\pm0.00003$ & $1.89\pm0.06$ & $0.8\pm0.3$ & 12.5 &  & x\\
    \\
	2 & $0.76867\pm0.00001$ & $1.30095\pm0.00002$ & $1.80\pm0.04$ & $0.91\pm0.03$ & 13.1 &  & \\
	& $0.90025\pm0.00001$ & $1.11080\pm0.00002$ & $1.44\pm0.04$ & $0.28\pm0.04$ & 12.1 &  & \\
	& $1.179185\pm0.000003$ & $0.848043\pm0.000002$ & $5.74\pm0.04$ & $0.11\pm0.01$ & 26.5 &  & x\\
	& $1.209836\pm0.000008$ & $0.826558\pm0.000005$ & $2.39\pm0.04$ & $0.52\pm0.03$ & 15.7 &  & x--\\
	& $1.28270\pm0.00002$ & $0.77961\pm0.00001$ & $0.97\pm0.04$ & $0.18\pm0.06$ & 10.1 &  & \\
	& $1.309379\pm0.000002$ & $0.763721\pm0.000001$ & $10.21\pm0.04$ & $0.138\pm0.006$ & 36.8 &  & x--\\
	& $1.436476\pm0.000007$ & $0.696148\pm0.000003$ & $2.82\pm0.04$ & $0.55\pm0.02$ & 16.2 &  & \\
	& $1.60952\pm0.00004$ & $0.62130\pm0.00002$ & $0.47\pm0.04$ & $1.0\pm0.1$ & 6.3 &  & x\\
	& $1.71990\pm0.00001$ & $0.581429\pm0.000003$ & $1.93\pm0.04$ & $0.37\pm0.03$ & 16.7 &  & x\\
	& $2.8991\pm0.0001$ & $0.34493\pm0.00002$ & $0.13\pm0.04$ & $0.1\pm0.4$ & 7.1 &  & \\
	& $4.9798\pm0.0004$ & $0.20081\pm0.00002$ & $0.04\pm0.04$ & $0\pm1$ & 5.3 &  & \\
    \\
	3 & $0.768658\pm0.000009$ & $1.30097\pm0.00002$ & $1.66\pm0.04$ & $0.94\pm0.03$ & 13.8 &  & \\
	& $0.89888\pm0.00001$ & $1.11250\pm0.00001$ & $1.29\pm0.04$ & $0.10\pm0.03$ & 12.3 &  & \\
	& $1.179188\pm0.000003$ & $0.848041\pm0.000002$ & $5.39\pm0.04$ & $0.100\pm0.008$ & 32.3 & x & x\\
	& $1.209859\pm0.000007$ & $0.826543\pm0.000005$ & $2.13\pm0.04$ & $0.45\pm0.02$ & 16.5 &  & x--\\
	& $1.28141\pm0.00002$ & $0.78039\pm0.00001$ & $0.93\pm0.04$ & $0.73\pm0.05$ & 10.6 &  & \\
	& $1.309381\pm0.000002$ & $0.7637197\pm0.0000009$ & $9.49\pm0.04$ & $0.130\pm0.005$ & 43.5 & x & x--\\
	& $1.436468\pm0.000006$ & $0.696152\pm0.000003$ & $2.56\pm0.04$ & $0.58\pm0.02$ & 18.4 & x & x--\\
	& $1.719905\pm0.000009$ & $0.581428\pm0.000003$ & $1.77\pm0.04$ & $0.36\pm0.02$ & 17.6 & x & x\\
	& $2.8991\pm0.0001$ & $0.34494\pm0.00002$ & $0.11\pm0.04$ & $0.3\pm0.4$ & 6.9 &  & \\
    \hline
    \end{tabular}
    \label{tab:app_5941844}
\end{table*}

\begin{table*}
    \caption{Same as Table~\ref{tab:app_3240411} but for KIC~6352430.}
    \centering
    \begin{tabular}{ccccccc}
    \hline
    no. chunks & $f$ / d$^{-1}$ & $P$ / d & $A$ / mmag & $\phi$ & SNR & Pedersen pattern\\
    \hline
	1 & $0.4654\pm0.0002$ & $2.149\pm0.001$ & $0.46\pm0.04$ & $1.0\pm0.9$ & 5.6 & \\
	& $1.2848\pm0.0002$ & $0.7784\pm0.0001$ & $0.52\pm0.04$ & $0.5\pm0.8$ & 6.7 & x--\\
	& $1.36118\pm0.00003$ & $0.73466\pm0.00002$ & $3.5\pm0.2$ & $0.0\pm0.1$ & 50.4 & x--\\
	& $1.36252\pm0.00003$ & $0.73393\pm0.00002$ & $3.6\pm0.2$ & $0.1\pm0.1$ & 14.9 & \\
	& $1.46383\pm0.00006$ & $0.68314\pm0.00003$ & $1.99\pm0.04$ & $0.8\pm0.2$ & 25.3 & x\\
	& $1.51892\pm0.00002$ & $0.658363\pm0.000009$ & $5.63\pm0.04$ & $0.55\pm0.07$ & 28.9 & x--\\
	& $2.8810\pm0.0005$ & $0.34711\pm0.00006$ & $0.24\pm0.04$ & $1\pm1$ & 8.7 & \\
	& $2.9831\pm0.0003$ & $0.33522\pm0.00003$ & $0.44\pm0.04$ & $0.7\pm0.9$ & 15.2 & \\
	& $3.0867\pm0.0006$ & $0.32397\pm0.00007$ & $0.18\pm0.04$ & $1\pm1$ & 9.2 & \\
	& $3.247\pm0.001$ & $0.3080\pm0.0001$ & $0.10\pm0.04$ & $0\pm1$ & 6.5 & \\
	& $3.8223\pm0.0005$ & $0.26162\pm0.00003$ & $0.24\pm0.04$ & $0\pm1$ & 17.1 & \\
	& $5.713\pm0.001$ & $0.17504\pm0.00004$ & $0.09\pm0.04$ & $0\pm1$ & 13.0 & \\
	& $5.814\pm0.003$ & $0.17200\pm0.00009$ & $0.03\pm0.04$ & $0\pm1$ & 5.4 & \\

    \\
	2 & $0.46489\pm0.00008$ & $2.1510\pm0.0004$ & $0.43\pm0.02$ & $0.8\pm0.3$ & 7.2 & \\
	& $1.1517\pm0.0001$ & $0.86828\pm0.00009$ & $0.29\pm0.02$ & $0.5\pm0.4$ & 5.3 & x--\\
	& $1.28445\pm0.00006$ & $0.77854\pm0.00004$ & $0.53\pm0.02$ & $0.6\pm0.2$ & 9.1 & x--\\
	& $1.361901\pm0.000006$ & $0.734268\pm0.000003$ & $5.37\pm0.02$ & $0.37\pm0.02$ & 66.6 & x--\\
	& $1.46401\pm0.00002$ & $0.683054\pm0.000008$ & $1.93\pm0.02$ & $0.19\pm0.06$ & 33.1 & x\\
	& $1.518881\pm0.000006$ & $0.658379\pm0.000003$ & $5.41\pm0.02$ & $0.68\pm0.02$ & 38.4 & x\\
	& $1.5910\pm0.0001$ & $0.62854\pm0.00005$ & $0.26\pm0.02$ & $0.6\pm0.4$ & 5.4 & x--\\
	& $1.6211\pm0.0002$ & $0.61687\pm0.00006$ & $0.22\pm0.02$ & $0.2\pm0.5$ & 5.6 & \\
	& $1.95334\pm0.00009$ & $0.51194\pm0.00002$ & $0.38\pm0.02$ & $0.4\pm0.3$ & 7.7 & \\
	& $2.0954\pm0.0002$ & $0.47724\pm0.00004$ & $0.20\pm0.02$ & $0.4\pm0.6$ & 5.6 & \\
	& $2.8808\pm0.0001$ & $0.34712\pm0.00002$ & $0.23\pm0.02$ & $0.4\pm0.5$ & 12.5 & \\
	& $2.98292\pm0.00008$ & $0.335242\pm0.000009$ & $0.42\pm0.02$ & $0.2\pm0.3$ & 19.7 & \\
	& $3.0865\pm0.0002$ & $0.32399\pm0.00002$ & $0.17\pm0.02$ & $0.6\pm0.7$ & 13.2 & \\
	& $3.2470\pm0.0003$ & $0.30798\pm0.00003$ & $0.11\pm0.02$ & $0\pm1$ & 10.7 & \\
	& $3.8222\pm0.0001$ & $0.26163\pm0.00001$ & $0.23\pm0.02$ & $0.6\pm0.5$ & 19.6 & \\
	& $5.7132\pm0.0004$ & $0.17503\pm0.00001$ & $0.08\pm0.02$ & $0\pm1$ & 17.6 & \\
	& $5.816\pm0.001$ & $0.17194\pm0.00003$ & $0.03\pm0.02$ & $1\pm1$ & 7.2 & \\
    \\
	3 & $0.40773\pm0.00003$ & $2.4526\pm0.0002$ & $0.28\pm0.02$ & $0.7\pm0.1$ & 6.5 & \\
	& $0.46478\pm0.00002$ & $2.15158\pm0.00009$ & $0.45\pm0.02$ & $0.22\pm0.06$ & 9.5 & \\
	& $0.99783\pm0.00003$ & $1.00218\pm0.00003$ & $0.28\pm0.02$ & $0.8\pm0.1$ & 6.1 & x--\\
	& $1.04825\pm0.00004$ & $0.95397\pm0.00003$ & $0.23\pm0.02$ & $0.4\pm0.1$ & 5.7 & x--\\
	& $1.12857\pm0.00004$ & $0.88608\pm0.00003$ & $0.21\pm0.02$ & $0.0\pm0.1$ & 5.3 & x\\
	& $1.15210\pm0.00004$ & $0.86798\pm0.00003$ & $0.20\pm0.02$ & $0.1\pm0.1$ & 5.6 & x--\\
	& $1.28447\pm0.00002$ & $0.77853\pm0.00001$ & $0.53\pm0.02$ & $0.54\pm0.05$ & 11.3 & x--\\
	& $1.361918\pm0.000002$ & $0.7342586\pm0.0000008$ & $5.50\pm0.02$ & $0.309\pm0.005$ & 86.6 & x--\\
	& $1.464018\pm0.000004$ & $0.683052\pm0.000002$ & $1.96\pm0.02$ & $0.17\pm0.01$ & 42.8 & x\\
	& $1.518881\pm0.000002$ & $0.6583795\pm0.0000007$ & $5.54\pm0.02$ & $0.682\pm0.005$ & 47.7 & x\\
	& $1.59112\pm0.00003$ & $0.62849\pm0.00001$ & $0.26\pm0.02$ & $0.2\pm0.1$ & 6.4 & x--\\
	& $1.62105\pm0.00004$ & $0.61688\pm0.00001$ & $0.23\pm0.02$ & $0.3\pm0.1$ & 6.4 & \\
	& $1.82979\pm0.00004$ & $0.54651\pm0.00001$ & $0.20\pm0.02$ & $0.2\pm0.1$ & 5.5 & \\
	& $1.95281\pm0.00005$ & $0.51208\pm0.00001$ & $0.21\pm0.02$ & $0.3\pm0.2$ & 7.4 & \\
	& $1.95908\pm0.00006$ & $0.51044\pm0.00001$ & $0.20\pm0.02$ & $0.3\pm0.2$ & 6.2 & \\
	& $2.10128\pm0.00005$ & $0.47590\pm0.00001$ & $0.16\pm0.02$ & $0.7\pm0.2$ & 5.9 & \\
	& $2.17570\pm0.00005$ & $0.45962\pm0.00001$ & $0.16\pm0.02$ & $0.4\pm0.2$ & 6.3 & \\
	& $2.88081\pm0.00004$ & $0.347125\pm0.000005$ & $0.23\pm0.02$ & $0.4\pm0.1$ & 14.6 & \\
	& $2.98292\pm0.00002$ & $0.335242\pm0.000002$ & $0.43\pm0.02$ & $0.22\pm0.06$ & 25.4 & \\
	& $3.08637\pm0.00005$ & $0.324005\pm0.000005$ & $0.17\pm0.02$ & $1.0\pm0.2$ & 14.5 & \\
	& $3.24695\pm0.00007$ & $0.307982\pm0.000007$ & $0.11\pm0.02$ & $0.4\pm0.2$ & 14.6 & \\
	& $3.82215\pm0.00004$ & $0.261633\pm0.000003$ & $0.23\pm0.02$ & $0.6\pm0.1$ & 24.9 & \\
	& $5.7133\pm0.0001$ & $0.175032\pm0.000003$ & $0.08\pm0.02$ & $0.7\pm0.3$ & 20.5 & \\
    \hline
    \end{tabular}
    \label{tab:app_6352430}
\end{table*}

\begin{table*}
    \caption{Same as Table~\ref{tab:app_3240411} but for KIC~7760680.}
    \centering
    \begin{tabular}{cccccccc}
    \hline
    no. chunks & $f$ / d$^{-1}$ & $P$ / d & $A$ / mmag & $\phi$ & SNR & our pattern & Pedersen pattern \\
    \hline
	1 & $0.81157\pm0.00003$ & $1.23218\pm0.00005$ & $1.97\pm0.02$ & $0.4\pm0.1$ & 19.9 &  & x--\\
	& $0.83813\pm0.00001$ & $1.19313\pm0.00002$ & $4.93\pm0.02$ & $0.61\pm0.05$ & 25.7 &  & x--\\
	& $0.862285\pm0.000008$ & $1.15971\pm0.00001$ & $7.63\pm0.02$ & $0.72\pm0.03$ & 32.6 &  & x--\\
	& $0.88971\pm0.00003$ & $1.12396\pm0.00004$ & $1.96\pm0.02$ & $1.0\pm0.1$ & 16.0 &  & x--\\
	& $0.90710\pm0.00005$ & $1.10241\pm0.00006$ & $1.21\pm0.02$ & $0.6\pm0.2$ & 12.3 &  & \\
	& $0.9665\pm0.0001$ & $1.0346\pm0.0001$ & $0.57\pm0.02$ & $0.9\pm0.4$ & 9.2 &  & x\\
	& $0.99316\pm0.00007$ & $1.00688\pm0.00007$ & $0.88\pm0.02$ & $0.1\pm0.3$ & 10.9 &  & x\\
	& $1.7266\pm0.0003$ & $0.5792\pm0.0001$ & $0.20\pm0.02$ & $0\pm1$ & 6.8 &  & \\
	& $1.7674\pm0.0004$ & $0.5658\pm0.0001$ & $0.16\pm0.02$ & $0\pm1$ & 6.4 &  & \\
	& $1.8065\pm0.0003$ & $0.55357\pm0.00009$ & $0.23\pm0.02$ & $0\pm1$ & 8.7 &  & \\
	& $4.6080\pm0.0007$ & $0.21701\pm0.00003$ & $0.10\pm0.02$ & $0\pm1$ & 9.2 &  & \\
    \\
	2 & $0.75132\pm0.00008$ & $1.3310\pm0.0002$ & $0.26\pm0.02$ & $0.4\pm0.3$ & 5.8 &  & x--\\
	& $0.78300\pm0.00007$ & $1.2771\pm0.0001$ & $0.35\pm0.02$ & $0.5\pm0.2$ & 5.8 &  & x--\\
	& $0.79940\pm0.00004$ & $1.25093\pm0.00005$ & $0.82\pm0.03$ & $0.1\pm0.1$ & 8.2 & x & x--\\
	& $0.80890\pm0.00002$ & $1.23624\pm0.00003$ & $1.95\pm0.02$ & $0.92\pm0.06$ & 16.1 & x & \\
	& $0.82688\pm0.00003$ & $1.20937\pm0.00005$ & $1.10\pm0.02$ & $0.4\pm0.1$ & 10.5 & x & x\\
	& $0.838614\pm0.000007$ & $1.19244\pm0.00001$ & $3.67\pm0.05$ & $0.81\pm0.02$ & 35.1 & x & x--\\
	& $0.84402\pm0.00001$ & $1.18481\pm0.00001$ & $2.81\pm0.07$ & $0.98\pm0.03$ & 14.5 &  & \\
	& $0.863058\pm0.000004$ & $1.158671\pm0.000006$ & $8.58\pm0.05$ & $0.03\pm0.01$ & 41.7 & & x\\
	& $0.874013\pm0.000009$ & $1.14415\pm0.00001$ & $3.9\pm0.1$ & $0.96\pm0.03$ & 11.9 & x & \\
	& $0.87935\pm0.00001$ & $1.13720\pm0.00001$ & $2.97\pm0.09$ & $0.58\pm0.04$ & 20.6 &  & x\\
	& $0.89516\pm0.00001$ & $1.11712\pm0.00002$ & $1.89\pm0.02$ & $0.74\pm0.04$ & 21.5 & x & x\\
	& $0.91712\pm0.00003$ & $1.09037\pm0.00004$ & $0.74\pm0.02$ & $0.2\pm0.1$ & 10.1 & x & x-\\
	& $0.94179\pm0.00007$ & $1.06180\pm0.00008$ & $0.29\pm0.02$ & $0.8\pm0.3$ & 5.9 & x & \\
	& $0.96858\pm0.00004$ & $1.03244\pm0.00004$ & $0.57\pm0.02$ & $0.6\pm0.1$ & 9.7 & x & x\\
	& $0.99226\pm0.00002$ & $1.00781\pm0.00002$ & $1.00\pm0.01$ & $0.34\pm0.07$ & 11.8 & x & x\\
	& $1.01899\pm0.00008$ & $0.98136\pm0.00008$ & $0.27\pm0.01$ & $0.9\pm0.3$ & 5.6 & x & x\\
	& $1.7013\pm0.0002$ & $0.58777\pm0.00006$ & $0.12\pm0.01$ & $0.3\pm0.6$ & 5.4 &  & \\
	& $1.7255\pm0.0001$ & $0.57955\pm0.00004$ & $0.18\pm0.01$ & $0.5\pm0.4$ & 5.8 &  & \\
	& $1.80839\pm0.00009$ & $0.55298\pm0.00003$ & $0.23\pm0.01$ & $0.6\pm0.3$ & 8.7 &  & \\
	& $4.6080\pm0.0003$ & $0.21701\pm0.00002$ & $0.06\pm0.01$ & $0\pm1$ & 8.1 &  & \\
    \\
	3 & $0.79805\pm0.00002$ & $1.25306\pm0.00003$ & $0.39\pm0.01$ & $0.74\pm0.07$ & 7.8 & x & \\
	& $0.80097\pm0.00001$ & $1.24849\pm0.00002$ & $0.50\pm0.01$ & $0.77\pm0.05$ & 7.4 &  & x--\\
	& $0.808837\pm0.000005$ & $1.236343\pm0.000008$ & $1.75\pm0.01$ & $0.14\pm0.02$ & 17.6 & x & \\
	& $0.82596\pm0.00001$ & $1.21071\pm0.00002$ & $0.65\pm0.01$ & $0.57\pm0.04$ & 10.8 & x & x\\
	& $0.838573\pm0.000002$ & $1.192502\pm0.000002$ & $4.64\pm0.01$ & $0.044\pm0.006$ & 38.4 & x & x\\
	& $0.85082\pm0.00001$ & $1.17533\pm0.00002$ & $0.63\pm0.01$ & $0.15\pm0.04$ & 10.7 & x & x\\
	& $0.862849\pm0.000001$ & $1.158952\pm0.000001$ & $7.00\pm0.01$ & $0.734\pm0.004$ & 47.7 & x & x\\
	& $0.87567\pm0.00002$ & $1.14198\pm0.00002$ & $0.88\pm0.02$ & $0.57\pm0.05$ & 12.1 & x & x\\
	& $0.87761\pm0.00003$ & $1.13945\pm0.00004$ & $0.51\pm0.02$ & $0.27\pm0.09$ & 8.2 &  & \\
	& $0.894868\pm0.000003$ & $1.117483\pm0.000004$ & $2.26\pm0.01$ & $0.75\pm0.01$ & 24.1 & x & x\\
	& $0.912105\pm0.000009$ & $1.09636\pm0.00001$ & $0.70\pm0.01$ & $0.90\pm0.03$ & 11.1 & x & x\\
	& $0.93205\pm0.00002$ & $1.07290\pm0.00002$ & $0.36\pm0.01$ & $0.30\pm0.06$ & 6.8 & x-- & x--\\
	& $0.96812\pm0.00001$ & $1.03293\pm0.00001$ & $0.56\pm0.01$ & $0.17\pm0.04$ & 10.0 & x & x\\
	& $0.992257\pm0.000007$ & $1.007804\pm0.000007$ & $0.99\pm0.01$ & $0.33\pm0.02$ & 13.5 & x & x\\
	& $1.01888\pm0.00003$ & $0.98147\pm0.00003$ & $0.25\pm0.01$ & $0.31\pm0.09$ & 5.7 & x & x\\
	& $1.72569\pm0.00004$ & $0.57948\pm0.00001$ & $0.18\pm0.01$ & $0.8\pm0.1$ & 6.5 &  & \\
	& $1.80852\pm0.00003$ & $0.552939\pm0.000008$ & $0.23\pm0.01$ & $0.15\pm0.09$ & 9.6 &  & \\
	& $4.60794\pm0.00009$ & $0.217017\pm0.000004$ & $0.07\pm0.01$ & $0.6\pm0.3$ & 9.7 &  & \\
    \hline
    \end{tabular}
    \label{tab:app_7760680}
\end{table*}

\begin{table*}
    \caption{Same as Table~\ref{tab:app_3240411} but for KIC~8766405.}
    \centering
    \begin{tabular}{ccccccc}
    \hline
    no. chunks & $f$ / d$^{-1}$ & $P$ / d & $A$ / mmag & $\phi$ & SNR & Pedersen pattern\\
    \hline
	1 & $0.1475\pm0.0002$ & $6.780\pm0.009$ & $0.26\pm0.02$ & $0.4\pm0.7$ & 5.8 & \\
	& $0.8497\pm0.0002$ & $1.1769\pm0.0002$ & $0.29\pm0.02$ & $0.0\pm0.6$ & 6.7 & x\\
	& $0.86677\pm0.00004$ & $1.15371\pm0.00006$ & $1.22\pm0.02$ & $0.4\pm0.2$ & 23.0 & x\\
	& $1.83208\pm0.00002$ & $0.545827\pm0.000005$ & $3.26\pm0.02$ & $0.59\pm0.06$ & 43.3 & \\
	& $1.97703\pm0.00005$ & $0.50581\pm0.00001$ & $1.06\pm0.02$ & $0.2\pm0.2$ & 34.6 & \\
	& $2.8438\pm0.0003$ & $0.35164\pm0.00004$ & $0.16\pm0.02$ & $1\pm1$ & 7.7 & \\
	& $3.6666\pm0.0001$ & $0.272735\pm0.000007$ & $0.52\pm0.02$ & $0.8\pm0.3$ & 27.2 & \\
	& $3.7111\pm0.0007$ & $0.26946\pm0.00005$ & $0.07\pm0.02$ & $1\pm1$ & 5.9 & \\
	& $3.8104\pm0.0004$ & $0.26244\pm0.00003$ & $0.13\pm0.02$ & $1\pm1$ & 9.5 & \\
	& $3.9543\pm0.0007$ & $0.25289\pm0.00004$ & $0.07\pm0.02$ & $1\pm1$ & 6.1 & \\
	& $5.6852\pm0.0007$ & $0.17590\pm0.00002$ & $0.07\pm0.02$ & $1\pm1$ & 8.4 & \\
    \\
	2 & $0.14916\pm0.00009$ & $6.704\pm0.004$ & $0.20\pm0.01$ & $0.5\pm0.3$ & 5.6 & \\
	& $0.85017\pm0.00004$ & $1.17623\pm0.00005$ & $0.47\pm0.01$ & $0.4\pm0.1$ & 13.6 & x\\
	& $0.86488\pm0.00001$ & $1.15623\pm0.00002$ & $1.40\pm0.01$ & $0.05\pm0.04$ & 32.1 & x\\
	& $1.7349\pm0.0001$ & $0.57639\pm0.00004$ & $0.15\pm0.01$ & $0.2\pm0.4$ & 7.2 & \\
	& $1.81701\pm0.00004$ & $0.55036\pm0.00001$ & $0.43\pm0.01$ & $0.0\pm0.1$ & 20.0 & \\
	& $1.832376\pm0.000005$ & $0.545739\pm0.000002$ & $3.15\pm0.01$ & $0.55\pm0.02$ & 57.0 & \\
	& $1.97716\pm0.00002$ & $0.505776\pm0.000004$ & $1.06\pm0.01$ & $0.78\pm0.06$ & 42.6 & \\
	& $2.8421\pm0.0001$ & $0.35186\pm0.00001$ & $0.17\pm0.01$ & $1.0\pm0.3$ & 11.5 & \\
	& $3.6514\pm0.0002$ & $0.27387\pm0.00001$ & $0.11\pm0.01$ & $0.9\pm0.6$ & 12.7 & \\
	& $3.66461\pm0.00003$ & $0.272880\pm0.000003$ & $0.53\pm0.01$ & $0.7\pm0.1$ & 37.3 & \\
	& $3.6787\pm0.0002$ & $0.27184\pm0.00002$ & $0.09\pm0.01$ & $0.2\pm0.7$ & 9.7 & \\
	& $3.7071\pm0.0002$ & $0.26976\pm0.00002$ & $0.08\pm0.01$ & $0.3\pm0.8$ & 9.5 & \\
	& $3.8096\pm0.0001$ & $0.262497\pm0.000009$ & $0.13\pm0.01$ & $0.4\pm0.4$ & 12.8 & \\
	& $3.9541\pm0.0002$ & $0.25290\pm0.00002$ & $0.07\pm0.01$ & $0.7\pm0.8$ & 8.0 & \\
	& $5.4832\pm0.0003$ & $0.18238\pm0.00001$ & $0.05\pm0.01$ & $1\pm1$ & 7.8 & \\
	& $5.4975\pm0.0004$ & $0.18190\pm0.00001$ & $0.04\pm0.01$ & $1\pm1$ & 6.3 & \\
	& $5.6844\pm0.0002$ & $0.175920\pm0.000007$ & $0.08\pm0.01$ & $0.8\pm0.8$ & 11.7 & \\
    \\
	3 & $0.14929\pm0.00003$ & $6.698\pm0.001$ & $0.16\pm0.01$ & $0.0\pm0.1$ & 5.3 & \\
	& $0.84986\pm0.00001$ & $1.17666\pm0.00002$ & $0.41\pm0.01$ & $0.49\pm0.04$ & 15.3 & x\\
	& $0.864969\pm0.000004$ & $1.156111\pm0.000005$ & $1.46\pm0.01$ & $0.75\pm0.01$ & 42.0 & x\\
	& $1.73525\pm0.00003$ & $0.57628\pm0.00001$ & $0.15\pm0.01$ & $0.1\pm0.1$ & 8.5 & \\
	& $1.81725\pm0.00001$ & $0.550283\pm0.000004$ & $0.40\pm0.01$ & $0.17\pm0.04$ & 20.3 & \\
	& $1.832309\pm0.000002$ & $0.5457596\pm0.0000006$ & $3.09\pm0.01$ & $0.785\pm0.007$ & 65.5 & \\
	& $1.83973\pm0.00003$ & $0.543557\pm0.000008$ & $0.23\pm0.01$ & $0.69\pm0.09$ & 13.2 & \\
	& $1.977268\pm0.000005$ & $0.505748\pm0.000001$ & $1.12\pm0.01$ & $0.40\pm0.01$ & 56.6 & \\
	& $2.84221\pm0.00003$ & $0.351839\pm0.000004$ & $0.18\pm0.01$ & $0.46\pm0.09$ & 13.6 & \\
	& $3.65848\pm0.00004$ & $0.273337\pm0.000003$ & $0.21\pm0.01$ & $0.2\pm0.1$ & 26.9 & \\
	& $3.66482\pm0.00002$ & $0.272865\pm0.000002$ & $0.36\pm0.01$ & $0.05\pm0.07$ & 33.3 & \\
	& $3.80963\pm0.00004$ & $0.262493\pm0.000003$ & $0.13\pm0.01$ & $0.2\pm0.1$ & 14.2 & \\
	& $3.95460\pm0.00008$ & $0.252870\pm0.000005$ & $0.07\pm0.01$ & $0.8\pm0.3$ & 8.3 & \\
	& $5.4830\pm0.0001$ & $0.182380\pm0.000004$ & $0.05\pm0.01$ & $0.3\pm0.4$ & 9.2 & \\
	& $5.68449\pm0.00006$ & $0.175917\pm0.000002$ & $0.08\pm0.01$ & $0.4\pm0.2$ & 15.3 & \\
    \hline
    \end{tabular}
    \label{tab:app_8766405}
\end{table*}

\begin{table*}
    \caption{Same as Table~\ref{tab:app_3240411} but for KIC~11360704.}
    \centering
    \begin{tabular}{ccccccc}
    \hline
    no. chunks & $f$ / d$^{-1}$ & $P$ / d & $A$ / mmag & $\phi$ & SNR & Pedersen pattern\\
    \hline
	1 & $0.3405\pm0.0002$ & $2.937\pm0.002$ & $0.45\pm0.03$ & $1.0\pm0.7$ & 7.3 & \\
	& $0.3876\pm0.0001$ & $2.5797\pm0.0009$ & $0.68\pm0.03$ & $0.8\pm0.5$ & 8.9 & \\
	& $0.5452\pm0.0002$ & $1.8341\pm0.0007$ & $0.47\pm0.03$ & $0.9\pm0.7$ & 7.6 & \\
	& $1.9119\pm0.0002$ & $0.52303\pm0.00004$ & $0.59\pm0.03$ & $0.8\pm0.6$ & 11.1 & \\
	& $2.0525\pm0.0001$ & $0.48721\pm0.00003$ & $0.71\pm0.03$ & $0.2\pm0.5$ & 11.6 & x\\
	& $2.07345\pm0.00007$ & $0.48229\pm0.00002$ & $1.25\pm0.03$ & $0.0\pm0.3$ & 17.9 & x\\
	& $2.46349\pm0.00008$ & $0.40593\pm0.00001$ & $1.19\pm0.03$ & $0.6\pm0.3$ & 15.3 & \\
	& $3.9827\pm0.0003$ & $0.25109\pm0.00002$ & $0.31\pm0.03$ & $1\pm1$ & 5.5 & \\
	& $4.1029\pm0.0002$ & $0.24373\pm0.00001$ & $0.46\pm0.03$ & $0.8\pm0.7$ & 7.1 & \\
	& $4.1301\pm0.0003$ & $0.24213\pm0.00001$ & $0.37\pm0.03$ & $0.5\pm0.9$ & 5.6 & \\
	& $4.1638\pm0.0003$ & $0.24016\pm0.00001$ & $0.37\pm0.03$ & $0.3\pm0.9$ & 6.2 & \\
	& $4.3263\pm0.0003$ & $0.23115\pm0.00002$ & $0.30\pm0.03$ & $1\pm1$ & 5.8 & \\
	& $4.3724\pm0.0002$ & $0.228707\pm0.000009$ & $0.60\pm0.04$ & $0.8\pm0.6$ & 10.0 & \\
	& $4.3789\pm0.0001$ & $0.228368\pm0.000006$ & $0.90\pm0.04$ & $0.6\pm0.4$ & 14.7 & \\
	& $4.5154\pm0.0004$ & $0.22146\pm0.00002$ & $0.26\pm0.03$ & $1\pm1$ & 5.4 & \\
	& $4.9271\pm0.0001$ & $0.202958\pm0.000005$ & $0.78\pm0.03$ & $0.5\pm0.4$ & 12.7 & \\
    \\
	2 & $0.34222\pm0.00008$ & $2.9221\pm0.0007$ & $0.41\pm0.02$ & $0.9\pm0.3$ & 10.0 & \\
	& $0.38871\pm0.00005$ & $2.5726\pm0.0003$ & $0.62\pm0.02$ & $0.0\pm0.2$ & 12.2 & \\
	& $0.54321\pm0.00006$ & $1.8409\pm0.0002$ & $0.52\pm0.02$ & $0.1\pm0.2$ & 12.3 & \\
	& $1.7780\pm0.0001$ & $0.56244\pm0.00004$ & $0.25\pm0.02$ & $0.5\pm0.4$ & 6.8 & \\
	& $1.90995\pm0.00008$ & $0.52357\pm0.00002$ & $0.41\pm0.02$ & $0.8\pm0.3$ & 10.5 & \\
	& $1.9722\pm0.0002$ & $0.50704\pm0.00004$ & $0.21\pm0.02$ & $0.8\pm0.5$ & 6.2 & \\
	& $1.9825\pm0.0001$ & $0.50440\pm0.00003$ & $0.27\pm0.02$ & $0.9\pm0.4$ & 7.4 & \\
	& $2.05190\pm0.00007$ & $0.48735\pm0.00002$ & $0.45\pm0.02$ & $0.3\pm0.2$ & 11.0 & x\\
	& $2.0676\pm0.0001$ & $0.48364\pm0.00002$ & $0.36\pm0.02$ & $0.6\pm0.3$ & 10.2 & \\
	& $2.07643\pm0.00004$ & $0.481596\pm0.000009$ & $1.10\pm0.03$ & $0.5\pm0.1$ & 22.1 & x\\
	& $2.08239\pm0.00009$ & $0.48022\pm0.00002$ & $0.40\pm0.03$ & $1.0\pm0.3$ & 12.1 & \\
	& $2.46343\pm0.00003$ & $0.405938\pm0.000005$ & $1.10\pm0.02$ & $0.8\pm0.1$ & 23.6 & \\
	& $3.8303\pm0.0002$ & $0.26107\pm0.00001$ & $0.17\pm0.02$ & $0.8\pm0.6$ & 5.5 & \\
	& $3.9816\pm0.0001$ & $0.251157\pm0.000006$ & $0.33\pm0.02$ & $0.6\pm0.3$ & 8.0 & \\
	& $4.0010\pm0.0002$ & $0.24993\pm0.00001$ & $0.20\pm0.02$ & $0.5\pm0.6$ & 5.8 & \\
	& $4.10261\pm0.00008$ & $0.243747\pm0.000005$ & $0.40\pm0.02$ & $0.7\pm0.3$ & 9.3 & \\
	& $4.1284\pm0.0001$ & $0.242226\pm0.000006$ & $0.32\pm0.02$ & $0.6\pm0.3$ & 8.0 & \\
	& $4.1642\pm0.0001$ & $0.240143\pm0.000007$ & $0.28\pm0.02$ & $0.1\pm0.4$ & 7.5 & \\
	& $4.3282\pm0.0001$ & $0.231041\pm0.000006$ & $0.27\pm0.02$ & $0.6\pm0.4$ & 7.3 & \\
	& $4.36822\pm0.00008$ & $0.228926\pm0.000004$ & $0.43\pm0.02$ & $0.7\pm0.3$ & 10.9 & \\
	& $4.37874\pm0.00004$ & $0.228376\pm0.000002$ & $0.92\pm0.02$ & $0.2\pm0.1$ & 19.3 & \\
	& $4.3892\pm0.0001$ & $0.227833\pm0.000006$ & $0.26\pm0.02$ & $0.9\pm0.4$ & 7.3 & \\
	& $4.5346\pm0.0001$ & $0.220528\pm0.000007$ & $0.24\pm0.02$ & $0.0\pm0.5$ & 6.7 & \\
	& $4.92693\pm0.00004$ & $0.202966\pm0.000002$ & $0.73\pm0.02$ & $0.2\pm0.1$ & 18.0 & \\
    \\
	3 & $0.12105\pm0.00003$ & $8.261\pm0.002$ & $0.65\pm0.03$ & $1.0\pm0.1$ & 5.4 & \\
	& $0.12604\pm0.00003$ & $7.934\pm0.002$ & $0.59\pm0.03$ & $0.8\pm0.1$ & 5.6 & \\
	& $0.34218\pm0.00003$ & $2.9224\pm0.0003$ & $0.48\pm0.03$ & $1.00\pm0.09$ & 5.9 & \\
	& $0.38851\pm0.00002$ & $2.5740\pm0.0001$ & $1.19\pm0.03$ & $0.78\pm0.05$ & 10.8 & \\
	& $0.39362\pm0.00003$ & $2.5405\pm0.0002$ & $0.62\pm0.03$ & $0.2\pm0.1$ & 7.4 & \\
	& $1.90999\pm0.00002$ & $0.523564\pm0.000006$ & $0.71\pm0.03$ & $0.74\pm0.07$ & 13.1 & \\
	& $1.91686\pm0.00003$ & $0.521686\pm0.000009$ & $0.46\pm0.03$ & $0.2\pm0.1$ & 9.3 & \\
	& $2.02430\pm0.00004$ & $0.49400\pm0.00001$ & $0.35\pm0.03$ & $0.4\pm0.1$ & 7.5 & \\
	& $2.05612\pm0.00003$ & $0.486354\pm0.000006$ & $0.60\pm0.03$ & $0.20\pm0.08$ & 11.7 & x\\
	& $2.07148\pm0.00002$ & $0.482747\pm0.000005$ & $0.77\pm0.03$ & $0.00\pm0.06$ & 12.9 & x\\
	& $2.46311\pm0.00001$ & $0.405991\pm0.000002$ & $1.02\pm0.03$ & $0.96\pm0.04$ & 17.0 & \\
	& $3.98236\pm0.00004$ & $0.251107\pm0.000002$ & $0.40\pm0.03$ & $0.8\pm0.1$ & 6.8 & \\
	& $4.09747\pm0.00003$ & $0.244053\pm0.000002$ & $0.43\pm0.03$ & $0.9\pm0.1$ & 6.8 & \\
	& $4.12869\pm0.00003$ & $0.242207\pm0.000002$ & $0.53\pm0.03$ & $0.48\pm0.09$ & 8.0 & \\
	& $4.15939\pm0.00004$ & $0.240420\pm0.000002$ & $0.38\pm0.03$ & $0.1\pm0.1$ & 6.6 & \\
	& $4.37307\pm0.00002$ & $0.2286724\pm0.0000009$ & $0.84\pm0.03$ & $0.33\pm0.06$ & 12.4 & \\
	& $4.38423\pm0.00003$ & $0.228090\pm0.000001$ & $0.52\pm0.03$ & $0.65\pm0.09$ & 7.8 & \\
	& $4.92771\pm0.00002$ & $0.2029339\pm0.0000009$ & $0.63\pm0.03$ & $0.52\pm0.07$ & 9.9 & \\
    \hline
    \end{tabular}
    \label{tab:app_11360704}
\end{table*}

\section{Amplitude spectra and modelling figures}
\label{sec:app_figures}

Figures showing pre-whitened spectra of the three chunk light curves for all of our stars are listed here, as well as for the two chunk light curve of KIC~7760680. The frequencies are shown as coloured dashed lines, with amplitudes given by the height of the overplotted solid lines. The colours are consistent between the frequency and period panels. For stars which we then modelled, period spacing patterns with fits from {\sc amigo} and an HR~diagram showing the positions of models within the 68$^{\rm th}$ percentile of the probability distribution, which we use to calculate the 1$\sigma$ confidence intervals on the best fit model. The best model is marked on the HR~diagram with a white cross. For KIC~4930889 and KIC~7760680, there are forward asteroseismic modelling results in addition to \citet{pedersen2021} that we compare to, which are shown in additional subplots for these stars.

\begin{figure*}
    \centering
    \includegraphics[width=0.8\linewidth]{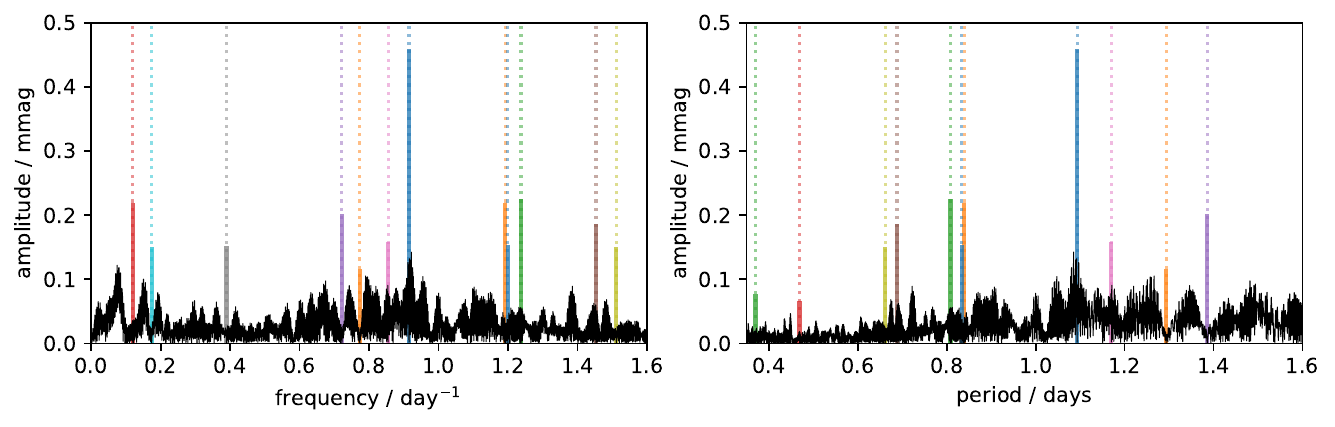}
    \caption{Amplitude spectrum of KIC~3240411 as a function of frequency (left panel) and as a function of period (right panel).}
    \label{fig:app_3240411}
\end{figure*}

\begin{figure*}
    \centering
    \includegraphics[width=0.8\linewidth]{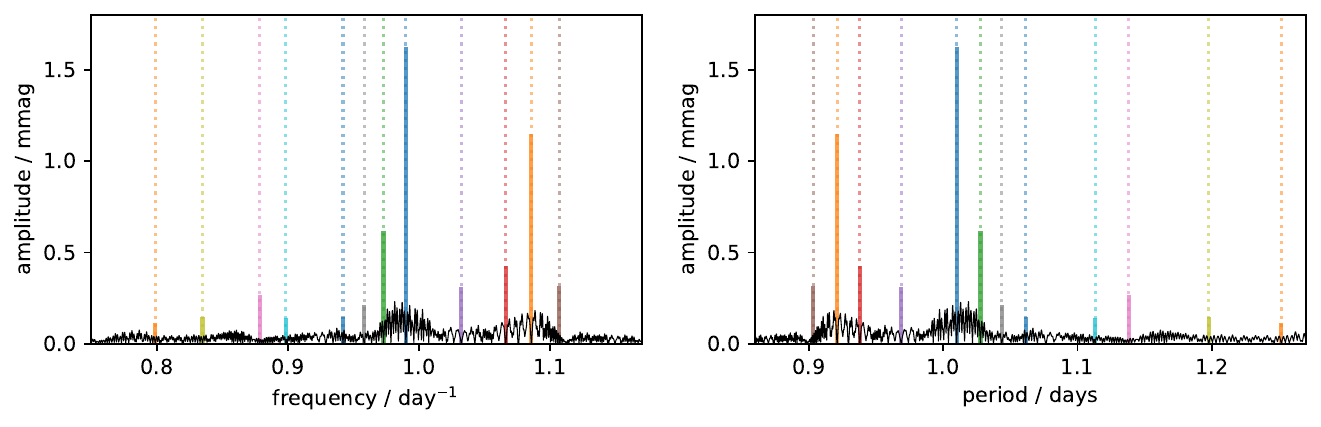}\\
    \includegraphics[width=0.4\linewidth]{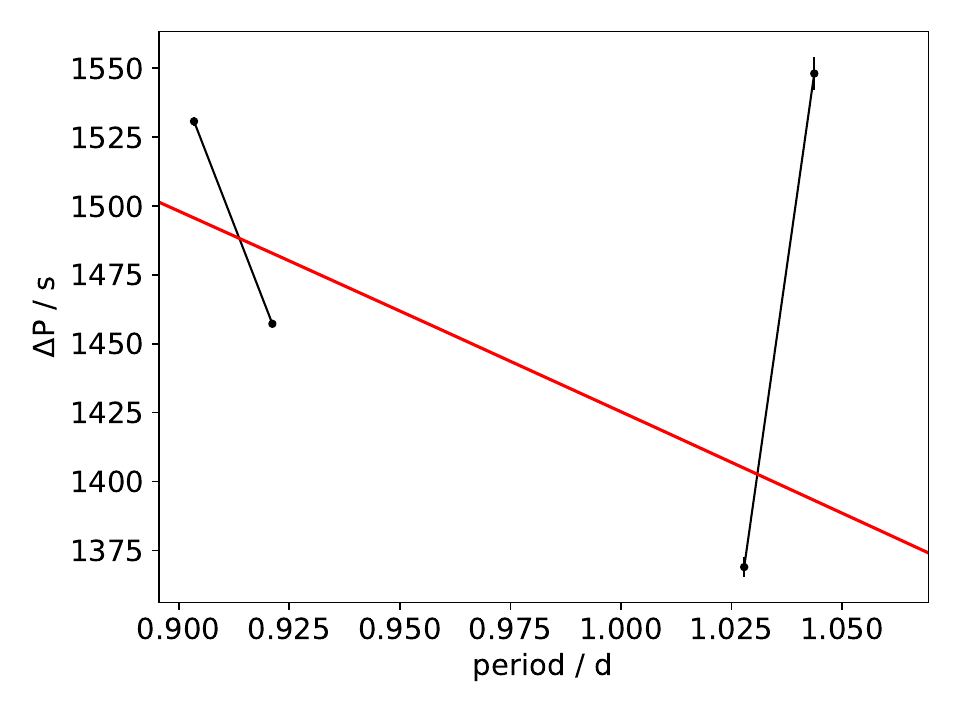}
    \includegraphics[width=0.45\linewidth]{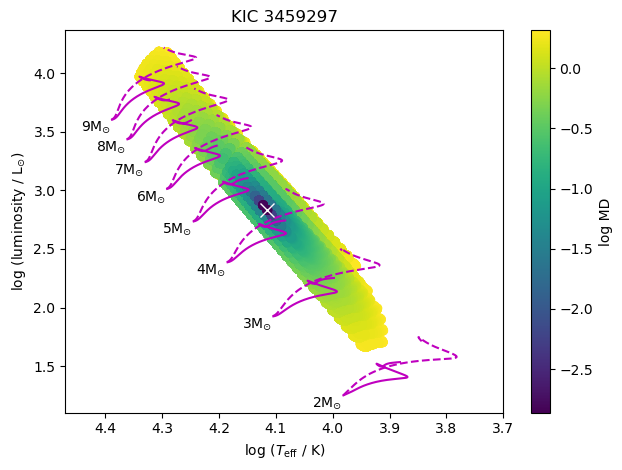}
    \caption{Summary of results for KIC~3459297. Top left: Amplitude spectrum versus frequency. Top right: Amplitude spectrum versus period. Bottom left: Period spacing pattern with the fit from \texttt{AMiGO}. Bottom right: HR~diagram showing the best fit model as a white cross. Also shown are the Mahalanobis distance of the grid models within the 68$^{\rm th}$ percentile of the probability distribution (colour bar) and evolutionary tracks with $f_{\rm{CBM}}=0.005$ and 0.04 (solid and dashed lines respectively).}
    \label{fig:app_3459297}
\end{figure*}

\begin{figure*}
    \centering
    \includegraphics[width=0.8\linewidth]{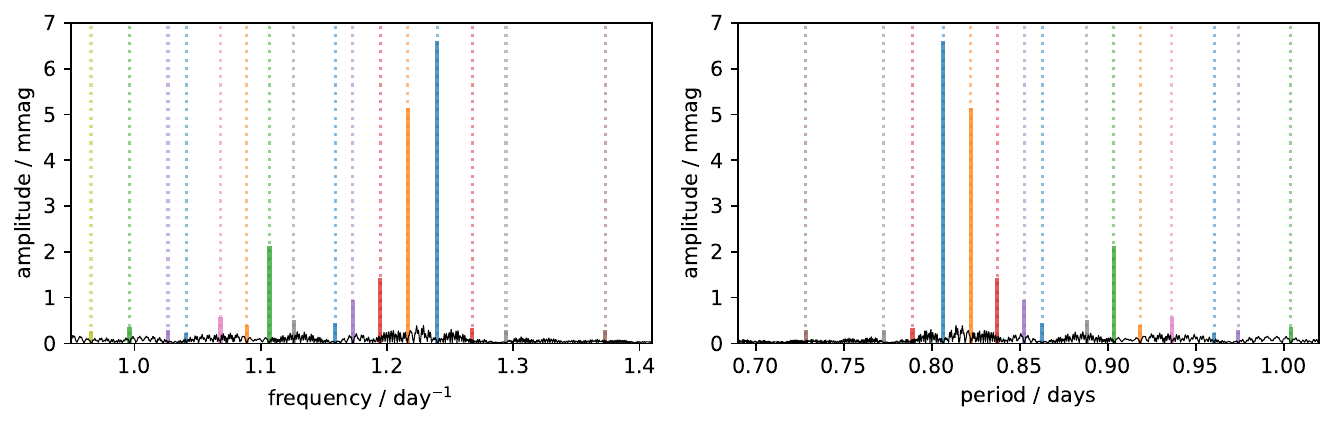}\\
    \includegraphics[width=0.4\linewidth]{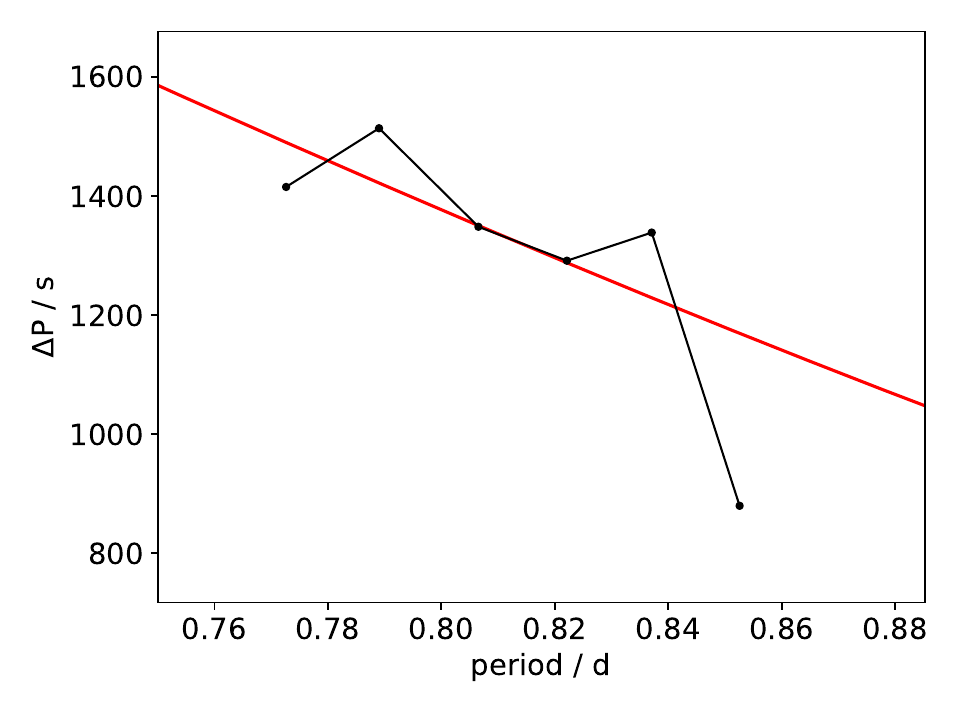}
    \includegraphics[width=0.45\linewidth]{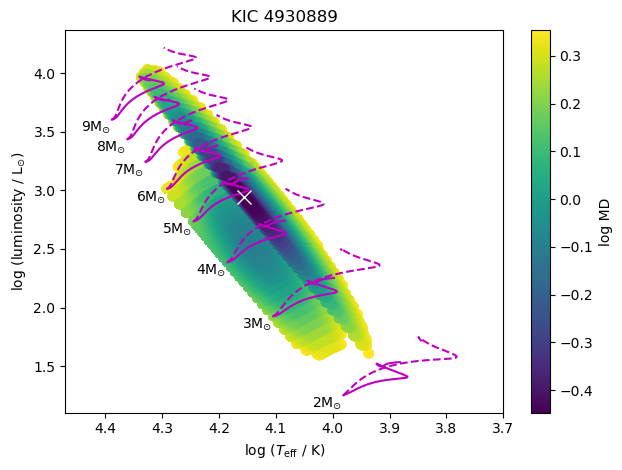}\\
    \includegraphics[width=\linewidth]{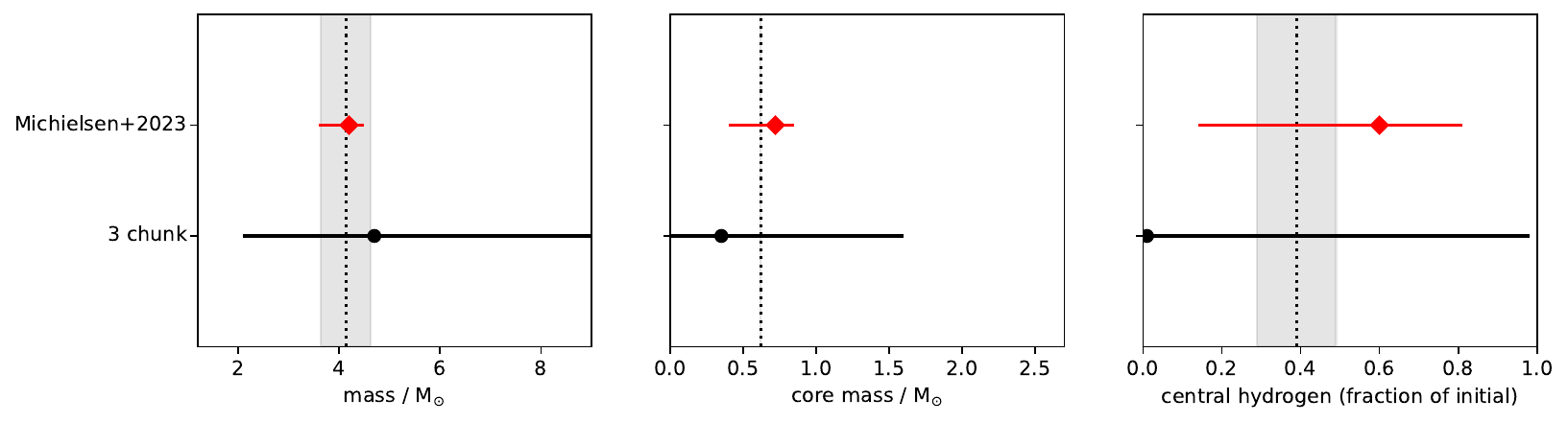}
    \caption{KIC\,4930889. Top left: Amplitude spectrum versus frequency. Top right: Amplitude spectrum versus period. Middle left: Period spacing pattern with the fit from \texttt{AMiGO}. Middle right: Hertzsprung-Russell diagram showing the best fit model as a white cross. Also shown are the Mahalanobis distance of the grid models within the 68$^{\rm th}$ percentile of the probability distribution (colour bar) and evolutionary tracks with $f_{\rm{CBM}}=0.005$ and 0.04 (solid and dashed lines respectively). Bottom panels: Model fitting results. Red diamonds come from \citet{michielsen2023} based on \textit{Kepler} data, and correspond to the lowest MD best fit model using period spacings (their table~5).}
    \label{fig:app_4930889}
\end{figure*}

\begin{figure*}
    \centering
    \includegraphics[width=0.8\linewidth]{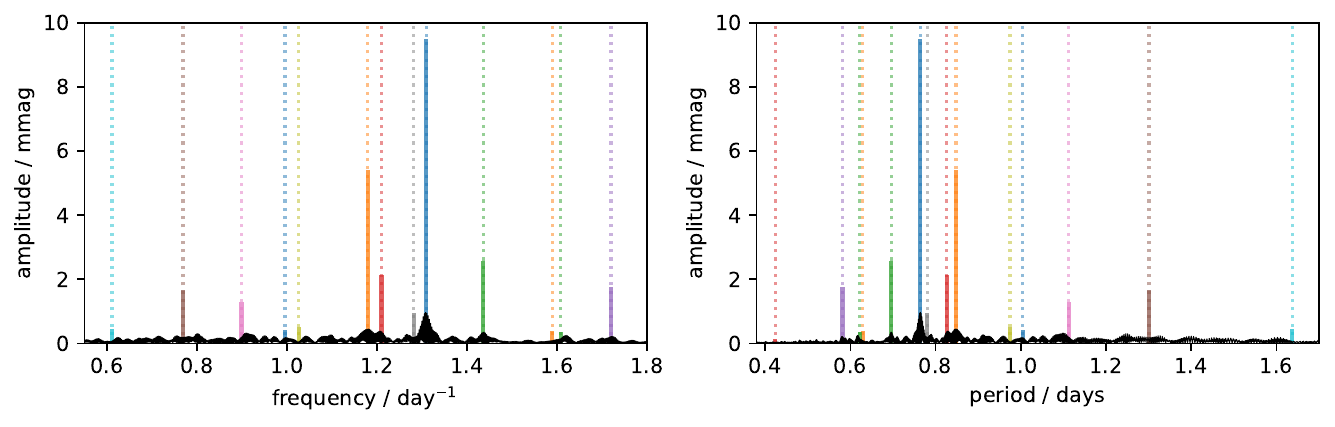}\\
    \includegraphics[width=0.4\linewidth]{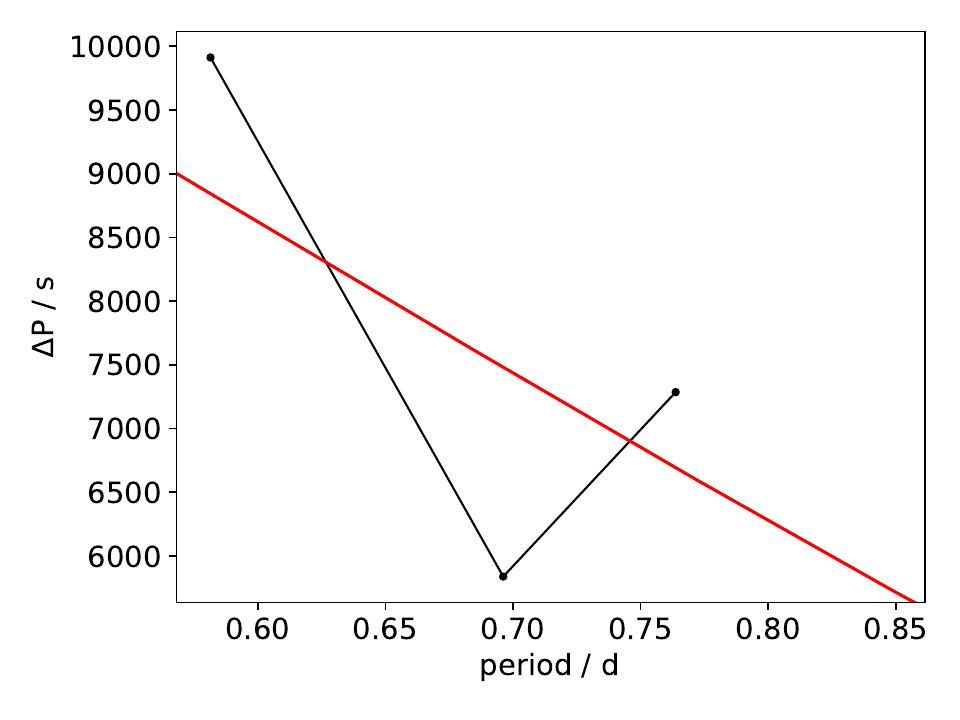}
    \includegraphics[width=0.45\linewidth]{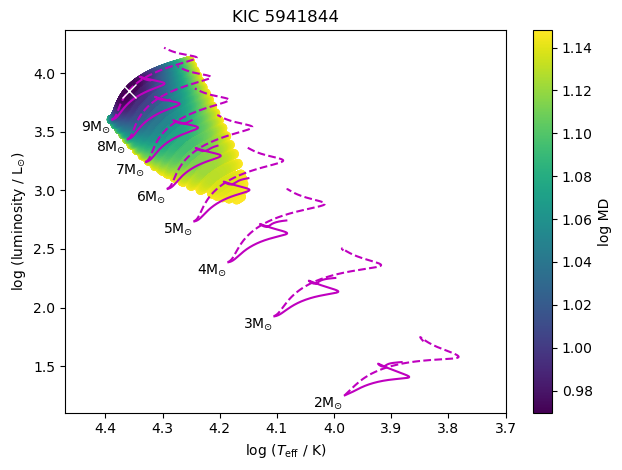}\\
    \caption{As Fig.~\ref{fig:app_3459297}, but for KIC~5941844.}
\end{figure*}

\begin{figure*}
    \centering
    \includegraphics[width=0.8\linewidth]{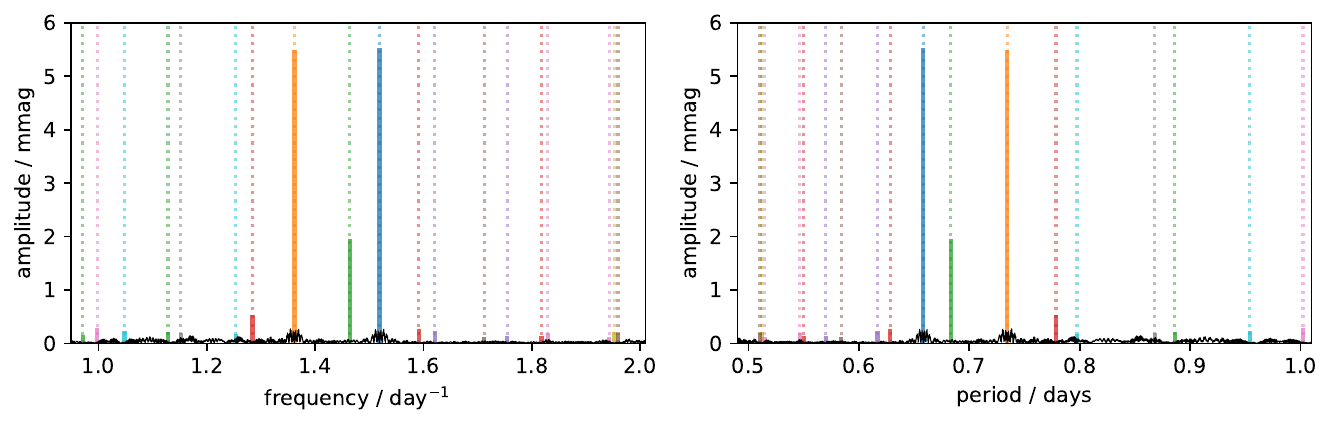}\\
    \includegraphics[width=0.4\linewidth]{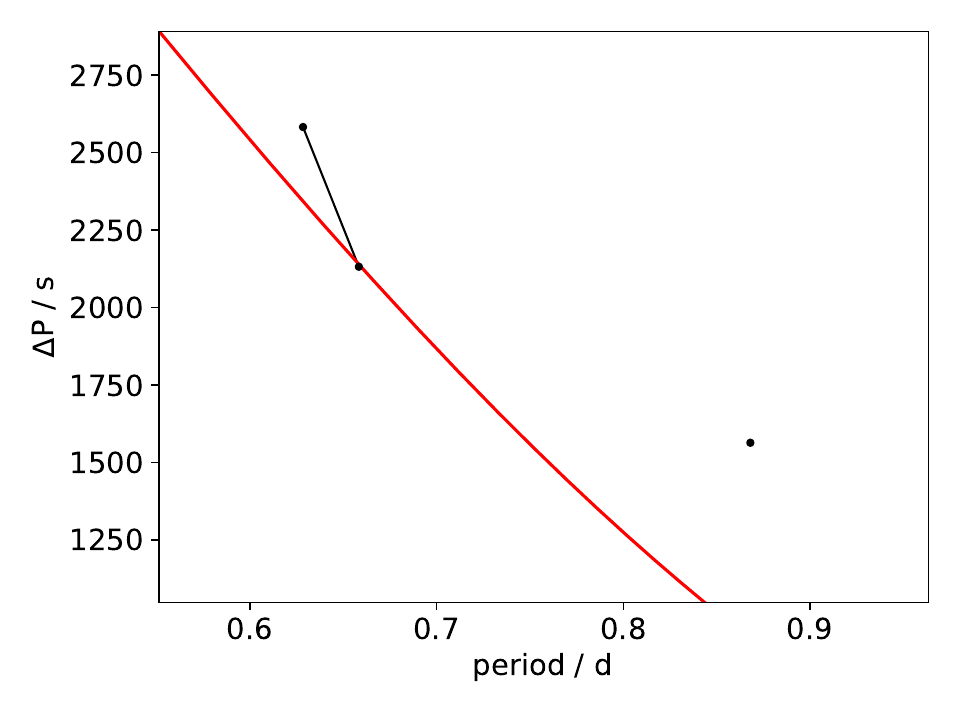}
    \includegraphics[width=0.45\linewidth]{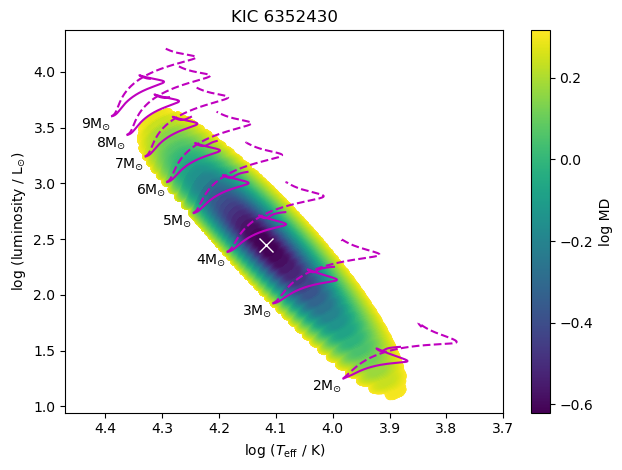}
    \caption{As Fig.~\ref{fig:app_3459297}, but for KIC~6352430.}
\end{figure*}

\begin{figure*}
    \centering
    \includegraphics[width=0.8\linewidth]{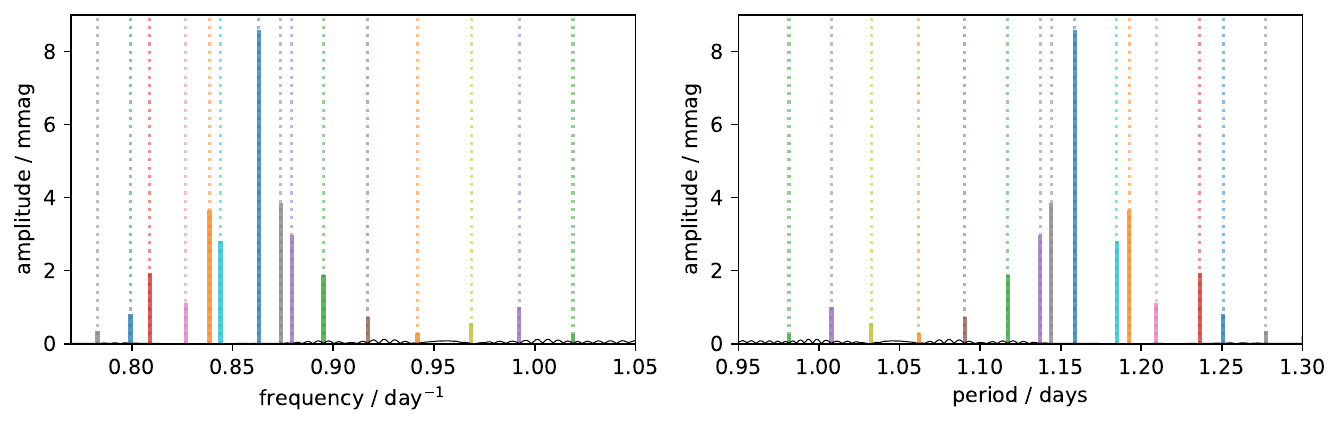}\\
    \includegraphics[width=0.4\linewidth]{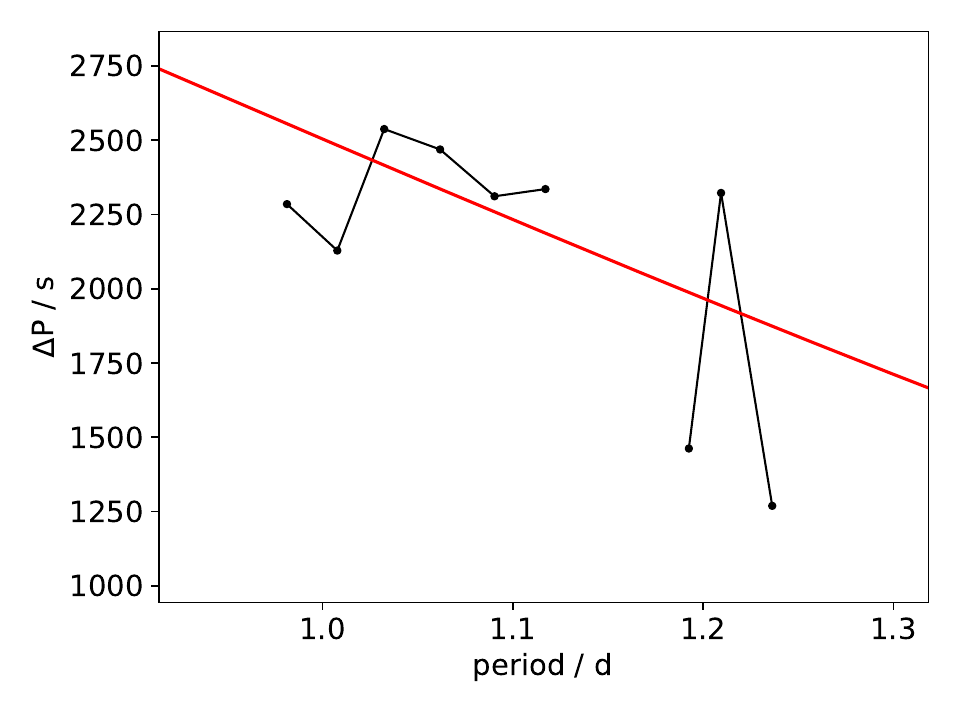}
    \includegraphics[width=0.45\linewidth]{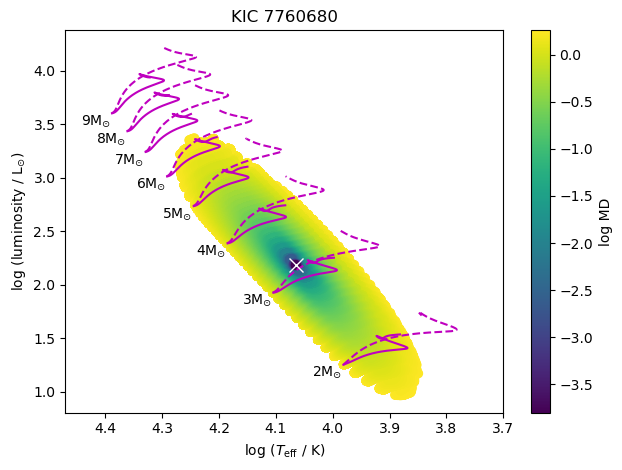}
    \caption{As Fig.~\ref{fig:app_3459297}, but for the 2 chunk KIC~7760680 light curve.}
    \label{fig:app_7760680_2c}
\end{figure*}

\begin{figure*}
    \centering
    \includegraphics[width=0.8\linewidth]{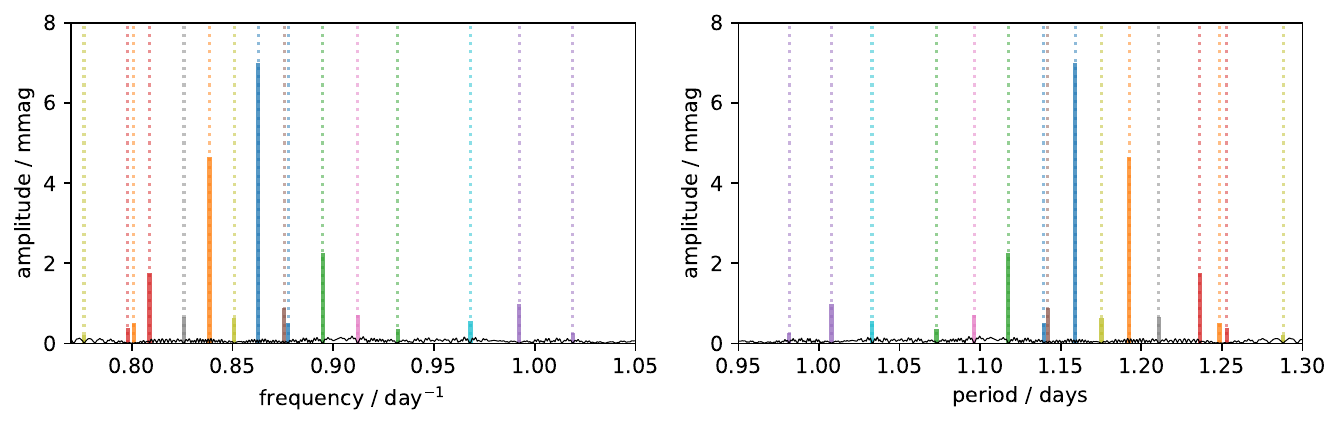}\\
    \includegraphics[width=0.4\linewidth]{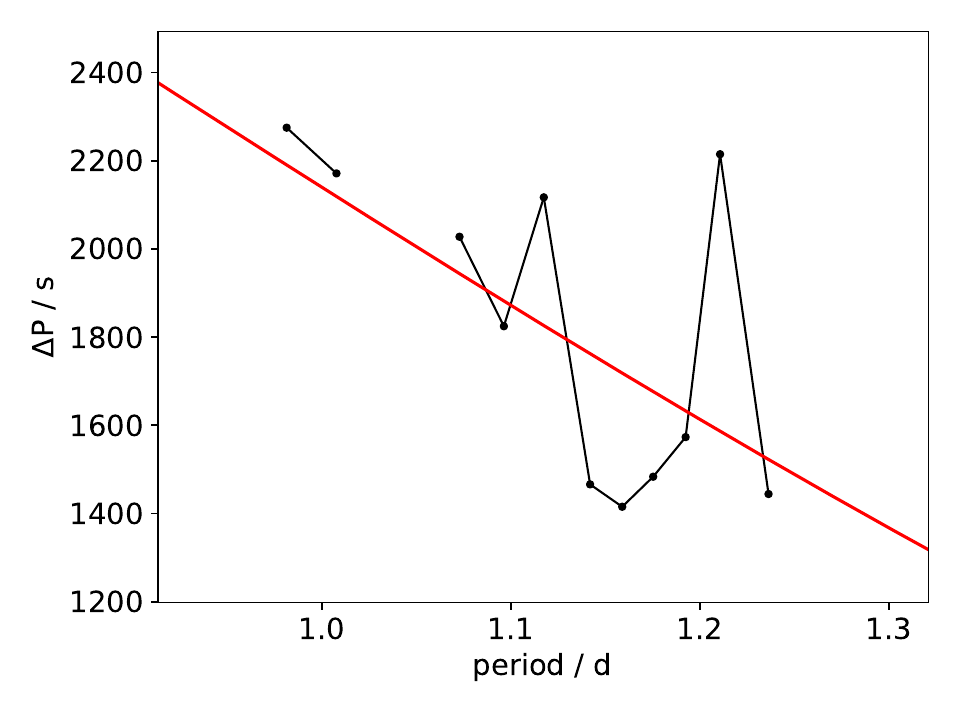}
    \includegraphics[width=0.45\linewidth]{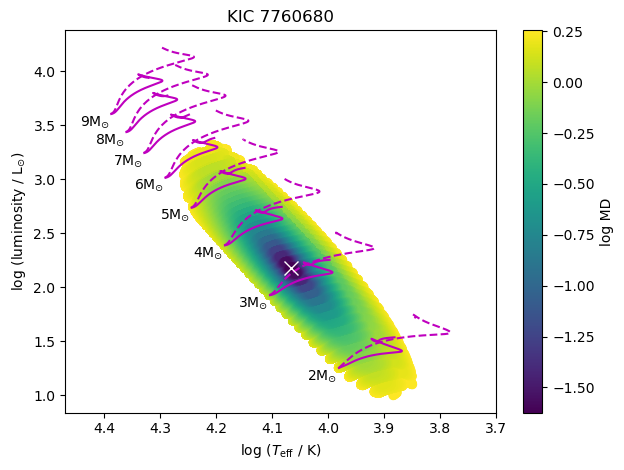}\\
    \includegraphics[width=\linewidth]{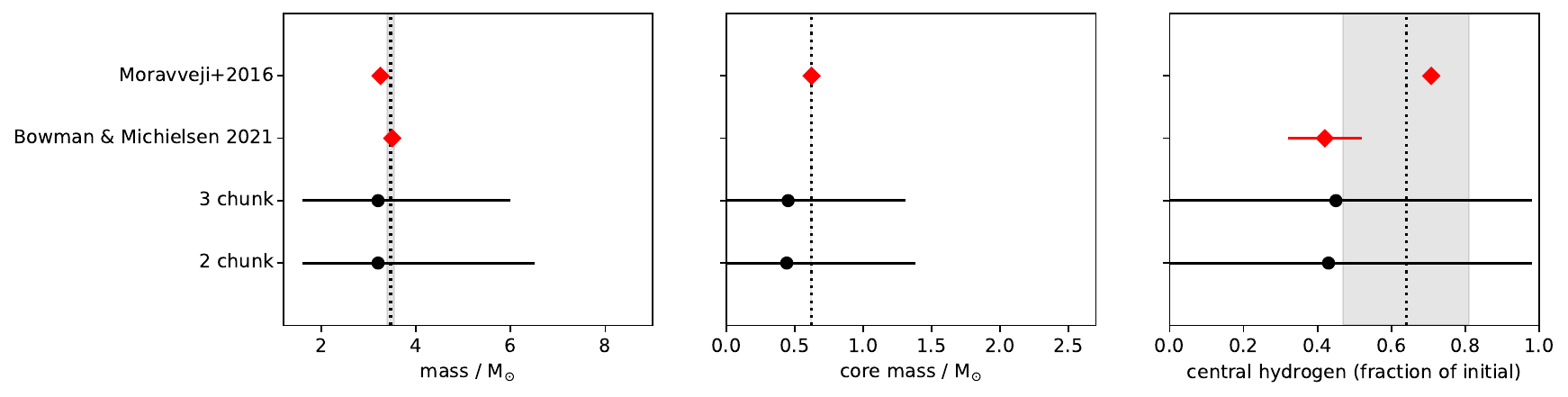}
    \caption{As Fig.~\ref{fig:app_4930889}, but for the 3 chunk KIC~7760680 light curve. Red diamonds in the lower panels are modelling results from \citet{moravveji2016} and \citet{bowman2021} based on {\it Kepler} light curves.}
    \label{fig:app_7760680}
\end{figure*}

\begin{figure*}
    \centering
    \includegraphics[width=0.8\linewidth]{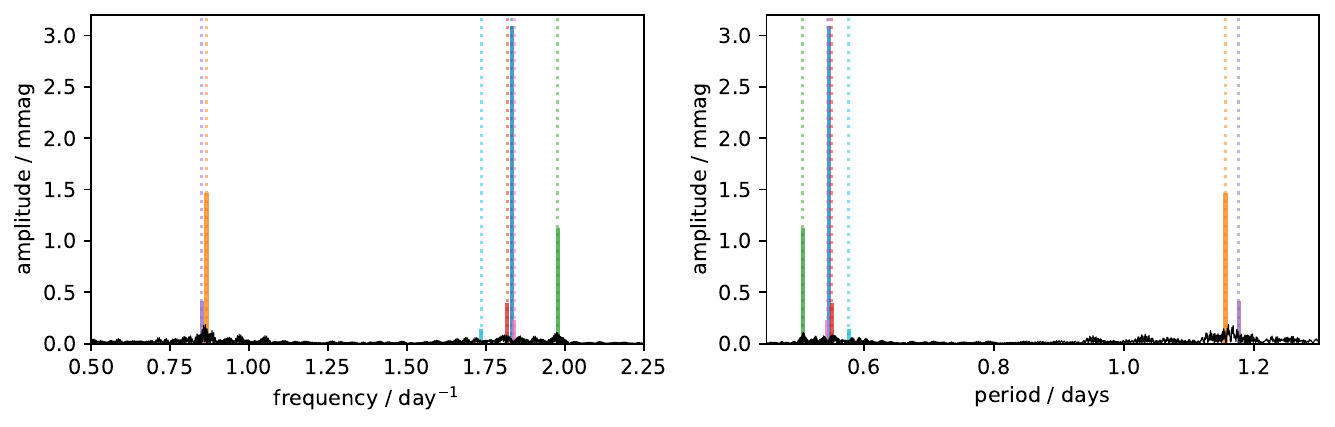}
    \caption{As Fig.~\ref{fig:app_3240411}, but for KIC~8766405.}
\end{figure*}

\begin{figure*}
    \centering
    \includegraphics[width=0.8\linewidth]{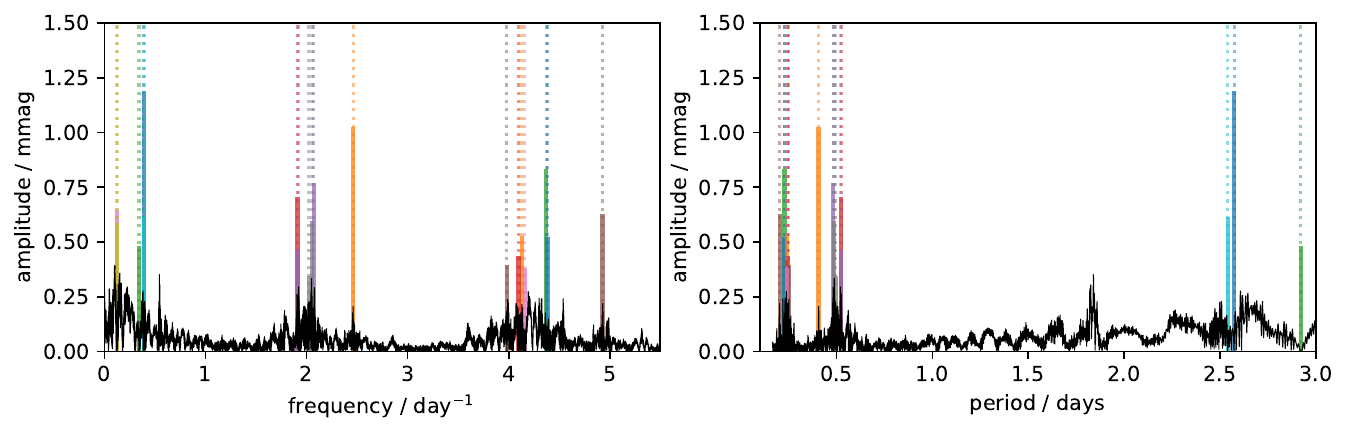}
    \caption{As Fig.~\ref{fig:app_3240411}, but for KIC 1136704.}
\end{figure*}

%%%%%%%%%%%%%%%%%%%%%%%%%%%%%%%%%%%%%%%%%%%%%%%%%%

% Don't change these lines
\bsp	% typesetting comment
\label{lastpage}
\end{document}